\documentclass[12pt,prd,a4paper,nofootinbib,preprintnumbers,floatfix]{revtex4}
\usepackage{graphicx,dcolumn}

\begin{document}

%%%%%%%%%%%%%%%%%%%%%%%%%%%%%%%%%%%%%%%%%%%%%%%%%%%%%%%%%%%%%%%%%%
\title{~~\\
Topology and chiral symmetry breaking in SU($N_c$) \\ 
gauge theories \\ ~~\\ }
%%%%%%%%%%%%%%%%%%%%%%%%%%%%%%%%%%%%%%%%%%%%%%%%%%%%%%%%%%%%%%%%%%

%%%%%%%%%%%%%%%%%%%%%%%%%%%%%%%%%%%%%%%%%%%%%%%%%%%%%%%%%%%%%%%%%%
\author{Nigel Cundy}
\author{Michael Teper}
\author{Urs Wenger}
\affiliation{Theoretical Physics, Oxford University, 1 Keble Road,\\
 Oxford OX1 3NP, United Kingdom}
%%%%%%%%%%%%%%%%%%%%%%%%%%%%%%%%%%%%%%%%%%%%%%%%%%%%%%%%%%%%%%%%%%

%\date{\today}

%%%%%%%%%%%%%%%%%%%%%%%%%%%%%%%%%%%%%%%%%%%%%%%%%%%%%%%%%%%%%%%%%%
\begin{abstract}
We study the low-lying eigenmodes of the lattice overlap Dirac
operator for SU$(N_c)$ gauge theories with $N_c=2,3,4$ and 5 colours.
We define a fermionic topological charge from the zero-modes of this
operator and show that, as $N_c$ grows, any disagreement with the
topological charge obtained by cooling the fields, becomes rapidly
less likely. By examining the fields where there is a disagreement,
we are able to show that the Dirac operator does not resolve
instantons below a critical size of about $\rho \simeq 2.5 a$, but
resolves the larger, more physical instantons. We investigate the
local chirality of the near-zero modes and how it changes as we go to
larger $N_c$. We observe that the local chirality of these modes,
which is prominent for SU(2) and SU(3), becomes rapidly weaker
for larger $N_c$ and is consistent with disappearing entirely
in the  limit of $N_c = \infty$. We find that this is not due to 
the observed disappearance of small instantons at larger $N_c$.
\end{abstract}
%%%%%%%%%%%%%%%%%%%%%%%%%%%%%%%%%%%%%%%%%%%%%%%%%%%%%%%%%%%%%%%%%%



\preprint{OUTP-02-14P}

\bibliographystyle{h-physrev4}

\maketitle

\newpage
%%%%%%%%%%%%%%%%%%%%%%%%%%%%%%%%%%%%%%%%%%%%%%%%%%%%%%%%%%%%%%%%%%
\section {Introduction} \label{sec:introduction}
%%%%%%%%%%%%%%%%%%%%%%%%%%%%%%%%%%%%%%%%%%%%%%%%%%%%%%%%%%%%%%%%%%

The chiral symmetry of QCD is broken in several different
ways. First, and most trivially, it is broken explicitly
by modest quark mass terms. In addition the symmetry
is broken spontaneously, leading to an octet of light
pseudoscalars in the observed spectrum. If there were
no quark mass terms these would be massless Goldstone
bosons. However the flavour singlet $\eta^{\prime}$ is much
too heavy to be a `near-Goldstone boson', indicating that the
associated U(1) axial symmetry is broken in some other
way; and in fact it is known to be anomalous.

The SU(3) gauge fields of QCD possess non-trivial topological
properties. In the semi-classical limit, on scales where one has
analytic control, we know that the topology is localised
in instantons. The anomaly is proportional to the topological 
charge density and one can show through an explicit calculation
\cite{'tHooft:1976fv:1986nc}
how instantons lead to a massive $\eta^{\prime}$. The mechanism
relies on the existence of zero-modes of the Dirac operator in 
a background instanton field. While this provides a
qualitative resolution of the U$_A$(1) problem, a complete
quantitative calculation is still lacking. 

In contrast to the anomalous breaking of the axial U(1) symmetry,
there is no necessary involvement of topology in the
spontaneous breaking of the remaining SU(3) chiral symmetry. 
Nonetheless it has long been speculated 
\cite{Caldi:1977rcCallan:1978gzCarlitz:1979yj}
that this might be so.
A simple motivation starts with the observation 
\cite{Banks:1980yr}
that the chiral
condensate is proportional to $\rho(\lambda=0)$,
the (normalised) density of eigenvalues of the Dirac
operator at zero eigenvalue. Now, we know that for a background
field with topological charge $Q$ the Dirac operator will
have (at least) $|Q|$ zero modes. These are however too few
to contribute to the chiral condensate in the infinite volume
limit, since a space-time volume $V$ will
contain  $O(V)$ topological charges of each sign and hence
(if $\theta=0$) a net topological charge $Q\sim O(\surd V)$.
However there will be $O(V)$ modes obtained through the
mixing of what would have been zero modes if the $O(V)$ 
topological charges had zero overlap with each other.
If this overlap is moderately small, the splitting from zero will
be small and one would expect the resulting mode
spectrum to have $\rho(\lambda=0)\not=0$, and hence to
contribute to chiral symmetry breaking. This idea receives
support from  specific instanton model calculations
\cite{Diakonov:1995eaSchafer:1998wv}.
(Indeed there is evidence that, in quenched QCD, $\rho(\lambda=0)$
is not just non-zero but may in fact be divergent
\cite{Dowrick:1995wrSharan:1998gj:1998vp:1999mg}.)

In practice, the topology of realistic gauge fields seems
to be carried not only by well-separated instantons but also,
and perhaps mainly, by larger
heavily overlapping fluctuations which are much less
tractable analytically. (See, for example, 
\cite{Smith:1998wt}.)
This should be no surprise in a 
theory with one length scale. Moreover there is a simple
argument 
\cite{Witten:1979kh}
that isolated instantons will not survive the
limit where the number of colours, $N_c$, becomes large:
at fixed 't Hooft coupling $g^2N_c$
\cite{'tHooft:1974jz},
the instanton weight includes a factor 
$\exp\{-8\pi^2/g^2\} = \exp\{-(8\pi^2/g^2N_c)N_c\}$
which disappears exponentially with increasing $N_c$.
This argument requires modification, but the conclusion
that small isolated instantons do not survive the 
$N_c\to\infty$ limit appears to remain valid
\cite{Teper:1980tq}.
On the other hand one believes that the chiral symmetry of 
QCD$_{N_c=\infty}$ is spontaneously broken. While it is
possible that the mechanisms for breaking chiral symmetry
are different for $N_c=3$ and $N_c=\infty$, it would
seem uneconomical of Nature to make use of such an option.

Whether topology drives part or all of the chiral symmetry
breaking and, if so, whether the relevant topological 
fluctuations are approximately instanton-like, are clearly
interesting questions and ones which it is natural to 
address by the simulation of the corresponding lattice field 
theories. Such calculations have long existed; for example in
\cite{Hands:1990wc}
the spectrum of the Dirac operator in the SU(2) gauge 
theory is shown to have a non-zero density at $\lambda=0$,
demonstrating the spontaneous breaking of the chiral symmetry.
In addition it is shown that one loses this chiral symmetry 
breaking if one excludes eigenmodes that have a topological
origin. Unfortunately
this interesting conclusion, that topology does indeed 
drive chiral symmetry breaking, can be criticised because
the staggered lattice fermions employed in 
\cite{Hands:1990wc}
suffer large lattice
spacing corrections, particularly where topological zero
modes are concerned.  Until recently a similar criticism could 
be made of any of the lattice fermions in common use. 
However we now know that lattice fermions which satisfy 
the Ginsparg-Wilson relation, such as overlap fermions
\cite{Narayanan:1993ss:1994sk:1995gwRandjbar-Daemi:1995sq:1995cq:1996mj:1997iz},
domain wall fermions
\cite{Kaplan:1992btShamir:1993zyFurman:1995ky}
or classically perfect fermions
\cite{Hasenfratz:1994spWiese:1993cbDeGrand:1995ji:1995jk:1997ncHasenfratz:2000xz}
have good chiral properties 
and satisfy exact index theorems at finite lattice spacing.
This opens the door to repeating the kind of calculation 
found in
\cite{Hands:1990wc}
but now on a much sounder theoretical footing. The present 
work is in this spirit and adds to a number of recent
papers on this topic
\cite{Horvath:2001ir,DeGrand:2000gq,Hip:2001hcEdwards:2001ndDeGrand:2001wxGattringer:2001mn,Horvath:2002gk}.

With staggered fermions, lattice artifacts shift the exact 
topological zero modes away from zero and so when the
would-be zero modes of neighbouring instantons and
anti-instantons mix, the mixing is, in general, far
from maximal. Thus, in practice, such 
mixed modes retain a substantial residual chirality,
$\langle \psi | \gamma_5 | \psi \rangle \not= 0$, which
can be used
\cite{Hands:1990wc}
to identify the modes that possess a topological origin.
With overlap lattice fermions, by contrast, the topological
zero modes remain zero on the lattice, so the mixing is
maximal and the chirality of non-zero modes is exactly zero,
$\langle \psi | \gamma_5 | \psi \rangle = 0$, just as it is 
in the continuum. It is thus not so simple to determine the
extent to which a $\lambda\not= 0$ mode has a topological origin.
If a mode is significantly influenced by an instanton of size
$\rho$, centered at the space-time point $x$, then one would 
expect that the chiral density would be positive in the
region of space-time, $v$, where the core of the instanton
is located, i.e.
\begin{equation} 
\int_v \psi^{\dagger}\gamma_5 \psi \not= 0.
\end{equation}
More generally, the chiral density of $\psi$ should be non-zero
in extended regions that match the locations of topological
charges. Moreover the size of any such region should be related to 
the size of the corresponding topological charge and the sign of 
the chiral density should match the sign of that charge. The
extent to which a nonzero mode can be described in this way
determines the extent to which it can be said to possess 
a topological origin.
  
How to make quantitative such an analysis of the low-lying 
eigenmodes of the Dirac operator is not evident. A simple
approach which has been used in much of the recent literature 
\cite{Horvath:2001ir,DeGrand:2000gq,Hip:2001hcEdwards:2001ndDeGrand:2001wxGattringer:2001mn,Horvath:2002gk}
focuses on the lattice sites at which the eigenmode
is large (using some reasonable cut-off) and asks how
chiral the mode is at those sites. Qualitatively, the more chiral 
it is, the more likely it is to have a topological origin.
A recently popular measure 
\cite{Horvath:2001ir,DeGrand:2000gq,Hip:2001hcEdwards:2001ndDeGrand:2001wxGattringer:2001mn,Horvath:2002gk}
is given by the quantity $X(x)$ \cite{Horvath:2001ir} defined via the
ratio of the positive and negative chiral densities
$\psi_{\uparrow}^\dagger \psi_{\uparrow}$ and
$\psi_{\downarrow}^\dagger \psi_{\downarrow}$ at the point $x$ through
\begin{equation}\label{eq:local_chirality_parameter_definition}
\tan \left( \frac{\pi}{4} (1+ X(x))\right) = 
\biggl( \frac{\psi_{\downarrow}^\dagger \psi_{\downarrow}}
{\psi_{\uparrow}^\dagger \psi_{\uparrow}} \biggr)^{1/2}\,.
\end{equation}
In this paper we shall present our results for this quantity,
and for some physically motivated variants,  
not only for the SU(3) gauge theory considered in
\cite{Horvath:2001ir,DeGrand:2000gq,Hip:2001hcEdwards:2001ndDeGrand:2001wxGattringer:2001mn,Horvath:2002gk}
but also for the SU(2), SU(4) and SU(5) gauge theories. In
this way we can simultaneously address the interesting question 
of what happens as $N_c \to \infty$.

A more sophisticated approach is to write a non-zero mode, $\psi$, 
as a linear combination of the would-be zero modes, $\psi^0_I$, 
associated with the (anti)instantons, $I$, in the background gauge 
field under consideration
\begin{equation}\label{eq:mode_mixing}
\psi =
\sum_I c_I \psi^0_I + \delta\psi.
\end{equation}
Such an approach has been pursued in  
\cite{DeGrand:2000gq}
using the semi-classical form of $\psi^0_I$ and a particular
algorithm to determine the topological charges in the gauge field.
Clearly there are ambiguities here both to do with what
to use for $\psi^0_I$ and how to determine the topological
structure of the gauge field. Our (ongoing) work on this approach
will be reported upon in another publication.

In addressing these questions one needs to know how
the locality properties of the overlap Dirac operator
\cite{Hernandez:1998et}
affects its ability to resolve the topological structure on typical
lattice gauge fields. Ideally one would like the situation
to be that the overlap operator possesses zero modes for
topological charges with sizes (in lattice units) 
$\rho \geq \rho_c$ where $\rho_c$ is small enough that
reliable calculations are accessible in practice. 
Because typical  SU(2) and SU(3) gauge fields contain a 
substantial number of small instantons, it has not been
easy to establish whether this is the case or not. In
typical SU(4) and SU(5) fields, by contrast, narrow topological
charges are rare and it becomes possible to address this
question with much less ambiguity. We shall be able to show 
explicitly that the ideal scenario described above, is what
actually occurs.

The calculations we perform involve a quark propagating
in the vacuum of an SU($N_c$) gauge theory rather than
in the vacuum of full QCD with $n_f \neq 0$ flavours
and $N_c$ colours. It is interesting to note, however,
that for any non-zero quark mass the
$N_c \to \infty$ limit of quenched QCD is precisely
the $N_c \to \infty$ limit of full QCD. Thus any statements 
we are able to make about the former apply to the latter
as well.

The paper is organised as follows. In section \ref{sec:setup} we
present the setup of our lattice calculation and we give some
technical details concerning the implementation of the chirally
symmetric overlap lattice Dirac operator. Section \ref{sec:results}
contains the results of the calculation. It is divided into several
subsections where we discuss the results for topology, local
chirality, correlation functions and instanton size distributions. A
summary and conclusions follow in section \ref{sec:conclusions}.

%%%%%%%%%%%%%%%%%%%%%%%%%%%%%%%%%%%%%%%%%%%%%%%%%%%%%%%%%%%%%%%%%%
\section {Setup}
\label{sec:setup}
%%%%%%%%%%%%%%%%%%%%%%%%%%%%%%%%%%%%%%%%%%%%%%%%%%%%%%%%%%%%%%%%%%
The explicit form of the standard massless overlap Dirac operator 
reads \cite{Neuberger:1998fp}
\begin{equation}\label{eq:overlap_operator}
D(0) = \frac{1}{2} (1  + 
    \gamma_5 \text{sgn} (H_{\text{W}}(m)), 
\end{equation}  
where $H_{\text{W}}(m)$ denotes the hermitian Wilson-Dirac operator
$\gamma_5 D_{\text{W}}(m)$ with its mass parameter $m = -1$ in the
supercritical mass region.  Since $H^2(0) = D^\dagger(0) D(0)$
commutes with $\gamma_5$ we can find eigenmodes of
$H^2(0)$ with definite chirality \cite{Edwards:1998yw}:
\begin{equation}
H^2(0) \psi_{\uparrow,\downarrow} = \lambda^2
\psi_{\uparrow,\downarrow}, \quad \gamma_5 \psi_{\uparrow,\downarrow} = \pm
\psi_{\uparrow,\downarrow}\, .
\end{equation}  
The non-zero eigenmodes of $H^2(0)$ with $0< \lambda < 1$ are
doubly degenerate with opposite chirality and the non-zero eigenmodes
of $D(0)$ can be obtained as
\begin{equation}
\psi_{\pm} = \frac{1}{\sqrt{2}} \left( \psi_{\uparrow} \pm i
\psi_{\downarrow} \right) 
\end{equation}
with eigenvalues
\begin{equation}
\lambda_\pm = \lambda^2 \pm i \sqrt{\lambda^2(1-\lambda^2)}\,. 
\end{equation}
In the following we count such an eigenmode pair as one mode.

The overlap Dirac operator possesses exact zero modes with $\lambda=0$
that are stable under small perturbations of the background gauge
field. They are used in the definition of the fermionic topological
charge $Q_f$ via the index of the overlap Dirac operator, i.e.,
\begin{equation}\label{eq:fermionic_index}
Q_f \equiv n_+ - n_-,
\end{equation}
where $n_\pm$ counts the number of zero modes with positive and
negative chirality, respectively.\footnote{It is observed that in any
given background gauge field configuration zero modes always appear in
one chiral sector only.}

The local chirality parameter is equal for the two related modes
$\psi_+$ and $\psi_-$ and can be calculated via the ratio of the
positive and negative chiral densities,
cf.~eq.~(\ref{eq:local_chirality_parameter_definition}).  For the
exact zero modes we obviously have $X(x) = \pm 1$ according to their
chirality.

We generated ensembles of gauge field configurations on $12^4$
lattices using the pure gauge Wilson action with four different
numbers of colours $N_c=2,3,4$ and 5 at a fixed lattice spacing $a
\simeq 0.12$ fm set by the string tension $a \sqrt{\sigma} = 0.261$
from \cite{Lucini:2001ej}. This corresponds to a physical volume $V
\simeq 4.3 \, \text{fm}^4$. The configurations are separated by 1000
update sweeps. The simulation parameters are collected in Table
\ref{tab:simulation_parameters}.
\begin{table}[hbt]
\begin{ruledtabular}
\begin{tabular}{ccc}
$N_c$ & $\beta$ & conf. \\
\hline
2 &  \phantom{1}2.40 & 30 \\
3 &  \phantom{1}5.90 & 50 \\    
4 &            10.84 & 30 \\ 
5 &            17.16 & 30 \\
\end{tabular}
\caption[]{Simulation parameters for the ensembles of gauge field
configurations on $12^4$ lattices. The $\beta$-values are interpolated
to match a fixed lattice spacing $a \simeq 0.12$ fm set by the string
tension $a \sqrt{\sigma} = 0.261$ from \cite{Lucini:2001ej}. The
number of colours, $N_c$, the values of $\beta$ and the number of
generated configurations are given.}
\label{tab:simulation_parameters}
\end{ruledtabular}
\end{table}

It is clear that setting the scale by the string tension is not unique
and there are other sensible ways to determine the scale in different
theories. Moreover, the interpolation which is needed for extracting
the $\beta$-values to match the physical scale introduces an
additional ambiguity. In order to assess these kind of uncertainties
in the scale we use the mass of the $2^{++}$ glueball to set the
lattice spacing for the different gauge groups instead of the string
tension. We use the $2^{++}$ rather than the $0^{++}$ glueball mass
because the latter suffers from large lattice corrections which are
quite different for the different gauge groups. This is due to the
different sensibility to the strong-to-weak coupling crossover at
finite lattice spacing. For the $\beta$-values quoted in table
\ref{tab:simulation_parameters} we obtain interpolated $2^{++}$
glueball masses of $a m_{2^{++}} = 1.34(4)$ for SU(5) and $a
m_{2^{++}} = 1.33(5)$ for SU(4) which has to be compared to the
directly measured $a m_{2^{++}} = 1.30(4)$ for SU(3)
\cite{Lucini:2001ej}. On the other hand, for SU(2) we obtain $a
m_{2^{++}} = 1.41(5)$ from an interpolation and $a m_{2^{++}} =
1.50(5)$ from a direct measurement \cite{Lucini:2001ej}. So, for SU(2)
the uncertainty in the lattice spacing is quite substantial while for
SU(4) and SU(5) it is completely negligible.

We then calculated all the eigenmodes $\psi_{\uparrow,\downarrow}$ of
$H^2(0)$ with $\lambda^2 < 0.1$ using the Ritz functional algorithm of
\cite{Kalkreuter:1996mm}. The cut-off amounts to $\text{Im}
\lambda_\pm < 0.3$ and corresponds roughly to a physical value of $520
\, \text{MeV}$. The average number of modes per configuration we found
below this cut-off (including zero modes) was 6.63(14), 5.46(11),
5.37(13) and 5.17(13) for $N_c=2, 3, 4$ and 5, respectively. The
eigenmodes were computed to an accuracy for which the gradient $g$ of
the Ritz functional had $|g| < 10^{-5}$.

In order to speed up the whole calculation we applied the following
two standard methods. Firstly, by working in a given chiral sector
where all vectors $b_\pm$ obey $\gamma_5 b_\pm = \pm b_\pm$ one can
write \cite{Edwards:1998wx}
\begin{equation}
H(0)^2 b_\pm =
\frac{1}{2} (1 \pm \frac{1}{2}(1\pm\gamma_5)
\text{sgn}(H_{\text{W}}(m))) b_\pm \, .
\end{equation}
One application of $H(0)^2$ on a vector $b_\pm$ therefore requires
only one application of the sign-function.  Secondly, we
calculated the 15 lowest eigenmodes of $H_{\text{W}}(m)$ to an
accuracy of $|g| < 10^{-8}$ and treated the 15 lowest ones exactly by
first projecting them out of $b_\pm$ and adding their contribution to
the sign-function analytically. The 16-th lowest and the highest
eigenvalue of $H_{\text{W}}(m)$ were used to define the interval over
which we approximated the sign-function using Legendre
polynomials \cite{Bunk:1998wj}.

The reason for employing Legendre polynomial approximations rather
than rational approximations was twofold. Firstly, it enabled us to
use a 'dynamical' accuracy for the sign-function, i.e.~we
could easily change the accuracy of the approximation during the
course of calculating the eigenvectors. To be more precise, we
started with a low order approximation of the sign-function at
the beginning of an eigenvector calculation and increased the order
successively so as to have the error of the sign-function
applied on the search vector in the Ritz-functional always $10^{-2}$
times smaller than the gradient of the search vector. For the
application of the sign-function on the eigenvector itself we
chose a higher order approximation in order to avoid destabilisation
of the procedure.

In this way we could manage to bring down the number of
$H_{\text{W}}(m)^2$ multiplications per $\text{sgn}$-function
application on a search vector, e.g.~for SU(2) to around 50.

Secondly, it turned out that the three term recursion relation
inherent in the Legendre polynomial approximation is much more
economic with respect to the required computer resources than the
multi-shift solver used for the partial fraction expansion of the
rational function approximation. As a consequence we found that the
Legendre polynomial approximation ran roughly two times faster than
the rational function approximation on our Alpha/Linux workstations
even when the converged systems were removed from the multi-shift
solver. We ascribe this fact to the large
memory requirement of the multi-shift solver resulting in a
substantially worse performance. In principle the large memory
requirement can be partially avoided by resorting to a double pass
multi-shift solver \cite{Neuberger:1998jk}, however, we did not employ
this variant here.

These two observations in favour of the Legendre polynomials apply
equally well to Chebysheff polynomials. Moreover, the latter allow
polynomial approximations which are optimal with respect to the
uniform norm and are regarded to be superior to Legendre polynomial
approximations. For all practical purposes, i.e.~counting the number
of $H_{\text{W}}$ applications, it turns out that the difference
between Chebysheff and Legendre polynomials are negligible.

Finally we note that our implementation is based on the QCDF90-package
\cite{Dasgupta:1996nj}.

%%%%%%%%%%%%%%%%%%%%%%%%%%%%%%%%%%%%%%%%%%%%%%%%%%%%%%%%%%%%%%%%%%
\section {Results}
\label{sec:results}
%%%%%%%%%%%%%%%%%%%%%%%%%%%%%%%%%%%%%%%%%%%%%%%%%%%%%%%%%%%%%%%%%%

\subsection{Topology}\label{sec:topology}

Since the overlap Dirac operator possesses exact zero modes 
we can use its index, via equation (\ref{eq:fermionic_index}), 
to obtain the total topological charge of the background
gauge field. We shall refer to this as the fermionic
topological charge $Q_f$. In table \ref{tab:top_charge_results}
we list our results for the expectation values of various
quantities involving $Q_f$ for the different $N_c$ ensembles.
Since our ensembles are small, the statistical errors are
large, but it is already clear that the topological susceptibility
$\chi_t$, when expressed in units of the string tension $\sigma$,
varies little with $N_c$. This confirms similar claims made using
other methods
\cite{Lucini:2001ej}.
\begin{table}[hbt]
\begin{ruledtabular}
\begin{tabular}{ccccccc}
$N_c$ & $\langle Q_f \rangle$ & $\langle |Q_f| \rangle$ & $\langle Q_f^2
\rangle - \langle Q_f \rangle^2$ & $\chi_t/\sigma^2$ & $\langle Q_g^2
\rangle - \langle Q_g \rangle^2$ & $\langle (Q_f -
Q_g)^2 \rangle /\langle Q_f^2 \rangle$ \\ \hline 
2 & 0.13(29) & 1.20(19) & 2.52(63) & 0.0261(65) & 2.49(51) & 0.318(69) \\ 
3 & 0.08(21) & 1.08(14) & 2.03(51) & 0.0211(53) & 2.36(49) & 0.314(70) \\ 
4 & 0.23(25) & 1.17(16) & 2.05(45) & 0.0213(47) & 2.36(52) & 0.079(33) \\ 
5 & 0.53(28) & 1.27(17) & 2.25(49) & 0.0234(51) & 2.45(57) & 0.039(22) \\
\end{tabular}
\caption[]{Expectation values of the fermionic topological charge, its
modulus and variance  and the topological susceptibility
in units of the string tension as well as the variance of the gluonic
topological charge. The last column provides a measure of the
violation of the index theorem.}
\label{tab:top_charge_results}
\end{ruledtabular}
\end{table}

An important question, which we shall now address, 
is how reliable is the index theorem for
these particular lattice ensembles. As with any measure of
the topology of lattice gauge fields there must be ambiguities.
For example, if we have a smooth lattice gauge field
containing a single instanton whose size is much greater
than the lattice spacing, $\rho \gg a$, then any reasonable
measure of topology should assign the field $Q=1$. We can
now smoothly shrink the instanton and eventually we will
have $\rho \ll a$. If the instanton is centered well
within a lattice hypercube then the link gauge fields
will now be those corresponding to a gauge singularity
and the same reasonable measure of topology must assign the
field $Q=0$. So for any measure of lattice topology there 
will be some critical value of the size, $\rho_c$, at which 
its value will change from $Q=1$ to $Q=0$. We would like 
$\rho_c$ to be large enough that it excludes any very small
instanton with $\rho\sim a$ whose action is grossly distorted 
from its continuum value by the lattice discretisation, yet small
enough that it does not exclude larger continuum-like instantons.
As $a\to 0$ the latter requirement is achieved automatically since 
the usual $\exp\{-8\pi^2/g^2(\rho)\}$ factor suppresses very 
small instantons for any SU$(N_c)$ gauge theory. What happens
in our case, where $a \simeq 0.26/\surd\sigma \simeq 0.12$ fm,
is a question we shall address in this section. 

We also remark that in the case of the overlap Dirac operator, an 
explicit ambiguity arises from the fact
that when one varies the mass parameter $m$ within its theoretically 
acceptable range, a mode of $H_{\text{W}}(m)$ will 
occasionally change sign, thus changing the operator index
\cite{Edwards:1999bm}.
This creates a corresponding ambiguity in the value of $Q_f$ to
be assigned to the underlying gauge field.

The extent and nature of all these ambiguities will not
only affect the value of $\chi_t$ in 
table \ref{tab:top_charge_results} , but will also affect
any attempt to investigate the topological content of the
non-zero but low-lying modes that drive chiral symmetry 
breaking. 

Ideally, and optimistically, one would like the situation 
to be the following. For the overlap Dirac operator we
would like to have
$\rho_c \sim 2a$ so that it is large enough to exclude
lattice `dislocations' yet small enough to have a minimal
impact on physical topology. The ambiguity in the choice of
$m$ should simply reflect the fact that $\rho_c$ varies with
$m$, but this variation should be weak enough that for any 
reasonable choice of $m$, the value of $\rho_c(m)$ continues to 
be as described above. That is to say, the mode crossing of  
$H_{\text{W}}(m)$ with $m$, should involve very small instantons
which are at the margin of being resolved by the overlap
Dirac operator: for some values of $m$ they are resolved and 
give $Q_f=1$ while for other values of $m$ they remain
unresolved and contribute $Q_f=0$. We shall now provide some
evidence that this optimistic scenario is indeed the case.

What places us in a position to address these questions
is that as $N_c$ grows, the continuum density of small instantons
is rapidly suppressed
\cite{Lucini:2001ej}.
Thus most of our SU(4) and SU(5) lattice
field configurations do not contain any small instantons at all;
and where they do, they usually only contain one. Thus if
there is a correlation between small instantons and the ambiguities 
in the overlap index, it should be easy to see. In SU(3)
by contrast, the fields frequently possess several small instantons 
and patterns are much harder to discern.

In order to determine something about the topological charge
content of the lattice gauge fields, we analyse them using a 
standard cooling technique
\cite{Hoek:1987ndTeper:1999wp}.
This method also has its ambiguities but these are reasonably 
well understood. We now sketch the method and summarise its 
uncertainties.

To cool a given rough lattice
gauge field we perform a sequence of sweeps through the lattice,
always choosing the new link matrices so as to minimise the plaquette
action. After each sweep we calculate the lattice topological
charge density $Q_g(x)$ using a simple lattice version of the continuum
$F\widetilde{F}$ operator. The sum of this charge density over the
lattice gives us a measure of $Q$ that we label $Q_g$. In principle 
this depends on the number of cooling sweeps. In practice,
after a very few such sweeps, the fields become smooth enough 
that $Q_g$ becomes independent of the cooling. This is not
quite true: an instanton will (usually) slowly shrink
under cooling and will eventually shrink out of the lattice.
However such a narrowing instanton produces a prominent
$\propto 1/\rho^4$ peak in  $Q_g(x)$ which is easy to monitor
and its disappearance within a hypercube is accompanied by
characteristic signals. We monitor peaks in $Q_g(x)$ (and
also in the action density) so that we are able to unambiguously
detect the disappearance of any topological charge after about
4 cooling sweeps. Since a few cooling sweeps cannot undo
coherent long-range structures in the fields (any more than a few
Monte Carlo sweeps can create them) or indeed to change their
size by a large amount, we expect that the sizes of the
instantons after a few cooling sweeps, are closely related to 
their sizes in the original uncooled fields. This relationship
is however only qualitative, since what happens to an
individual instanton depends on its environment. For example,
a narrow instanton would normally shrink under cooling.
However if it is overlapping strongly with an anti-instanton
it might well grow in size instead (so that the two opposite
charges can more readily annihilate). Occasionally, if the
environment is extreme, one can see that the size increases
quite rapidly with cooling. 

We shall make the choice of 10 cooling sweeps at which to analyse the
topological structure. From the peaks in $Q_g(x)$ we can extract an
instanton size $\rho$ using the continuum formula
$Q_{peak}=6/\pi^2\rho^4$ (see also the more detailed discussion in
section \ref{sec:instanton_size_distributions}). This is obviously
approximate and can be partially corrected for discretisation effects.
However we shall not do so here since we are only aiming at uncovering
qualitative features. If $\rho$ is small then the peak is large and
its identification is unambiguous.  For large $\rho$, by contrast, the
peak is very small and could be a remnant of the roughness of the
original fields. Thus any attempt to identify individual large
instantons is not reliable. Instantons that are small after 10 cooling
sweeps will usually correspond to instantons that were small in the
original uncooled fields. However, as explained above, this will only
usually be the case.
 
The total topological charge $Q_g$ should, of course, be an 
integer but, because of discretisation effects, the sum of
$Q_g(x)$ will not be. Accordingly we transform this
sum to the appropriate nearby integer value. Occasionally
it is not evident which integer is appropriate, and in
that case it suffices to note the charges of those instantons
that are smallest, and therefore suffer the largest lattice 
corrections, and correct the value of $Q_g$ accordingly. 
(One can finesse this step by using improved 
lattice topological charge operators which give values of $Q_g$ 
very close to an integer. See
\cite{Bilson-Thompson:2001ca}
for a recent discussion.) 

Thus the above cooling procedure gives us an integer topological
charge $Q_g$ and also some detailed information on the smaller 
instantons. Cooling does not, however, provide us with any
information on `dislocations' -- fields with $\rho \sim a$ which
are close to the $Q=0,1$ boundary and whose action is typically
far from the continuum action -- since,
as one would expect, these appear to be erased in the first one 
or two cooling sweeps, i.e. before the topological charge
density has become smooth enough for topological charges 
to be identifiable. The density of these dislocations
depends on the action used.  That such dislocations can be
numerous has been confirmed in SU(2) calculations with the 
plaquette action 
\cite{Pugh:1989qd:1989ek}
using a geometric definition of $Q$
\cite{Phillips:1986qd}.
Such dislocations are present in small volumes that do not
contain any real physical instantons, and one sees explicitly
\cite{Pugh:1989qd:1989ek}
by cooling such configurations that dislocations 
immediately disappear. 

\begin{figure}[htb]
\includegraphics[width=6.8cm]{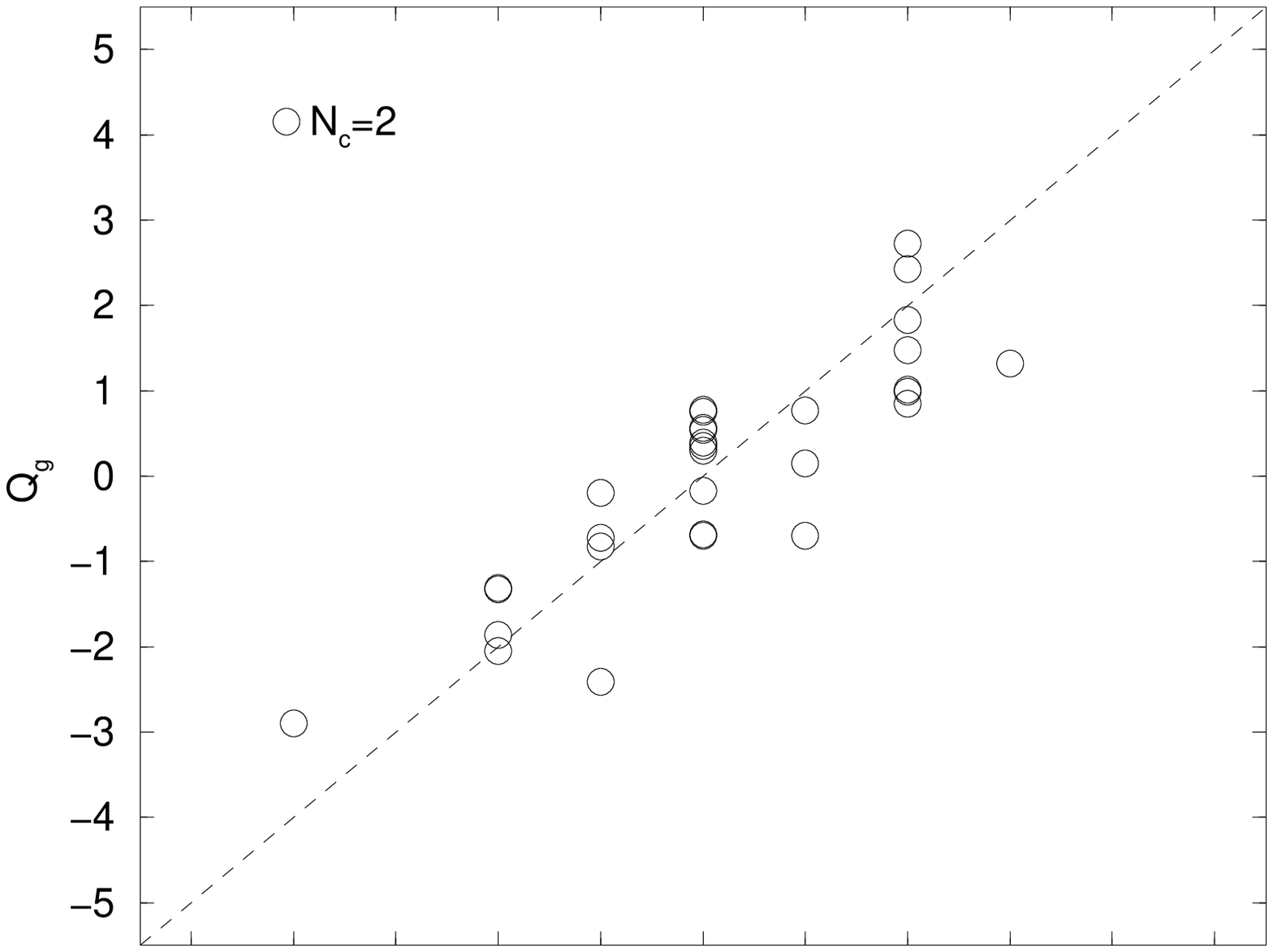}\\
\includegraphics[width=6.8cm]{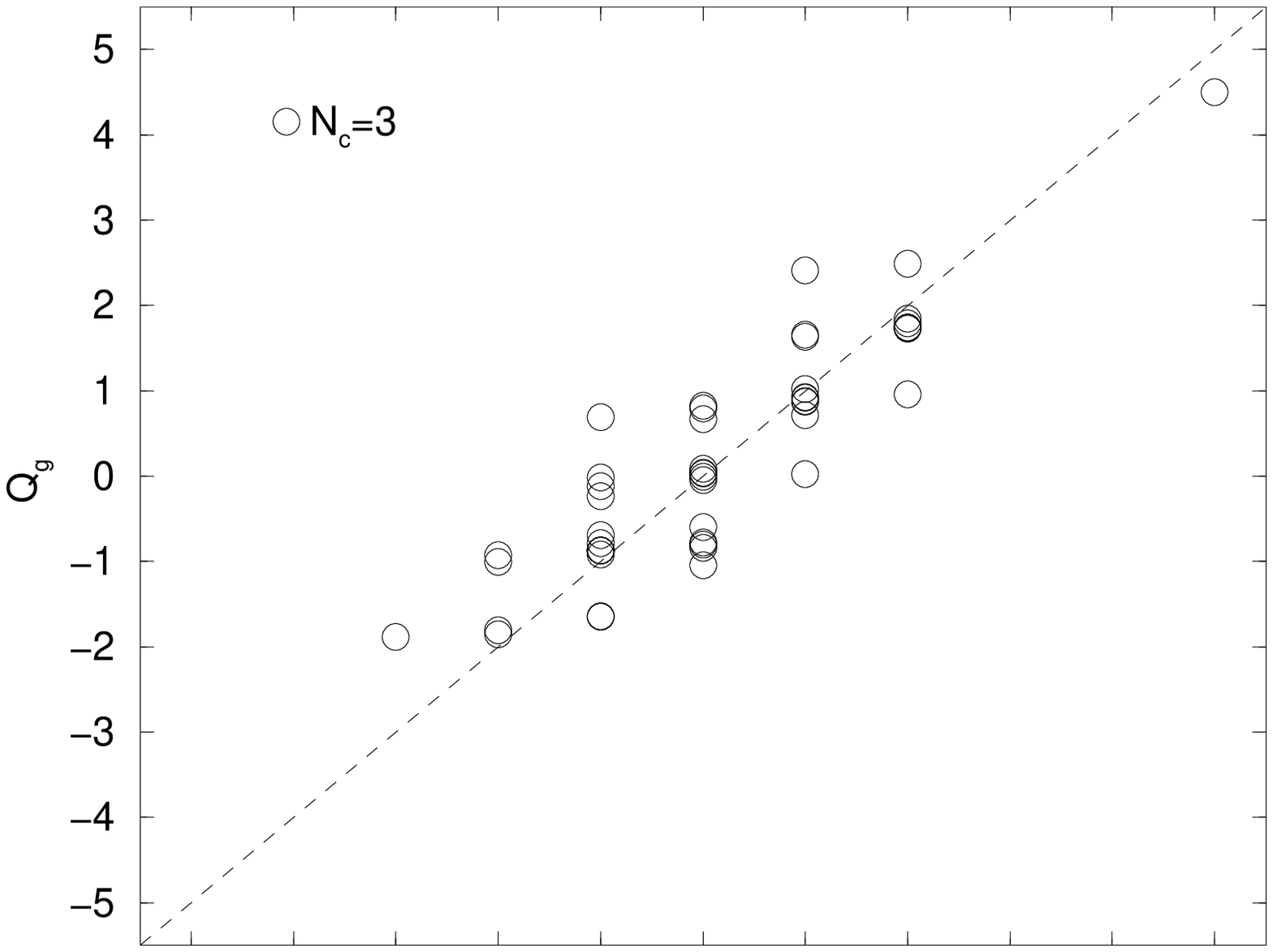}\\
\includegraphics[width=6.8cm]{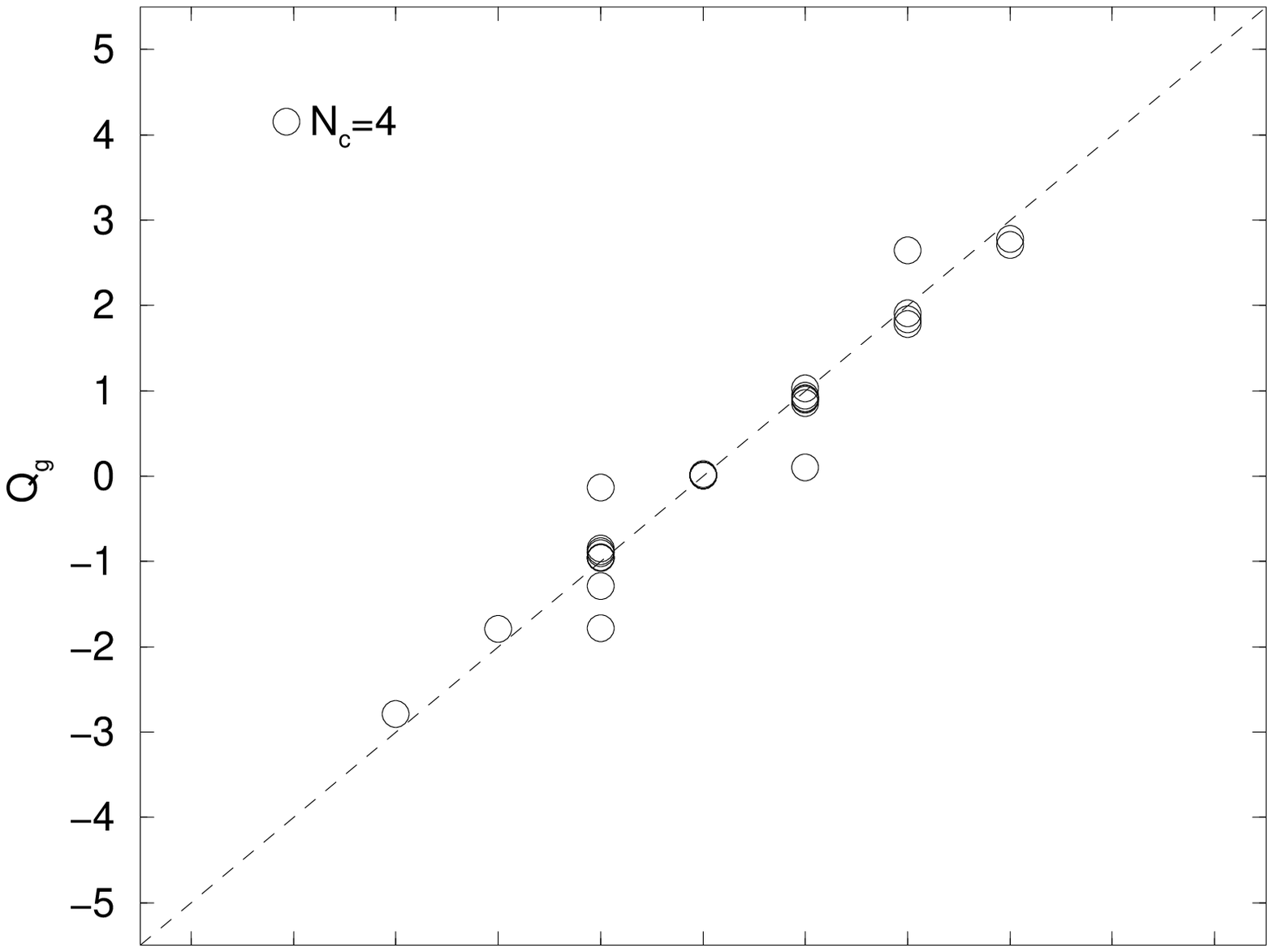}\\
\includegraphics[width=6.8cm]{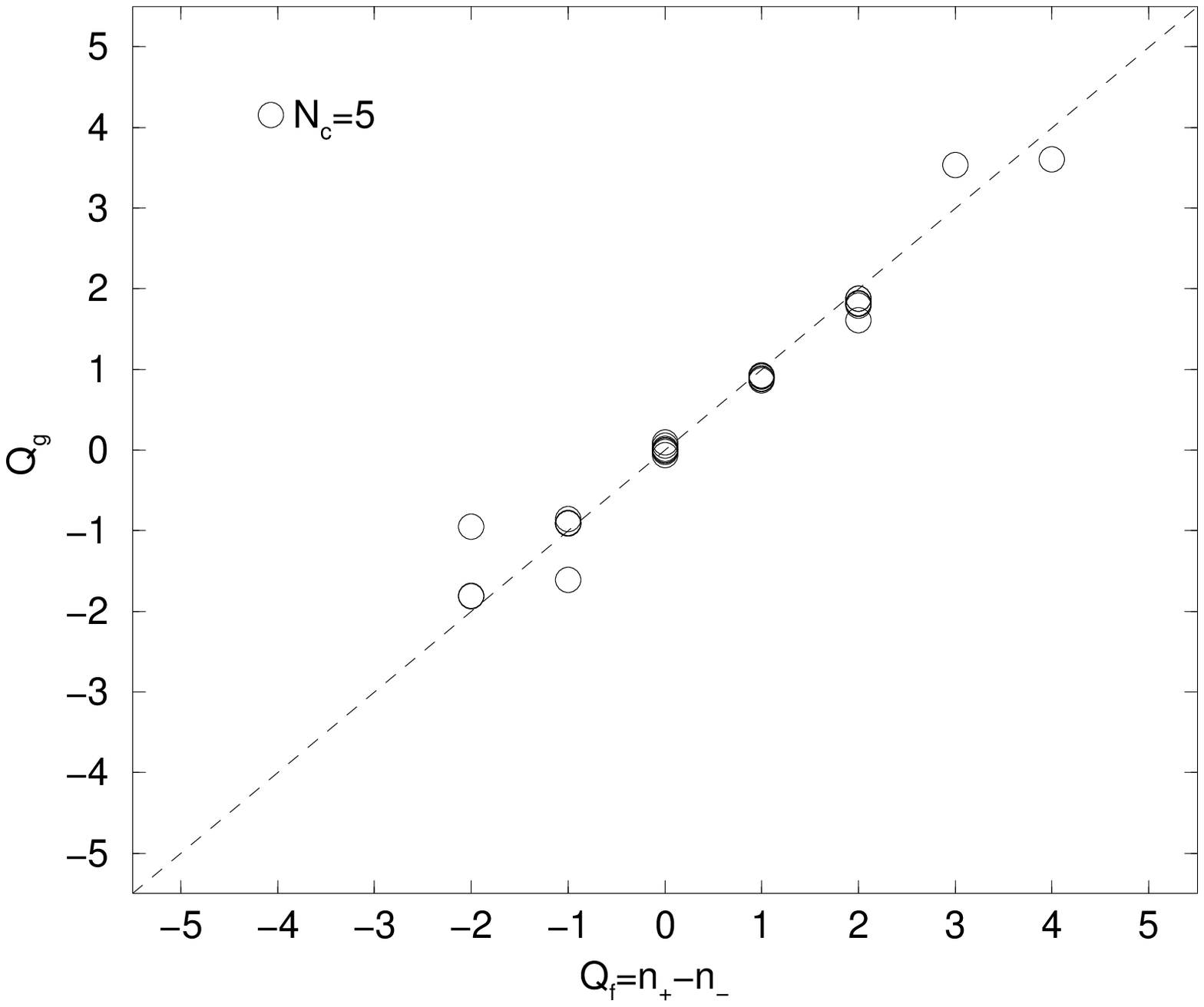}
\caption[]{The gluonic field theoretic topological charge $Q_g$ after
10 cooling sweeps vs.~the fermionic topological charge $Q_f = n_+ - n_-$ for
$N_c=2, 3, 4$ and 5.}
\label{fig:gauge_vs_fermi}
\end{figure}

We now turn to a comparison between the topological charge
$Q_g$ that one obtains by cooling and the fermionic
topological charge $Q_f$. In figure \ref{fig:gauge_vs_fermi} 
we compare the values of $Q_f$ and $Q_g$ as obtained for
each gauge field ensemble . (For presentational purposes, 
the values of $Q_g$ shown here are prior to any adjustment to 
integer values, so that the points corresponding to different 
lattice fields can be distinguished from each other.)
We see a dramatic change between these ensembles. For SU$(2)$
the differences are frequent even if a marked correlation is
evident. For SU$(5)$ it has become rare for the charges not 
to be in agreement.

\begin{figure}[htb]
\includegraphics[width=10cm]{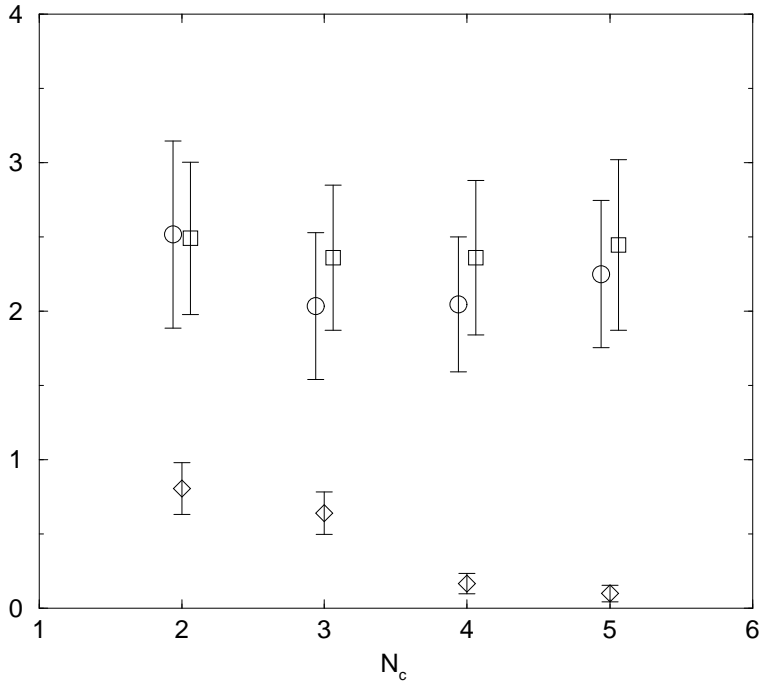}
\caption[]{$\langle Q_f^2\rangle - {\langle Q_f\rangle}^2$ (circles),
$\langle Q_g^2\rangle - {\langle Q_g\rangle}^2$ (squares) and $\langle
(Q_f - Q_g)^2 \rangle$ (diamonds) as a function of $N_c$.}
\label{fig:top_charge_difference}
\end{figure}

To make the comparison more quantitative we calculate the quantity
$\langle (Q_f - Q_g)^2 \rangle$ and plot it as a function of $N_c$ in
figure \ref{fig:top_charge_difference} together with the values of
$\langle Q_f^2\rangle$ and $\langle Q_g^2\rangle$.  (In fact we plot
$\langle Q^2\rangle - {\langle Q\rangle}^2$ in the latter two
cases. Of course the fact that $\langle Q\rangle \not= 0$ is an
artifact of our low statistics and happens to be only significant in
the case of SU$(5)$.) If the charges were uncorrelated, i.e. $\langle
Q_f Q_g \rangle = 0$, then we would expect to find $\langle (Q_f -
Q_g)^2 \rangle = \langle Q_f^2\rangle + \langle Q_g^2\rangle$. From
the values plotted in figure \ref{fig:top_charge_difference} (cf.~also
table \ref{tab:top_charge_results}) it is clear that the correlation
is strong even for SU(2), and rapidly increases as $N_c$ increases. We
also note that the susceptibility, when expressed in physical units as
$\chi_t/\sigma^2$, is independent of $N_c$ within our errors.

The fact that disagreement between $Q_f$ and $Q_g$ becomes
rare for SU(4) and SU(5), provides us with an opportunity
to determine the reason for such disagreements. We turn to 
this now.

As $N_c$ increases, a striking change is the rapid disappearance of
small instantons. This is a simple consequence of the $N_c$ dependence
of the $\exp\{-8\pi^2/g^2(\rho)\}$ factor that dominates the weight of
small (and therefore approximately `dilute') instantons. It has been
explicitly observed in the high statistics calculations of
\cite{Lucini:2001ej}
and we will confirm it through a somewhat different analysis later on
in this paper (see section \ref{sec:instanton_size_distributions}). We
know that cooling will remove very small instantons, but the SU(2)
instanton size distribution
\cite{Lucini:2001ej}
tells us that this must be confined to instantons with $\rho < 2a$. 
Thus the suppression we observe for increasing $N_c$ at larger
$\rho$ must indeed be a dynamical effect. 

\begin{figure}[htb]
\includegraphics[width=10cm]{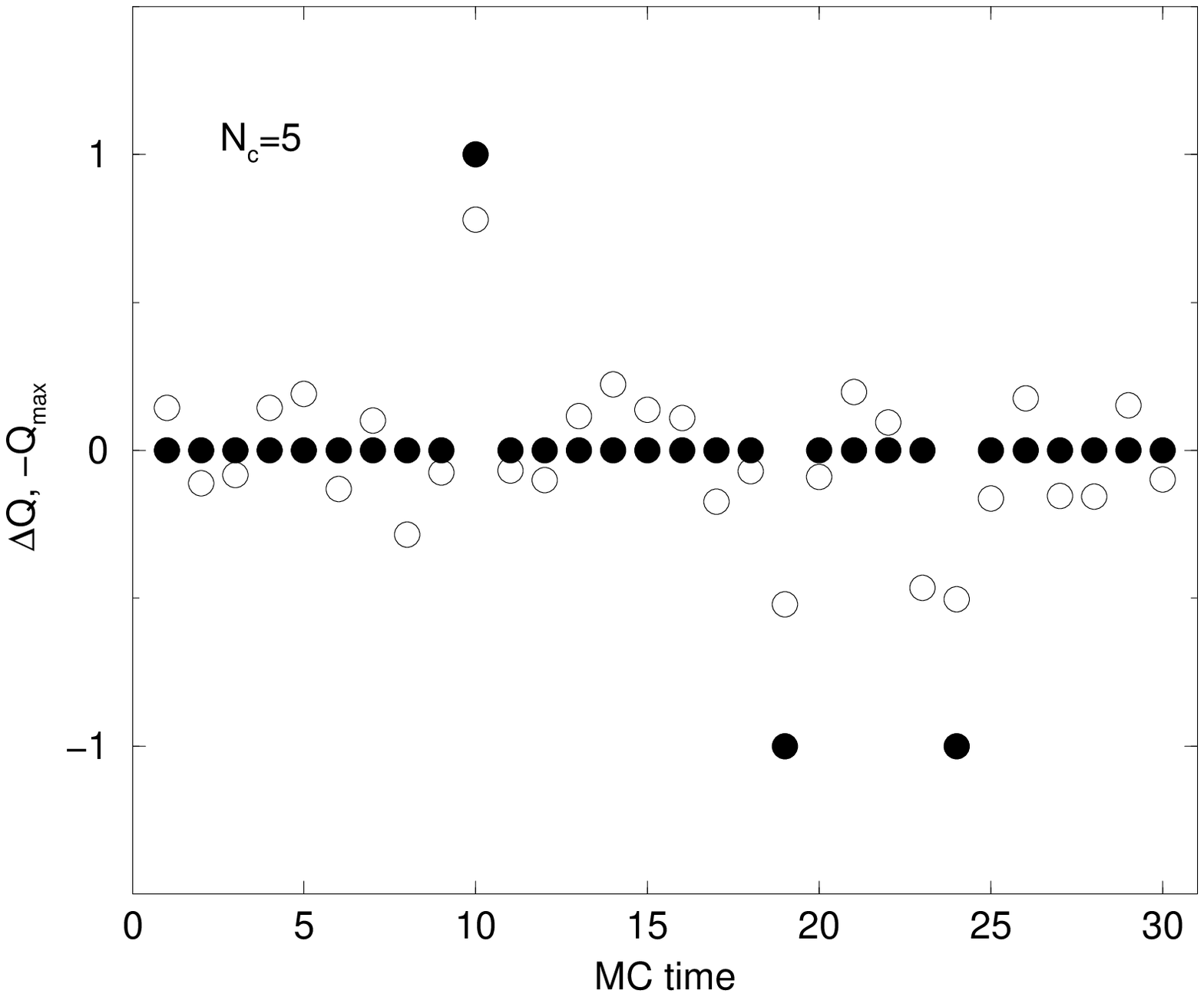}\\
\includegraphics[width=10cm]{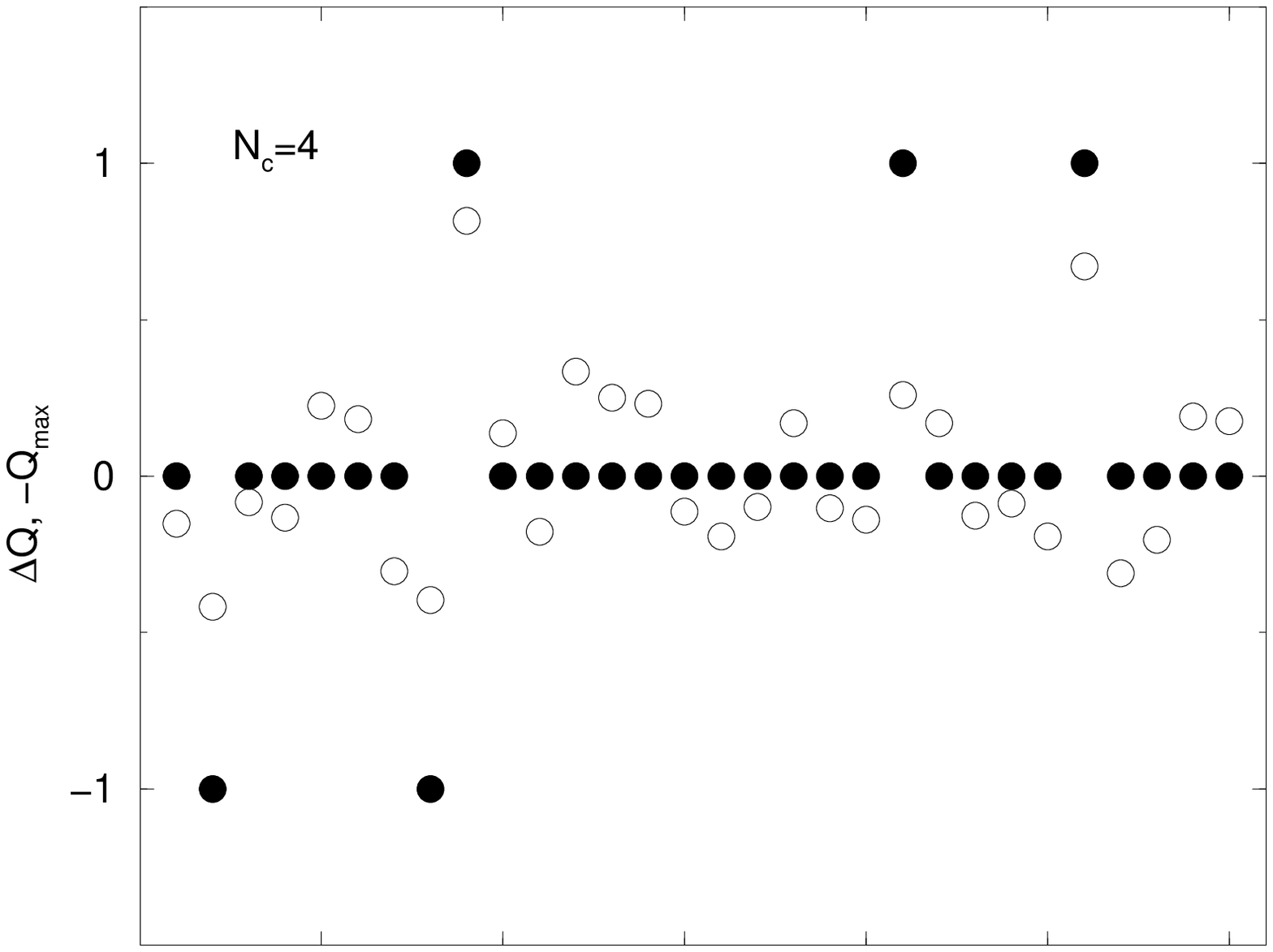}
\caption[]{Comparison of the maximum peak value $-Q_{\text{max}}$
(empty circles) with the difference between the fermionic and field
theoretic topological charge $\Delta Q = Q_f - Q_g$ (filled circles)
for $N_c=5$ and 4. Note that $Q_{\text{max}}$ is
rescaled to facilitate comparison. Monte Carlo (MC) time is in units
of $10^3$ sweeps.}
\label{fig:q_vs_peaks_nc4&5}
\end{figure}

So, as we increase $N_c$ small instantons are suppressed and so is any
disagreement between $Q_f$ and $Q_g$. It is natural to ask if these
two phenomena are related. To address this question we take each of
our lattice fields after 10 cooling sweeps, find the maximum value of
the topological charge density, $\max_x |Q_g(x)|$, and label by
$Q_{\text{max}}$ the corresponding value of $Q_g(x)$. If the reason
for $Q_f \not= Q_g$ is that the overlap fermionic operator does not
resolve small instantons, then we would expect that plotting $\Delta Q
\equiv Q_f - Q_g$ and $Q_{\text{max}}$ for each field configuration
will show that $\Delta Q \not= 0$ is associated with a large value of
$Q_{\text{max}}$ and, in addition, that the sign of $\Delta Q$ should
be the same as that of $-Q_{\text{max}}$. We display such plots in
figure \ref{fig:q_vs_peaks_nc4&5} for SU(4) and SU(5) respectively. We
observe that in the case of SU(5) $\Delta Q \not= 0$ is always
associated with a very large value of $|Q_{\text{max}}|$ and that the
sign of $\Delta Q$ is indeed the same as that of $-Q_{\text{max}}$. In
SU(4) that is also the case, except for one configuration where
$|Q_{\text{max}}|$ is not very large although the sign matches $\Delta
Q$. As we remarked earlier, occasionally a narrow instanton broadens
rapidly under cooling and indeed there is evidence from the earlier
cooling history of this particular field configuration that this is
what has happened here. However rather than appeal to such a much more
complicated analysis, we prefer to maintain our simpler analysis, and
simply accept occasional mismatches as being consistent with the
expected frequency of such effects.

\begin{figure}[htb]
\includegraphics[width=10cm]{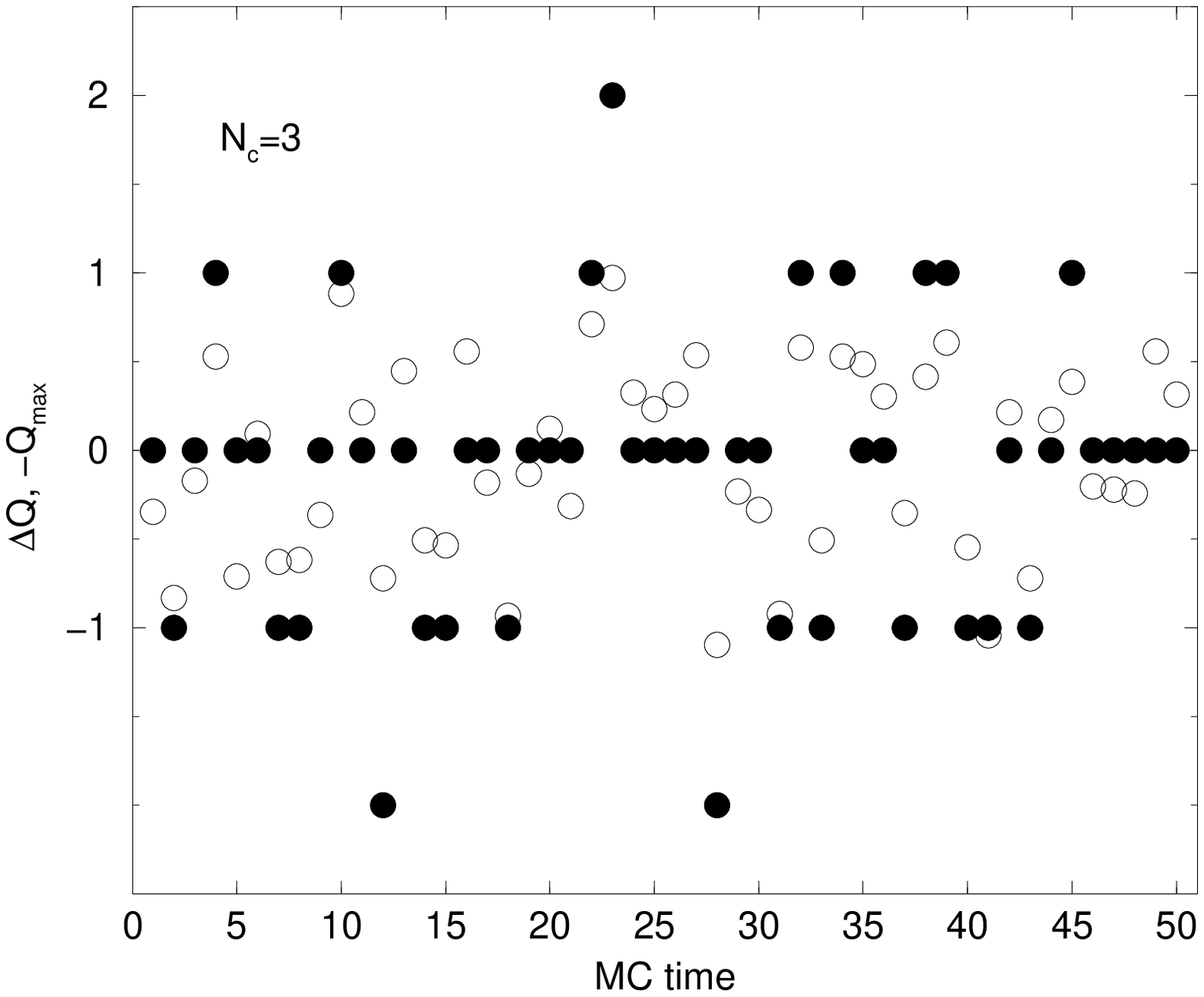}
\caption[]{Same as figure \ref{fig:q_vs_peaks_nc4&5} but for $N_c=3$.}
\label{fig:q_vs_peaks_nc3}
\end{figure}

What helps to make these simple plots so compelling is that small
instantons are rare in SU(4) and SU(5), so that if one of our lattice
fields has one such small instanton it is unlikely to have a second
one as well. Otherwise one could frequently have cancellations, and
hence $\Delta Q = 0$, undermining any simple visual inspection of the
kind that suffices here.  In SU(3), by contrast, there are more small
instantons and things are indeed considerably more complex, as we see
from figure \ref{fig:q_vs_peaks_nc3}.  Nonetheless even here careful
scrutiny of the figure will largely confirm what we have seen in the
cases of SU(4) and SU(5).

We believe that all this provides convincing evidence that the overlap
Dirac operator resolves all topological charges with, roughly, $\rho
\geq 2.5a$ -- and perhaps a little less than that. This means that for
reasonably small $a$ the operator will resolve all the physical
instantons and, at the same time, will exclude all the unphysical
(near-)dislocations.

\begin{figure}[htb]
\includegraphics[width=10cm]{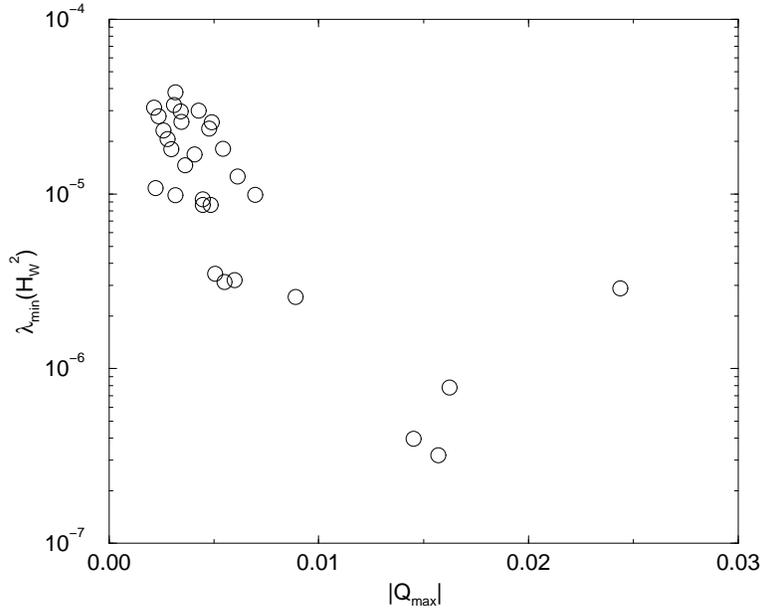}
\caption[]{The smallest eigenvalue of the hermitian Wilson Dirac
operator $\lambda_{\text{min}}(H_{\text{W}}^2)$ for each
field configuration in the $N_c=5$ ensemble against the value of $|Q_{\text{max}}|$ defined in the text.}
\label{fig:lambdamin_vs_qmax_nc5}
\end{figure}

We remarked earlier that an explicit ambiguity in $Q_f$ arises from
the fact that when we vary the $m$ of $H_{\text{W}}(m)$ in equation
(\ref{eq:overlap_operator}) within the supercritical mass region,
eigenvalues of $H_{\text{W}}(m)$ will sometimes change sign, at which
point $Q_f$ will change.  (All this for a given gauge field.) Is there
any evidence for our earlier suggestion that the mode that changes
sign may be associated with a very narrow instanton which is resolved
by the overlap Dirac operator for one range of $m$ and not for
another? To address this question directly one needs to vary $m$ which
we have not done in our calculations.  However one can go some way
following an indirect line of reasoning.  If a mode of
$H_{\text{W}}(m)$ passes through zero at some value of $m$, then the
mode will be very small at nearby values of $m$. Thus we would expect
the modes at $m=-1$ to be smaller, on the average, if we are examining
a field configuration for which there are modes that change sign at a
value of $m$ close to $m=-1$. So we can ask whether fields with small
instantons have smaller modes of $H_{\text{W}}(m=-1)$ than fields
without such small charges. We show the results of such an exercise
for SU(5) in figure \ref{fig:lambdamin_vs_qmax_nc5}.
Here we plot the value of the smallest non-zero eigenvalue
$\lambda_{\text{min}}$ of $H_{\text{W}}(m=-1)^2$ for each field
against the value of $|Q_{\text{max}}|$ defined earlier. We see from
figure \ref{fig:lambdamin_vs_qmax_nc5} that there is a significant
trend for field configurations with smaller instantons to be more
likely to have unusually small modes of $H_{\text{W}}(m=-1)$. This
provides some significant evidence that small instantons are indeed
the cause of this ambiguity. We also note that there is some evidence
that $\lambda_{\text{min}}$ actually has a minimum at
$|Q_{\text{max}}| \simeq 0.015$ and then increases again for the
narrower instantons that correspond to larger $|Q_{\text{max}}|$.
This fits our picture: the very smallest instantons are invisible for
any reasonable value of $m$ and so will not involve level crossings
near $m=-1$. The position of the minimum suggests that the critical
instanton size visible to the overlap operator is $\rho_c \simeq
2.5a$. Finally we would like to emphasise again that we owe the
possibility of the above argument entirely to the fact that the SU(5)
(and also the SU(4)) gauge field configurations are much 'cleaner'
than the ones for SU(3) or SU(2) in the sense that small instantons
are much rarer and usually unaccompanied by other small instantons in
the configuration. As a consequence it is no surprise that the trend
observed in figure \ref{fig:lambdamin_vs_qmax_nc5} is less pronounced
in an equivalent plot for SU(4) and almost completely distorted and
obscured for SU(3) and SU(2).

\begin{figure}[htb]
\includegraphics[angle=-90,width=10cm]{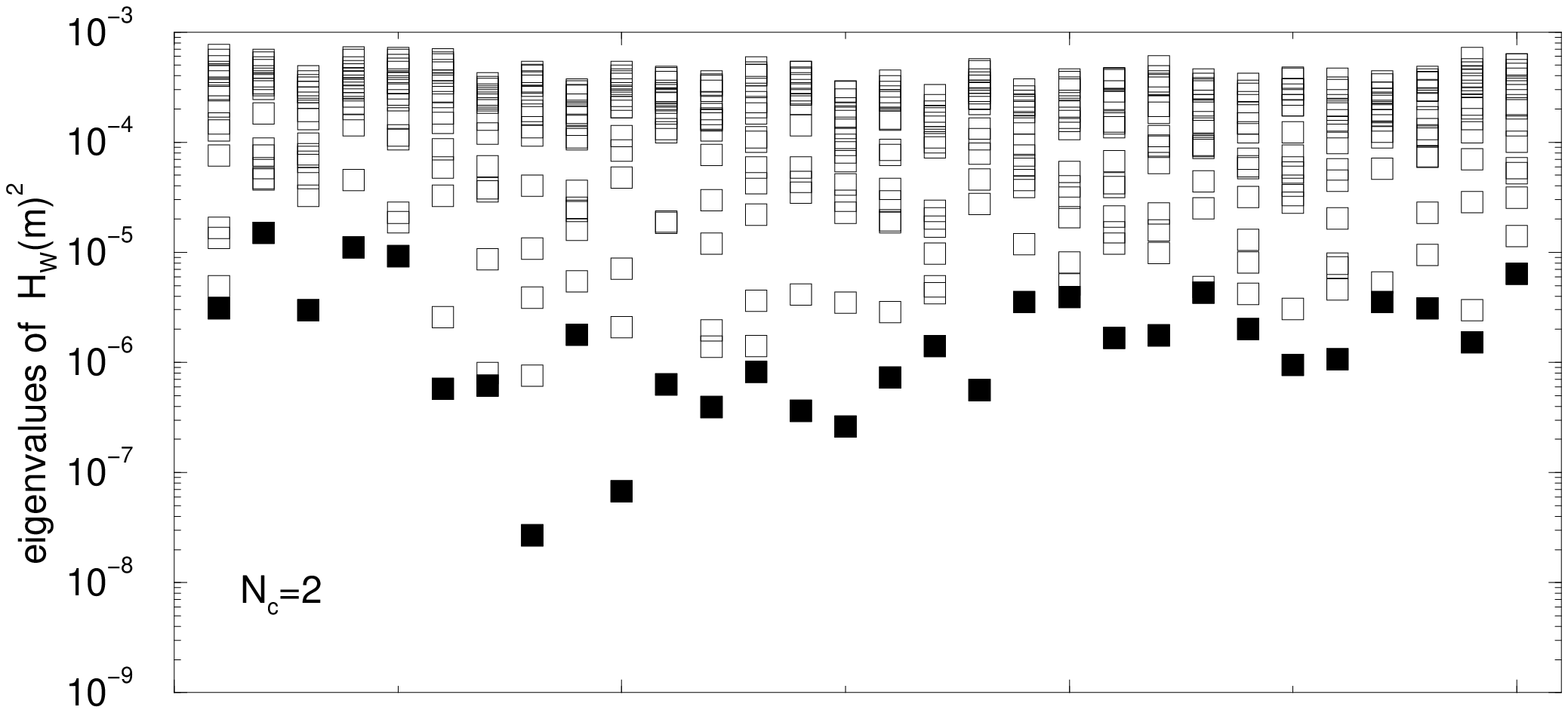}\\
\includegraphics[angle=-90,width=10cm]{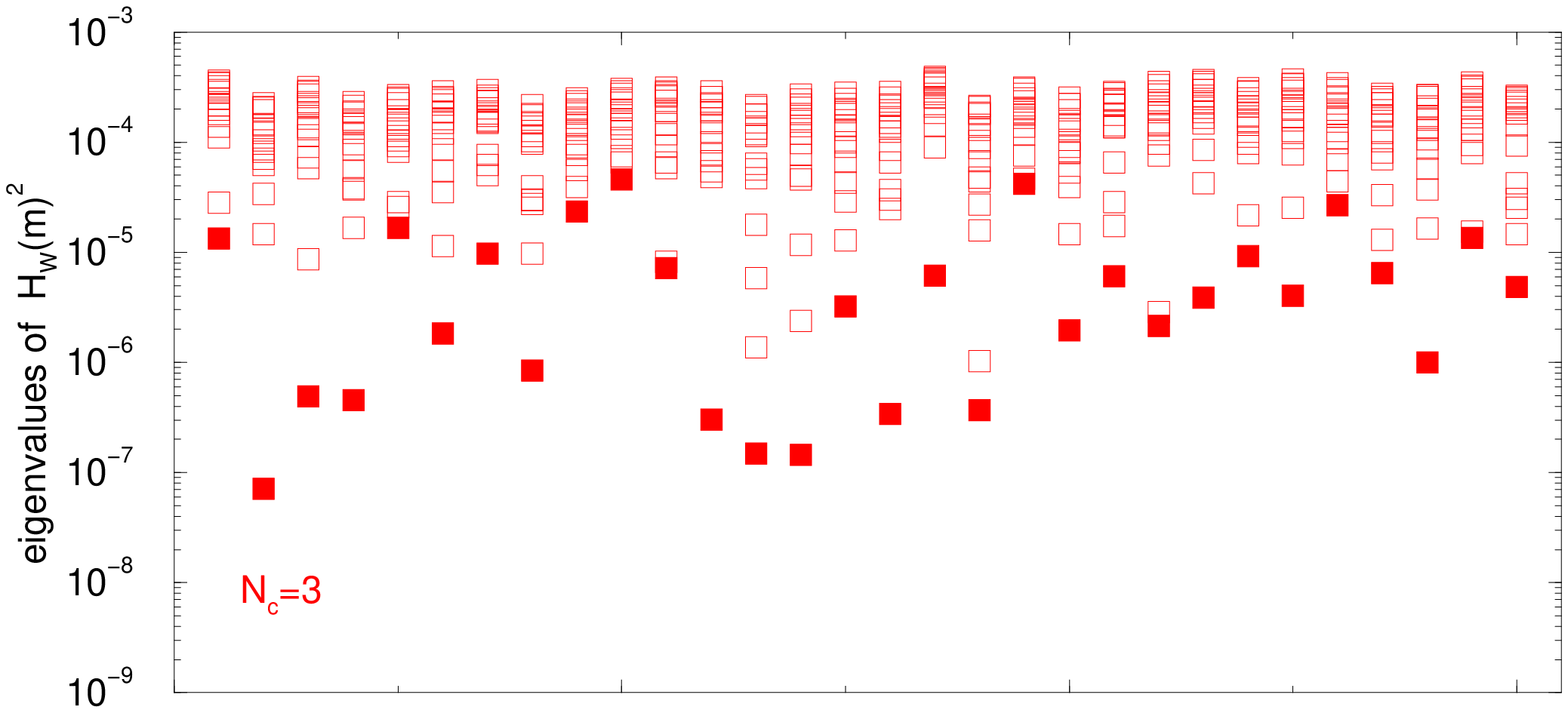}\\
\includegraphics[angle=-90,width=10cm]{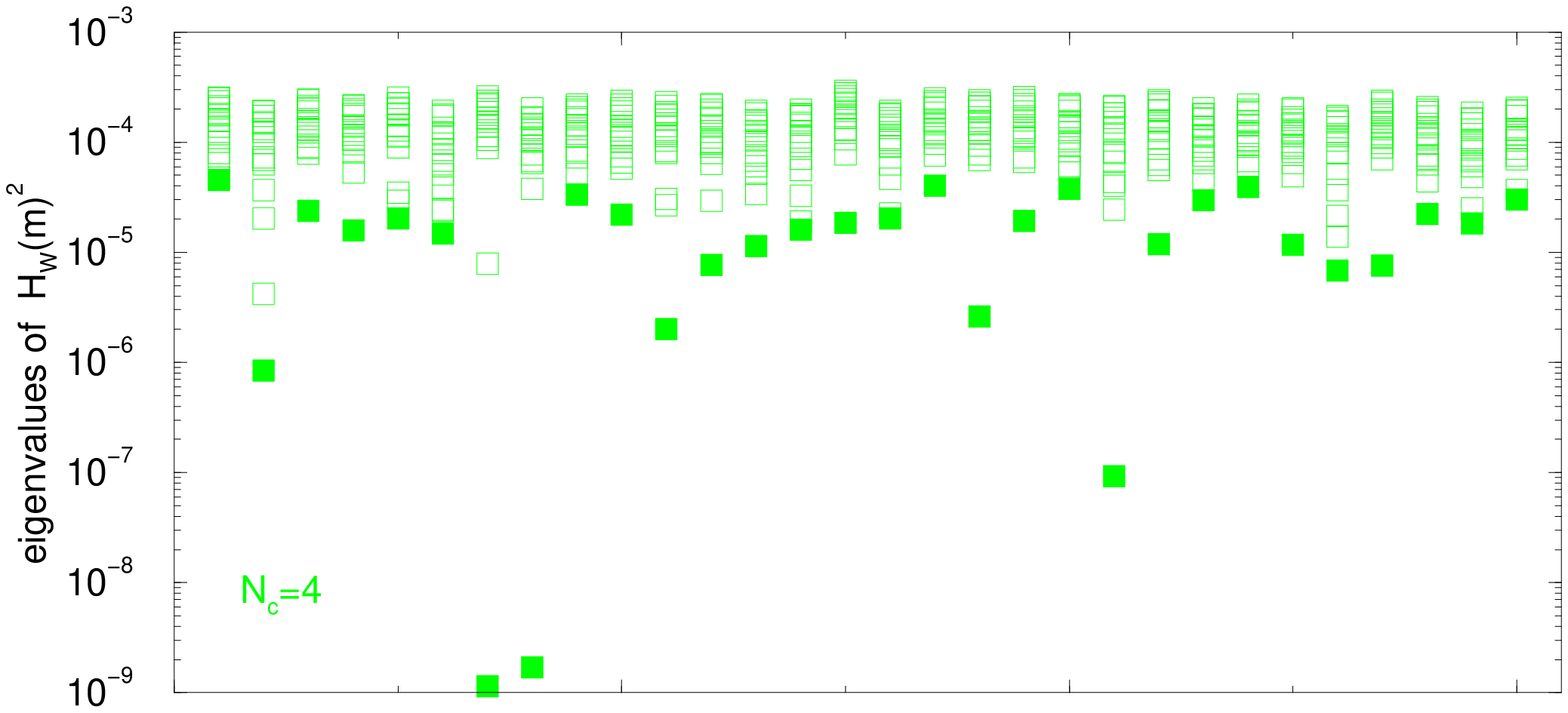}\\
\includegraphics[angle=-90,width=10cm]{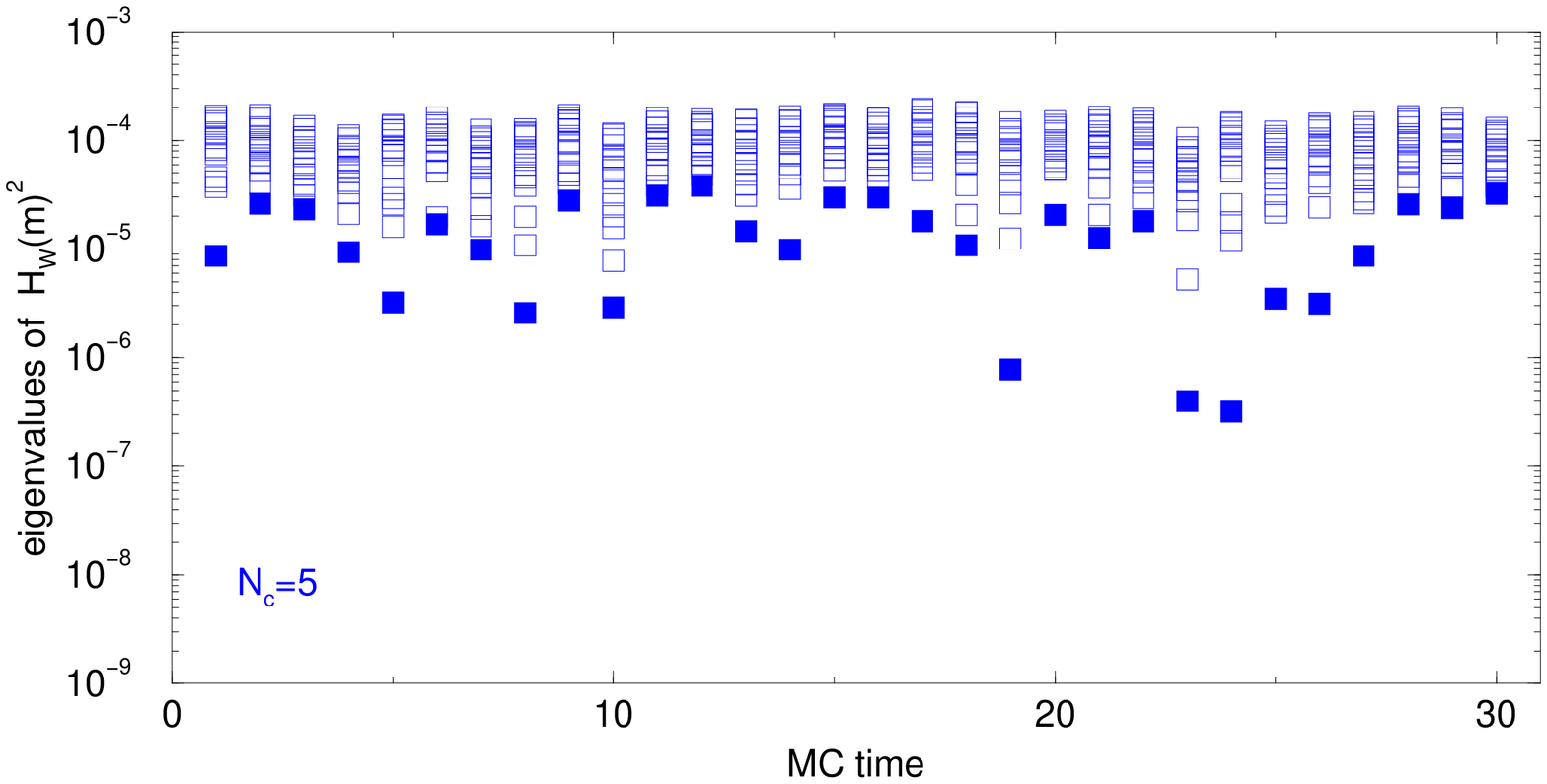}
\caption[]{The 15 lowest lying eigenvalues of $H_{\text{W}}(m)^2$ with
$m=-1$ for $N_c=2, 3, 4$ and 5. Monte Carlo (MC) time is in units of
$10^3$ sweeps. In the case of $N_c=3$ we only show the eigenvalues of
configurations 11 to 40 for ease of comparison. }
\label{fig:herm_Wilson_eigenvalues}
\end{figure}

More evidence for the smoothness of the gauge fields at larger $N_c$
is provided by figure \ref{fig:herm_Wilson_eigenvalues}. 
Here we plot the lowest 15 eigenvalues of $H_{\text{W}}(m)^2$ with
$m=-1$ as a function of Monte Carlo (MC) time. We clearly observe a
trend for larger $N_c$ field configurations to have less but more
isolated unusually small eigenvalues. In addition the lower bound of
the bulk of eigenvalues seems to be much more rigid for SU(5) and
SU(4) than it is for SU(3) or even SU(2).  Another observation which
is evident from just a visual inspection of the plots, is that the
density of small eigenvalues in the bulk grows with increasing $N_c$
just as one would expect. This is, however, in contrast to what we
observe for the density of small eigenvalues of the overlap operator
$D(0)$ (see section \ref{sec:setup}).  There we found that the number
of eigenvalues below the cut-off $\lambda^2 < 0.1$ shows a tendency to
decrease with increasing $N_c$. Although one expects the spectrum of
the Dirac operator at small eigenvalues to remain the same for
increasing $N_c$ the scale of the eigenvalue distribution is governed
by the quark condensate growing linearly with $N_c$ and thus the
spacing of the eigenvalues should become smaller when expressed in
terms of the mass of a bound state (or the string tension). 

\subsection{Local chirality}
After the discussion on how the overlap Dirac operator resolves the
total topological charge and how that compares to a standard analysis
using cooled gauge fields, we would now like to address the question 
how the low lying modes of the overlap Dirac operator are affected
by topology. As discussed in the introduction one possible approach is
to investigate the local chirality parameter $X(x)$ defined via the
ratio of the positive and negative chiral densities of the eigenmodes,
cf.~eq.~(\ref{eq:local_chirality_parameter_definition}). One can then
determine from the distribution of $X(x)$ for separate eigenmodes (or 
ensembles thereof) how chiral they are and so how likely they are to
have their origin in the mixing of zero modes. 

\begin{figure}[htb]
\begin{tabular}{cc}
\includegraphics[height=6.5cm]{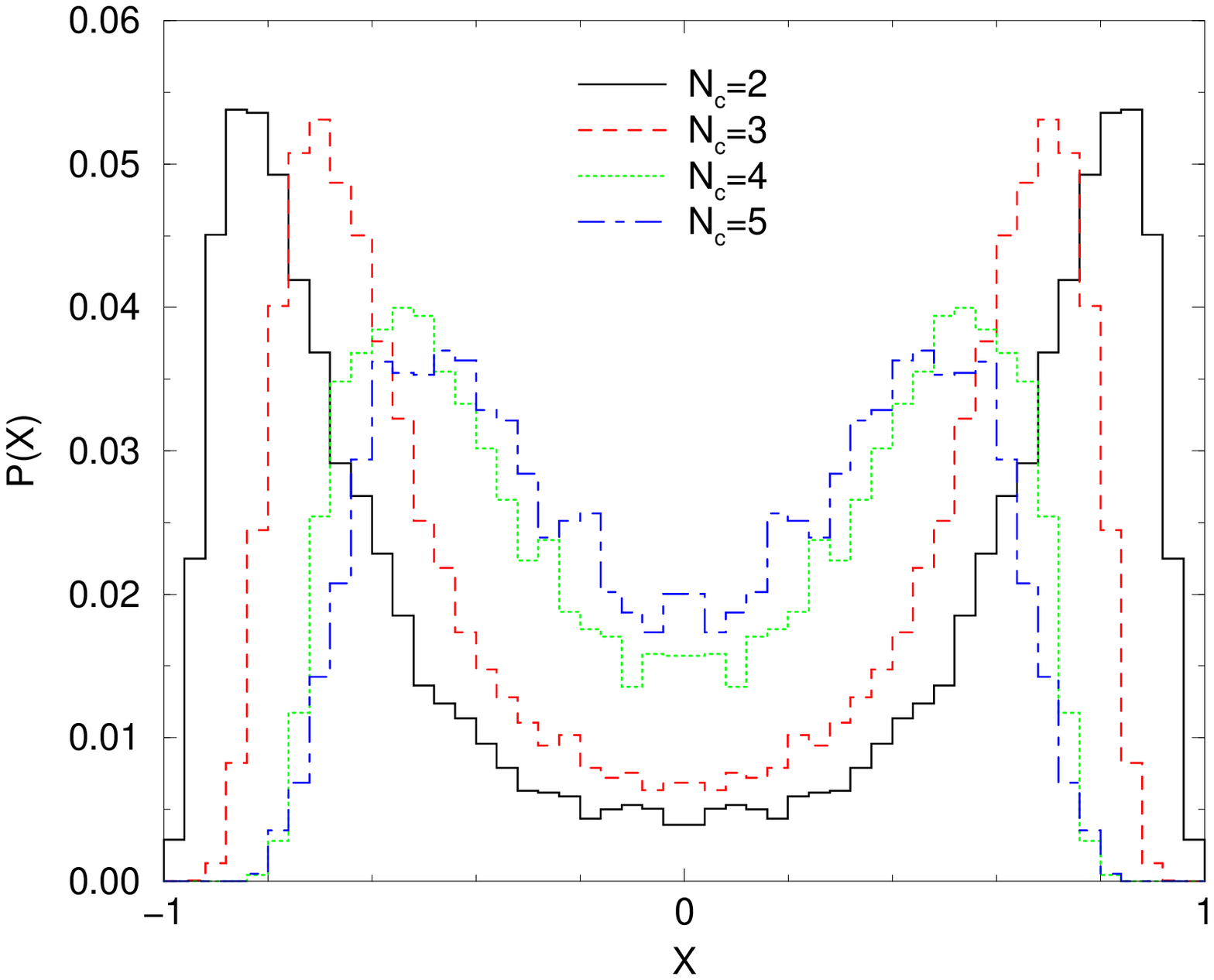}&
\includegraphics[height=6.5cm]{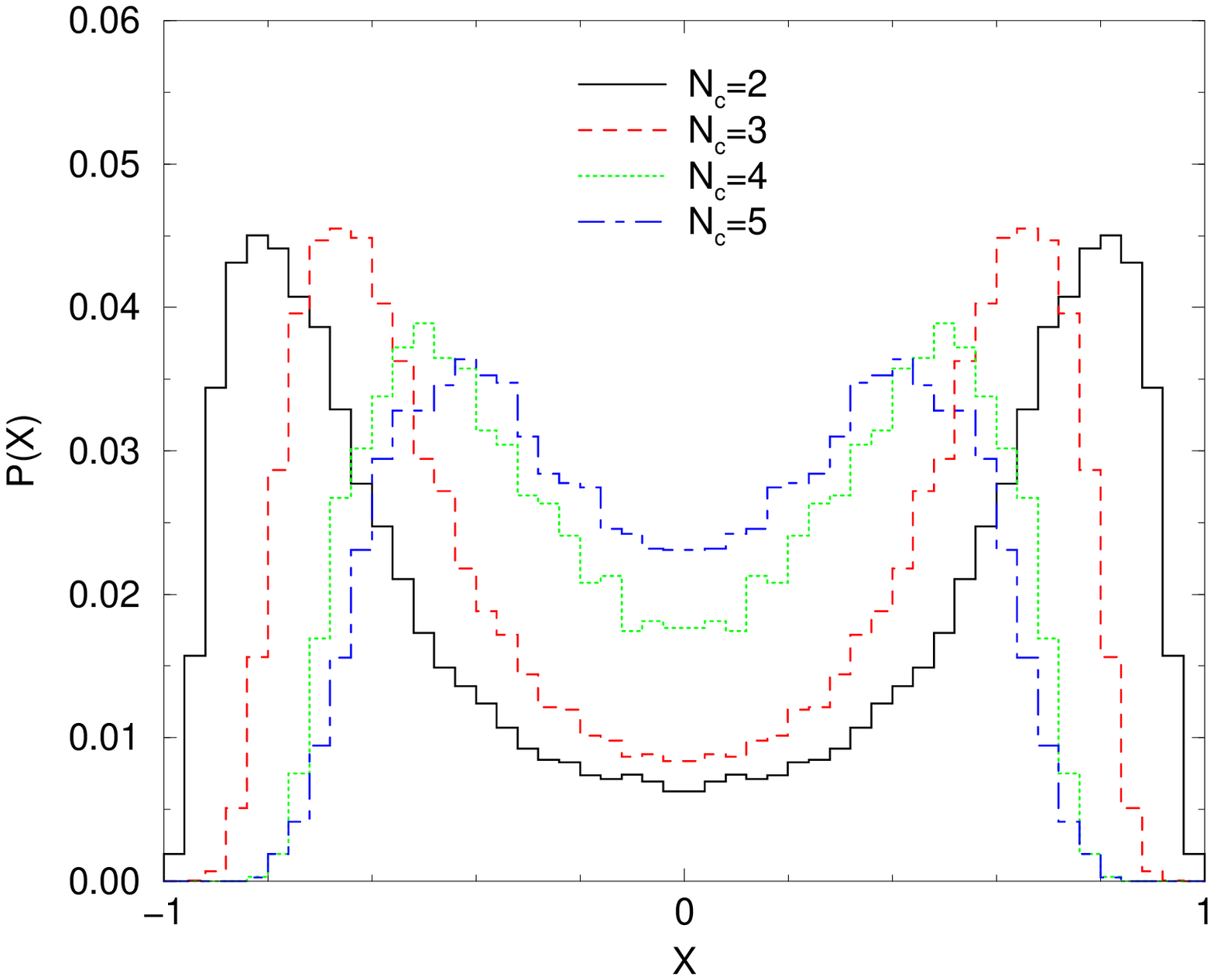}\\
\includegraphics[height=6.5cm]{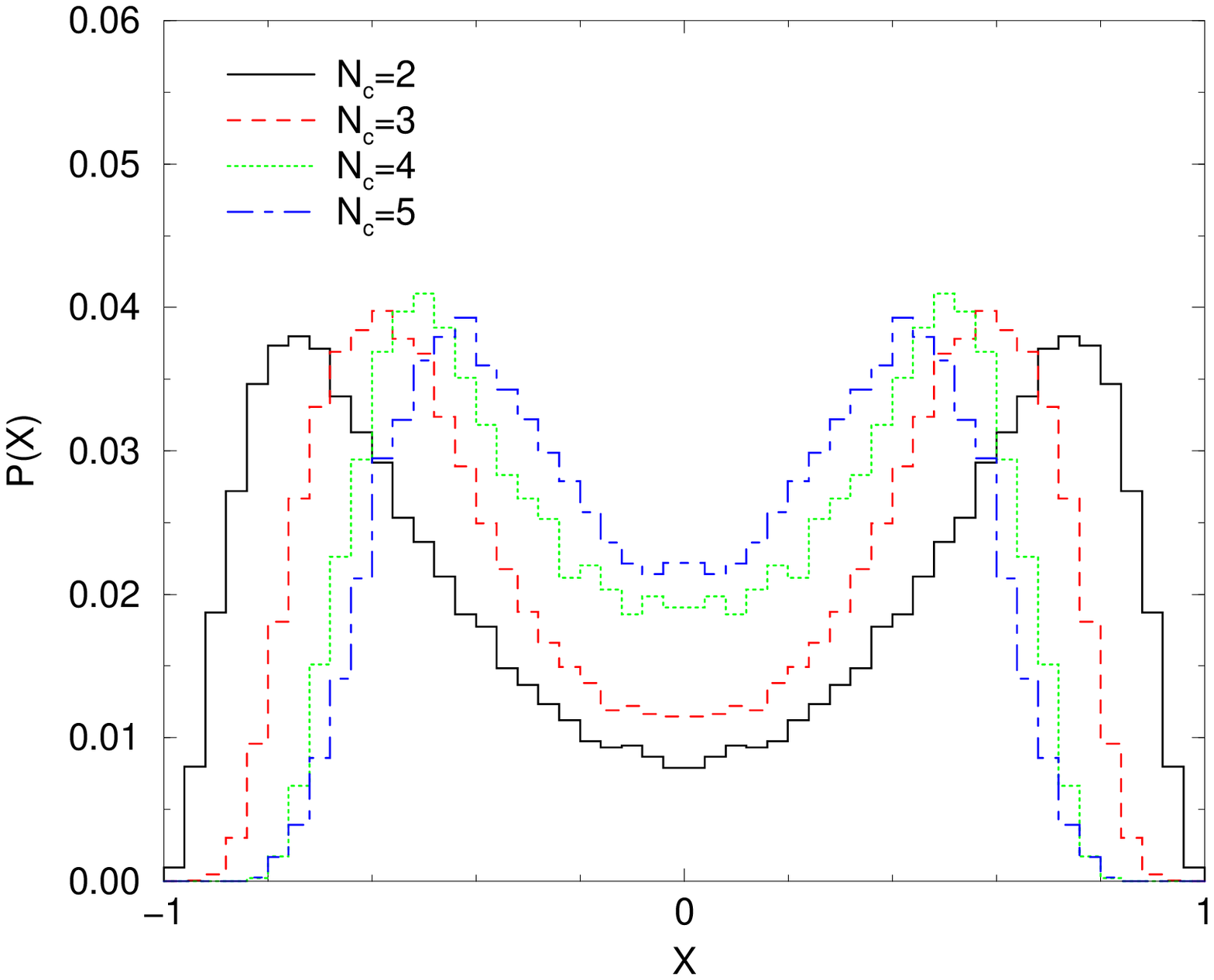}&
\includegraphics[height=6.5cm]{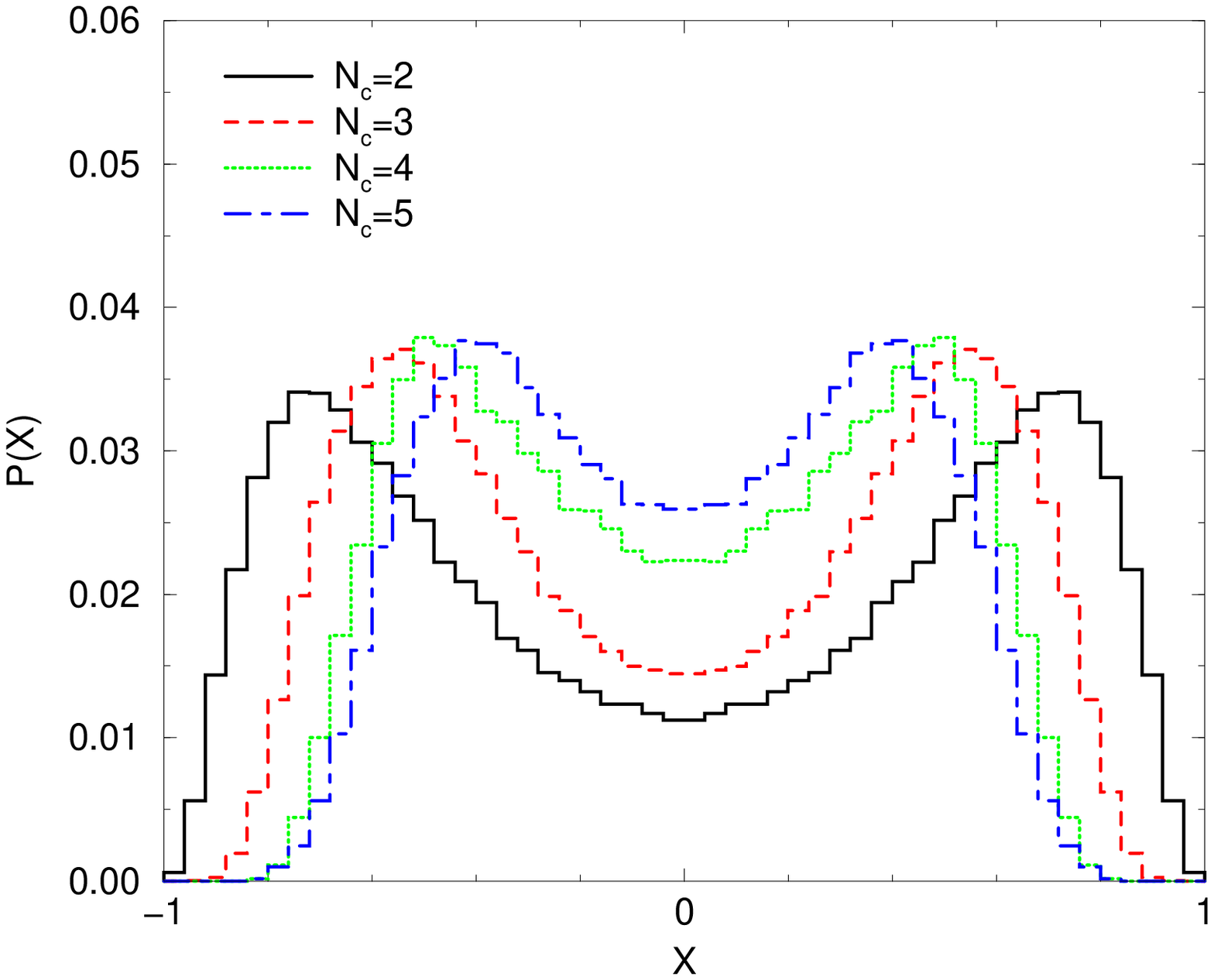}
\end{tabular}
\caption[]{Local chirality histograms for the lowest non-zero modes of
the overlap Dirac operator with $\lambda^2 < 0.03$ (top) and
$\lambda^2 < 0.1$ (bottom) at a volume fraction $f_V=1.0\%$ (left
column) and $f_V=2.0\%$ (right
column) of sites
with the largest wave function density $\psi^\dagger \psi(x)$.}
\label{fig:sun_2.0}
\end{figure}

\begin{figure}[htb]
\begin{tabular}{cc}
\includegraphics[height=6.5cm]{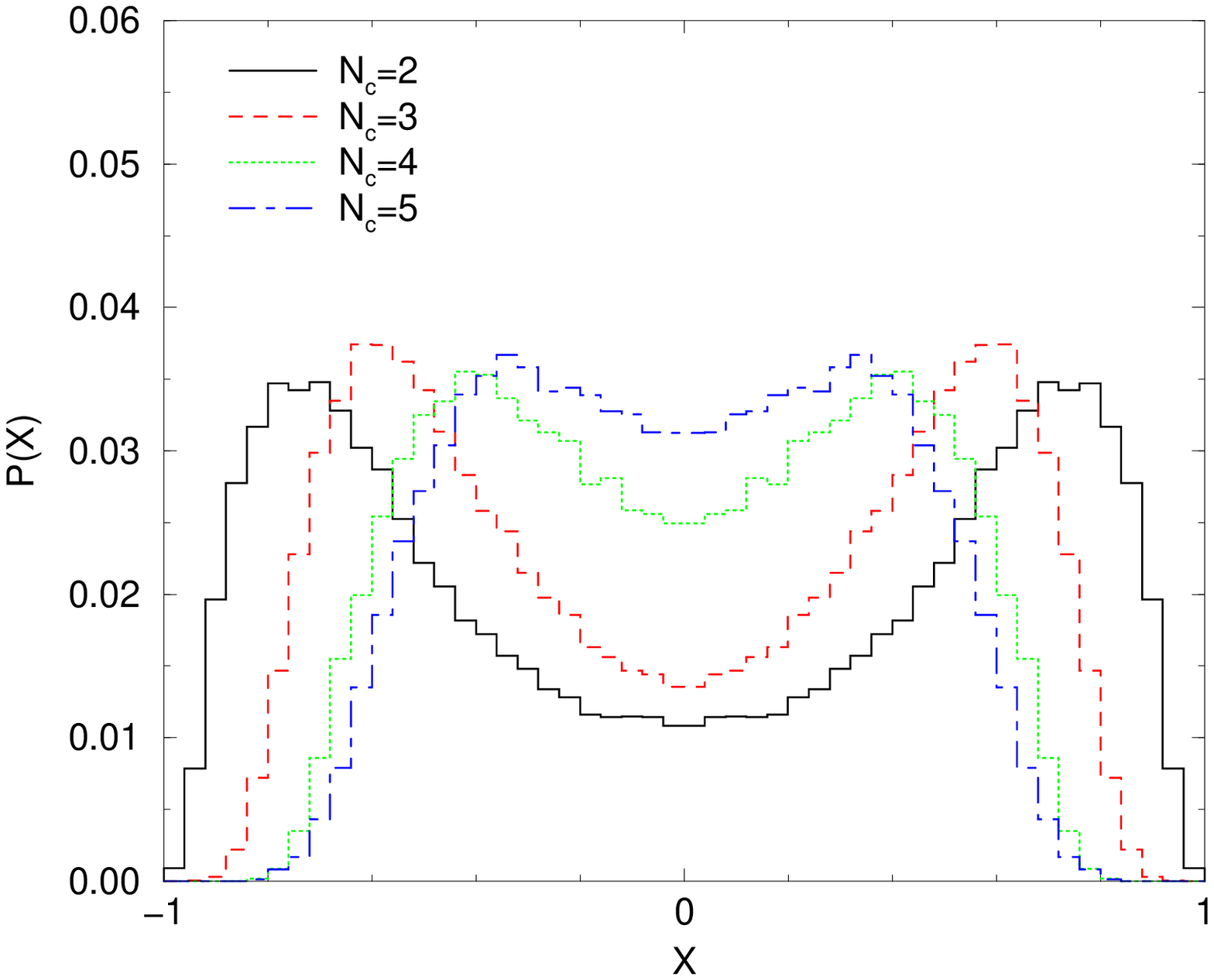}&
\includegraphics[height=6.5cm]{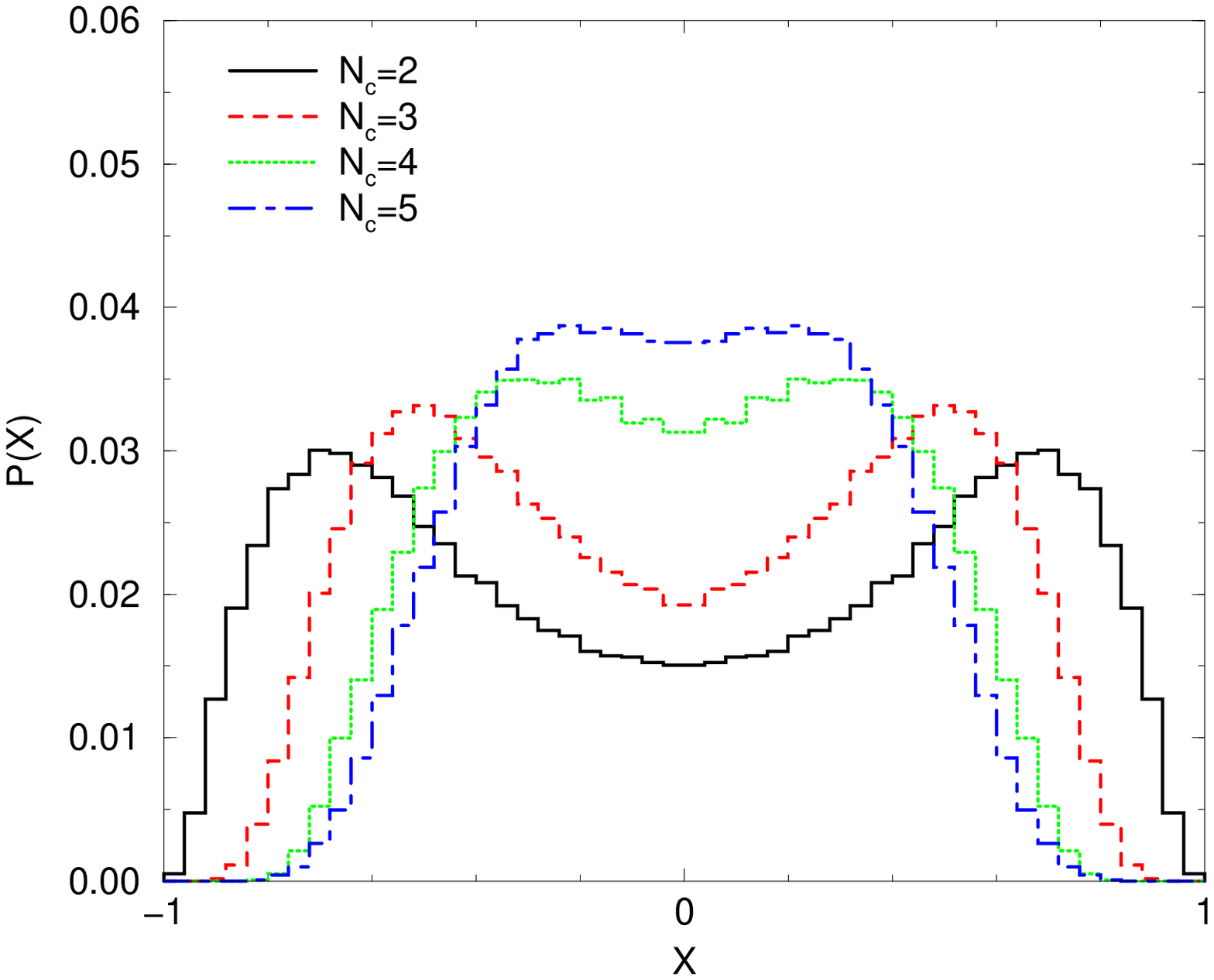}\\
\includegraphics[height=6.5cm]{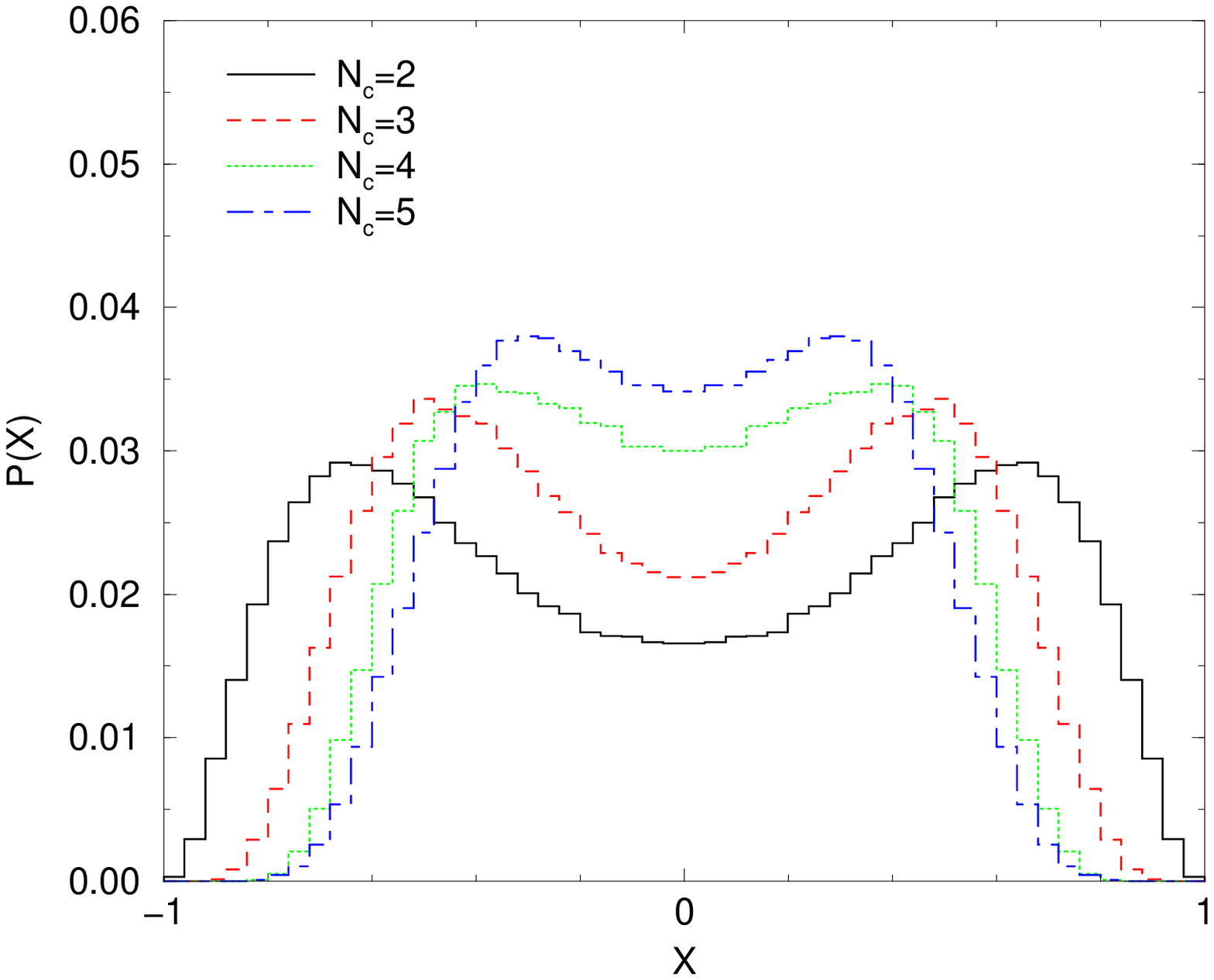}&
\includegraphics[height=6.5cm]{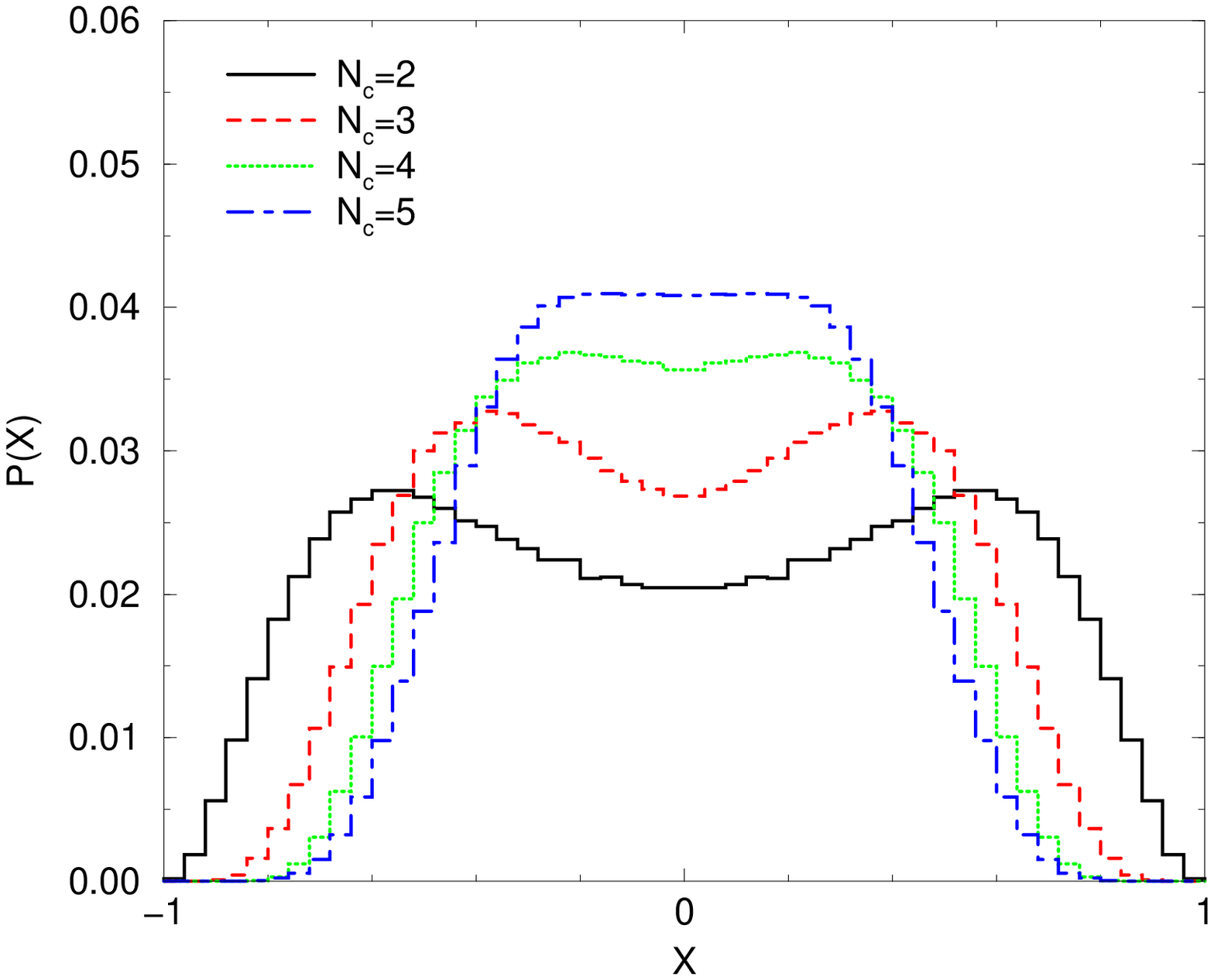}
\end{tabular}
\caption[]{Local chirality histograms for the lowest non-zero modes of
the overlap Dirac operator with $\lambda^2 < 0.03$ (top) and
$\lambda^2 < 0.1$ (bottom) at a volume fraction $f_V=6.25\%$ (left
column) and $f_V=12.5\%$ (right column) of
lattice sites with the largest wave function density $\psi^\dagger
\psi(x)$.}
\label{fig:sun_12.5}
\end{figure}

In figure \ref{fig:sun_2.0} and \ref{fig:sun_12.5} we show the results
of such a calculation. For each eigenmode we identify a volume
fraction $f_V$ of the lattice sites for which the wave function
density $\psi^\dagger \psi(x)$ is largest and include the
corresponding value $X(x)$ in the distribution. Figure
\ref{fig:sun_2.0} shows the local chirality probability distributions
$P(X)$ at a volume fraction $f_V=1.0\%$ (left plots) and $f_V=2.0\%$
(right plots), while figure \ref{fig:sun_12.5} shows the probability
distributions for $f_V=6.25\%$ (left plots) and $f_V=12.5\%$ (right
plots). In both figures the upper plots include all the non-zero
eigenmodes with $\lambda^2<0.03$ and the lower plots the ones with
$\lambda^2<0.1$, respectively. In order to make the plots more clear
we symmetrise the probability distributions $P(X)$ in all cases,
i.e.~$P(-X) = P(X) = 1/2 (\tilde P(X) + \tilde P(-X))$ where $\tilde
P(X)$ is the original unsymmetrised distribution.

In the upper plots of figure \ref{fig:sun_2.0} a clear double peak
structure is visible for all the gauge groups we studied.  Thus the
lowest non-zero eigenmodes of the overlap Dirac operator appear to be
chiral in their peaks, however, this is less pronounced as $N_c$
increases.  As we include more and more non-zero modes while keeping
$f_V$ fixed, cf.~lower plots in figure \ref{fig:sun_2.0}, the double
peaking becomes slightly weaker, but the modes appear still rather
chiral in their peaks. The weakening is most evident for SU(2) and
becomes gradually less apparent with increasing $N_c$. This suggests
that for SU(2) and SU(3) the non-zero eigenmodes become less chiral,
as the modes are lifted away from zero, while for SU(4) and SU(5) the
modes below $\lambda^2<0.1$ appear to be all similarly (but more
weakly) chiral in their peaks.

Of course one might object that comparing the distributions at a fixed
volume fraction $f_V$ is not the most sensible approach since it does
not take into account how big a fraction $f_W$ of the total wave
function density is considered. In order to assess this issue we
calculate for each volume fraction $f_V$ the corresponding
contribution to the total wave function. In table
\ref{tab:locality_measure_for_all_Nc} we collect the wave function
fraction $f_W$ for each ensemble of gauge fields as a function of
$f_V$ for all non-zero modes with $\lambda^2<0.1$. We observe that for
any given $f_V$ we account for less and less of the total wave
function as we increase $N_c$. In fact, the ratio $f_W/f_V$ in the
limit $f_V\rightarrow 0$, i.e.~the derivative of $f_W$ with respect to
$f_V$ at $f_V=0$, can serve as a measure of the smoothness or locality
of the eigenmodes.  It is obvious from table
\ref{tab:locality_measure_for_all_Nc} that the non-zero eigenmodes
become less localised with increasing $N_c$ and one should keep that
in mind when comparing the chirality distributions at a fixed volume
fraction.

Since for a volume fraction of $f_V=1.0\%$ or 2.0\% the wave function
fraction $f_W$ is rather small, we show in figure \ref{fig:sun_12.5}
the probability distributions for $f_V=6.25\%$ (plots in the left
column) and $f_V=12.5\%$ (plots in the right column). 

As we go to larger $N_c$ we observe two effects in the
$X$-distributions. First we note that the double peak structure for a
fixed volume fraction becomes less and less pronounced. This is due to
the fact that less of the most active lattice sites have a
distinguished chirality as we increase $N_c$. Secondly, the two peaks in
the distributions move further away from $|X|=1$ towards $X=0$,
suggesting that even for the lattice sites where there is a distinguished
chirality it is much weaker for larger $N_c$. This can for example be
seen by looking at the upper left plot in figure
\ref{fig:sun_2.0}. While the SU(2) probability distribution has its
maximum at $|X|\simeq 0.86$, we find that there is no lattice site at
all in the whole SU(5) ensemble contributing to such a high value of
$|X|$.

All this holds true independent of the volume fraction we consider. On
the other hand, it is clear from table
\ref{tab:locality_measure_for_all_Nc} that, again for a fixed volume
fraction, we account for less and less of the wave function fraction
when $N_c$ is increased. In other words, if we are to retain the
double peak structure in the $X$-distribution we have to choose a
smaller and smaller volume fraction corresponding to an even smaller
and smaller wave function fraction.  One might therefore ask whether
the double peak structure survives the large $N_c$ limit at all,
i.e.~whether the wave function fraction contributing to the double
peak structure remains non-zero.

In order to address this question in a quantitative way we resort to
the first, second and fourth moment of the chirality angle $X$, i.e.,
\begin{equation} \label{eq:average_Xn}
\langle |X|^n \rangle = \sum_i p_{X_i} |X_i|^n, \quad n=1,2,4.
\end{equation}
So for $n=1$ this is just the average value of the chirality angle
$|X|$ weighted with the probability distribution $p_{X_i}$
\cite{Horvath:2002gk} which approximately gives the position of the
peak (if there is one). In tables \ref{tab:average_X},
\ref{tab:average_X2} and \ref{tab:average_X4} we collect the values of
the moments $\langle |X|^n \rangle, n=1,2,4$ from all eigenmodes with
$\lambda^2<0.1$ for different $N_c$ and volume fractions $f_V$. From
the values of $\langle |X| \rangle$ it can be clearly seen that the
peaks move towards zero as we increase $N_c$ and/or the volume
fraction $f_V$. By comparing these numbers with the local chirality
histograms in figures \ref{fig:sun_2.0} and \ref{fig:sun_12.5}, one
can estimate a value $\langle |X| \rangle \sim 0.3$ for distributions
which are essentially flat, i.e.~for which the double peaking has
disappeared. Taking this value as a sharp criterion we find that for
the double peaking to survive in the large $N_c$ limit (assuming
$1/N_c^2$ corrections) one is not allowed to include more than a
volume fraction $f_V \lesssim 1 \%$. From table
\ref{tab:locality_measure_for_all_Nc} we find that for $N_c
\rightarrow \infty$ this correspond to a fraction of the wave function
$f_W \lesssim 2 \%$.

Another approach is to calculate from the moments of the chirality
angle the cumulant $C = 3 - \langle |X|^4 \rangle/\langle
|X|^2\rangle^2$. Such a quantity should tend towards two distinct
values depending on whether the probability distribution is single or
double peaked. The values of this cumulant are collected in table
\ref{tab:average_cumulant}. If the cumulant assumes its maximum value
$C=2.0$, the underlying distribution corresponds to one which is
peaked at just two values, say e.g.~$X = \pm 1$. On the other hand,
the minimum value $C=0.0$ is assumed for a single peaked gaussian-like
distribution centered around zero. This situation is present if the
values of $X$ are more or less randomly distributed, i.e.~if we
choose, e.g., a very high volume fraction $f_V$. Again from an
inspection of the plots in figures \ref{fig:sun_2.0} and
\ref{fig:sun_12.5} and the numbers in table \ref{tab:average_cumulant}
we find that a value of $C\simeq 1.0$ corresponds to a situation where
the double peaking has essentially disappeared.  One can now attempt
to find the minimum volume fraction $f_V$ for which the criterion $C
\gtrsim 1.0$ is still fulfilled in the limit $N_c \rightarrow
\infty$. As before, it seems that if the double peaking is to survive
in the large $N_c$ limit one can not include more than $f_V \gtrsim
2\%$ corresponding roughly to $f_W \lesssim 4\%$ and we are
essentially lead to the same conclusion as in the previous paragraph.

Although all these considerations look rather quantitative, they
should not distract from the fact that the whole discussion at this
stage is still understood to be on a very qualitative level.
Nevertheless, we think it is safe to conclude that local regions of
definite chirality in the low-lying fermion modes do not survive
the large $N_c$ limit .

At this point the more interesting question about what is the
mechanism behind the obvious suppression if not disappearance of local
chirality at large $N_c$ imposes itself. In the previous section we
have presented some evidence that the ambiguity in the topological
charge is caused by the presence of small instantons with $\rho
\lesssim 2.5 a$ and that, because these small instantons are
suppressed for larger $N_c$, this ambiguity quickly becomes rare. It
is therefore not far fetched to suggest that the suppression of local
chirality at large $N_c$ is also caused by that rarer appearance of
small instantons (dislocations). In order to check this proposition we
exclude from our SU(3) ensemble as an experiment all the field
configurations which contain one or several such small instantons. We
do so by monitoring for each configuration the value
$|Q_{\text{max}}|$ introduced before and extract from it an instanton
size $\rho$ using the continuum $Q_{\text{peak}} = 6/\pi^2 \rho^4$ as
discussed in section \ref{sec:topology}. We then produce probability
distributions for the local chirality parameter $X$ as before, but on
the reduced ensemble corresponding to a given cut-off size
$\rho_{\text{cut-off}}$ for the smallest allowed instanton. If our
proposition is correct we would expect to observe a gradual change
from the SU(3) distributions to the ones for SU(4) and eventually
SU(5) when $\rho_{\text{cut-off}}$ is slowly increased. However, the
result of such a calculation is disillusioning: we do not observe any
significant change in the probability distributions at
all. Nonetheless, this exercise leaves us with the realisation that
the suppression of local chirality in the lowest non-zero modes of the
overlap Dirac operator is not due to very small instantons
(dislocations) causing ambiguities in the topological charge and
therefore a more detailed investigation of the local structures in the
chiral densities of those eigenmodes is now in order.

%\clearpage
%%%%%%%%%%%%%%%%%%%%%%%%%%%%%%%%%%%%%%%%%%%%%%%%%%%%%%%%%%%%%%%%%%
\subsection{Correlation functions} 
\label{sec:correlation_functions}

\begin{figure}[htb]
\begin{tabular}{lr}
\includegraphics[height=6.5cm]{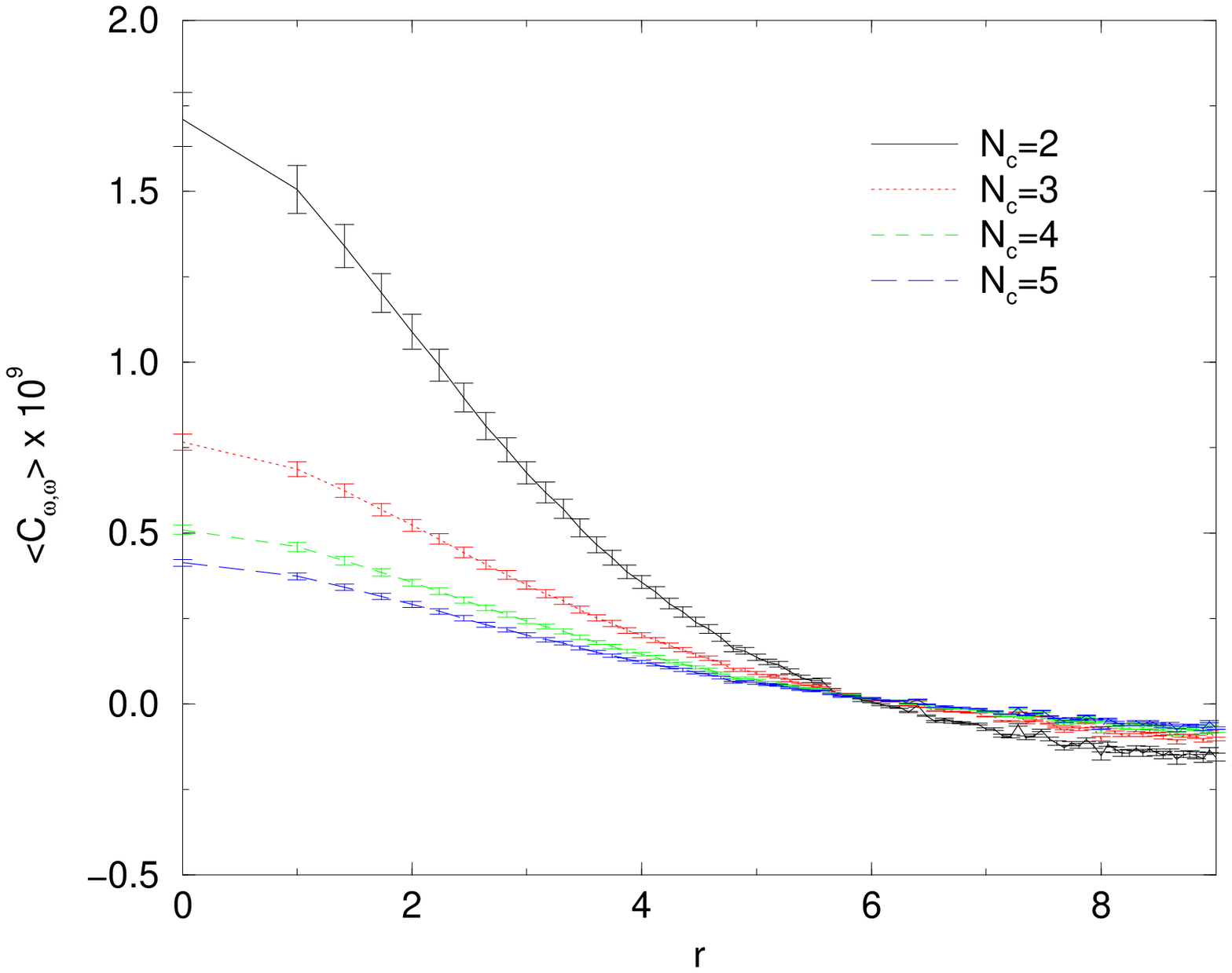} & 
\includegraphics[height=6.5cm]{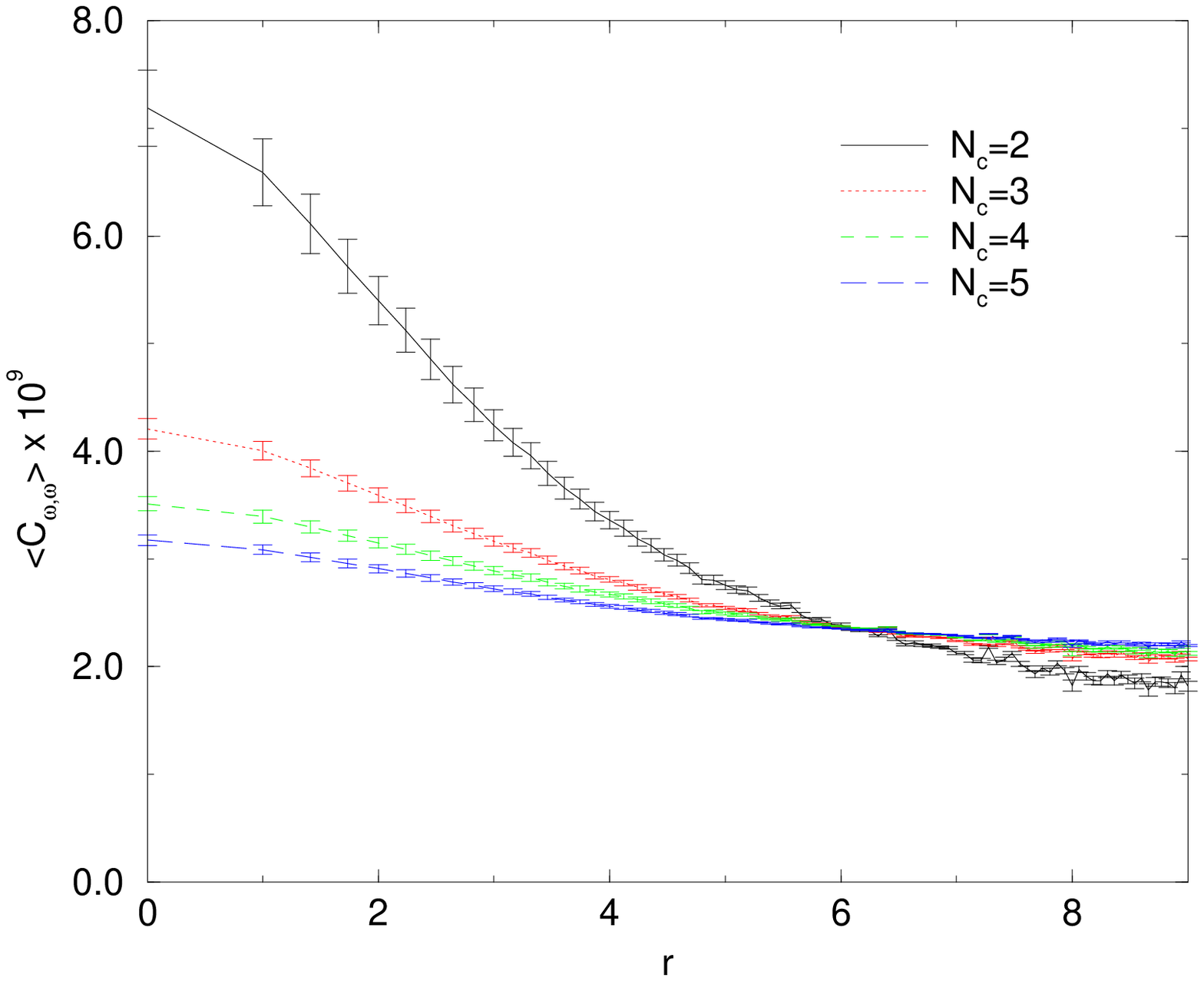} \\
\includegraphics[height=6.5cm]{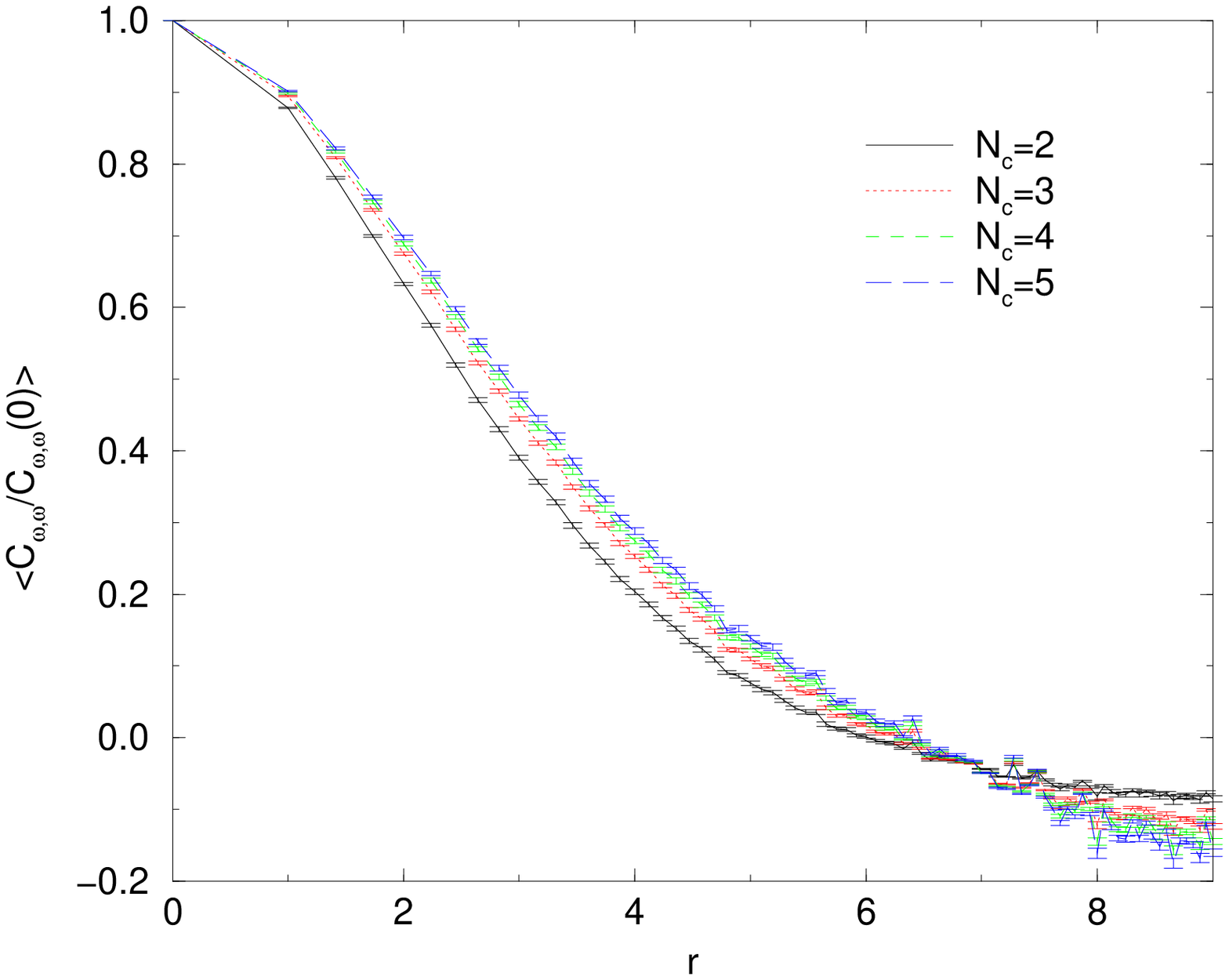} &
\includegraphics[height=6.5cm]{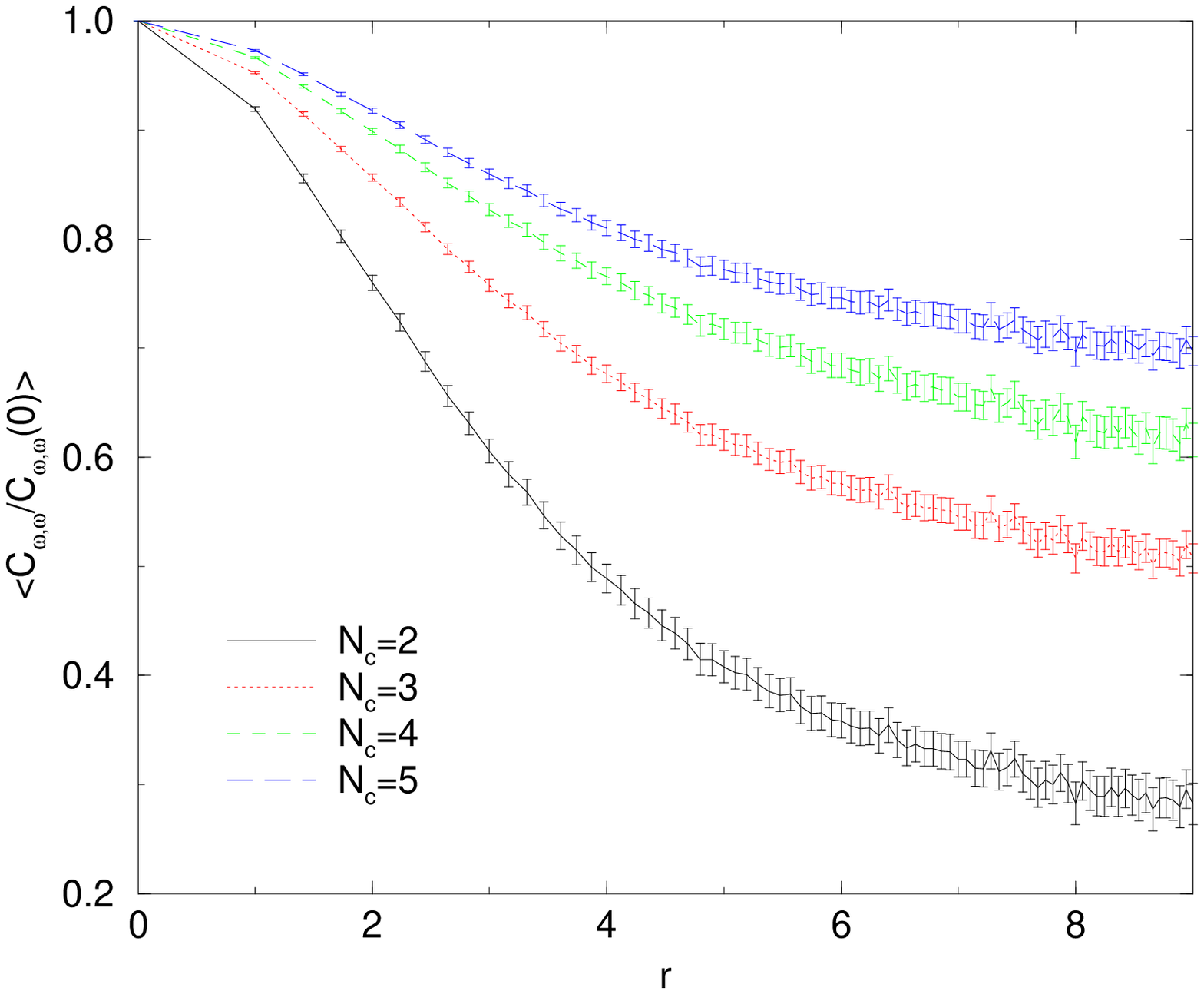}
\end{tabular}
\caption[]{Autocorrelation functions of the pseudoscalar density for
the near-zero modes (left column) and the zero modes (right column).}
\label{fig:ww_correlations}
\end{figure}

One way of studying local structures in the eigenmodes is to examine
correlation functions \cite{DeGrand:2000gq}. Thus we now consider the
autocorrelation function of the pseudoscalar (chiral) density
$\omega(x) = \psi^\dagger(x) \gamma_5 \psi(x)$ obtained from the
eigenmodes $\psi(x)$ of the overlap Dirac operator $D(0)$,
\begin{equation}\label{eq:Cww}
C_{\omega,\omega}(r) = \frac{1}{V} \int d^4x \omega(x)
 \frac{1}{\Omega_3(r)} \int_{|x-y|=r} d^4y \omega(y). 
\end{equation}
Here, $\Omega_3(r)$ is the surface of the four-dimensional sphere with
radius $r$. In figure \ref{fig:ww_correlations} we show the correlator
$C_{\omega,\omega}(r)$ averaged over all the non-zero modes below the
cutoff $\lambda^2 < 0.1$ in the left column and the average over all
the zero modes in the right column. The peak of the autocorrelation
functions at small $r$ indicates that the chiral density is localised
in the near-zero modes and even more so in the zero modes. However,
the amount of localisation is gradually decreasing for both the zero
modes and the non-zero modes for increasing $N_c$ as can be seen from
the values of the autocorrelation functions at the origin,
$C_{\omega,\omega}(0)$. For comparison we list the values in table
\ref{tab:correlation_normalisation} and
\ref{tab:correlation_normalisation_zeromodes} in the appendix. (Note
that the decrease of $C_{\omega,\omega}(0)$ with $N_c$ is essentially
a reflection of the loss of chirality discussed in the previous
subsection.)  By normalising the correlators of the non-zero modes by
their value at the origin we see that the typical size of a localised
region in the non-zero modes depends only slightly on $N_c$ suggesting
that the bulk of local chirality regions present in the pseudoscalar
density of these eigenmodes have roughly the same size for the
different gauge groups. On the other hand, from the normalised
autocorrelators of the zero modes (lower right plot in figure
\ref{fig:ww_correlations}) it becomes clear that the typical sizes of
local chirality regions dominating the chiral density of the zero
modes become larger quite distinctively for increasing $N_c$.

\begin{figure}[htb]
\begin{tabular}{lr}
\includegraphics[height=6.5cm]{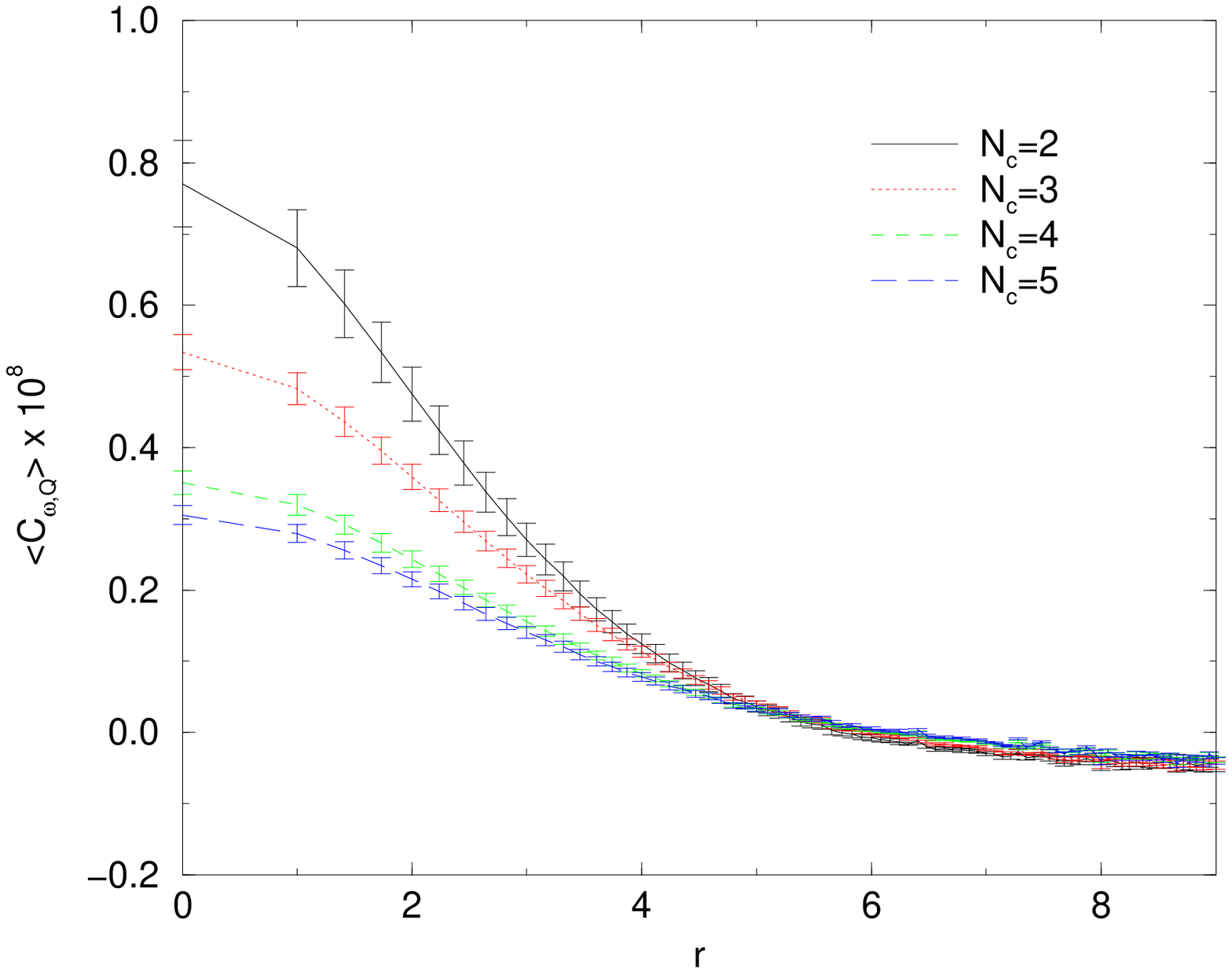} & 
\includegraphics[height=6.5cm]{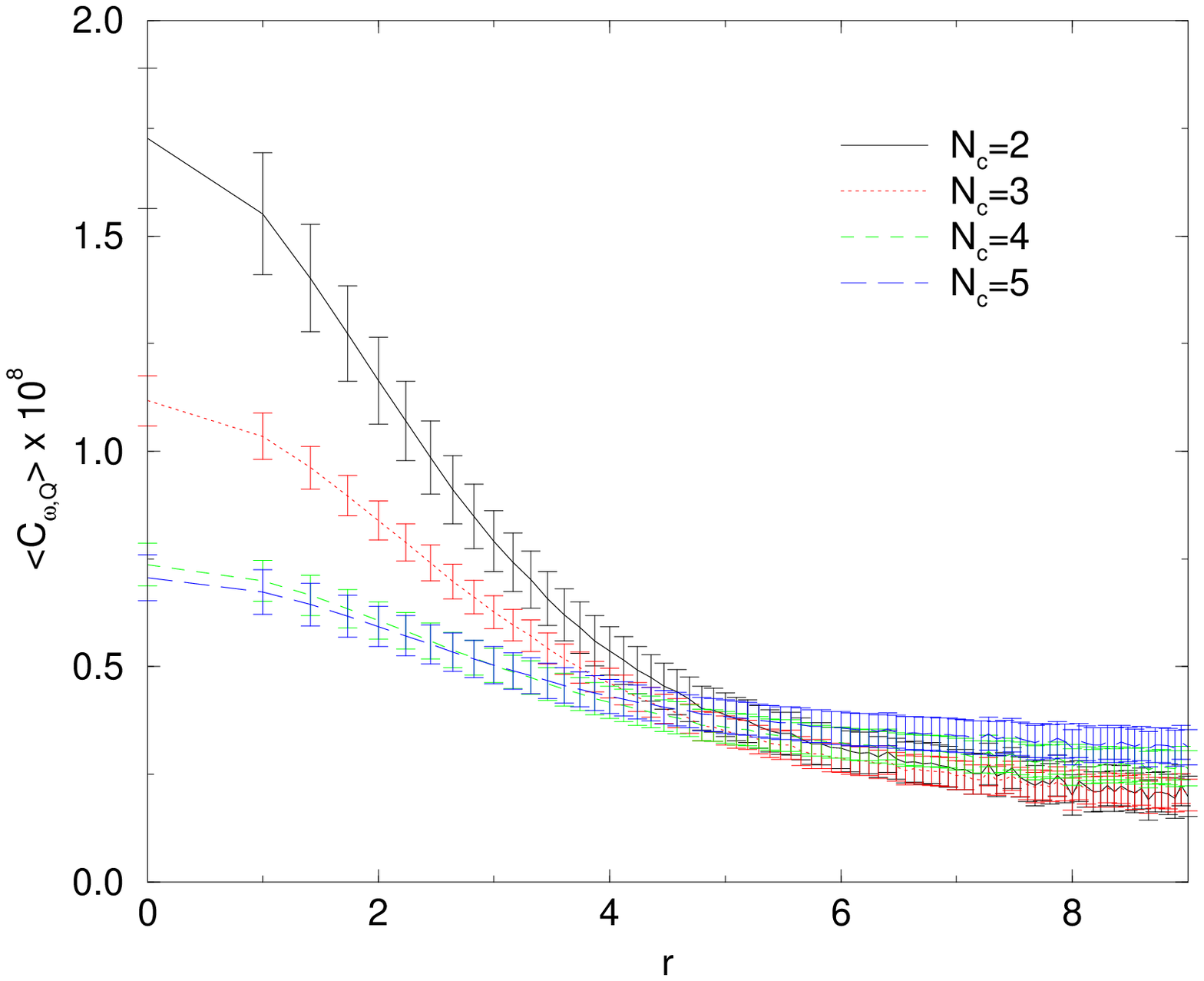} \\
\includegraphics[height=6.5cm]{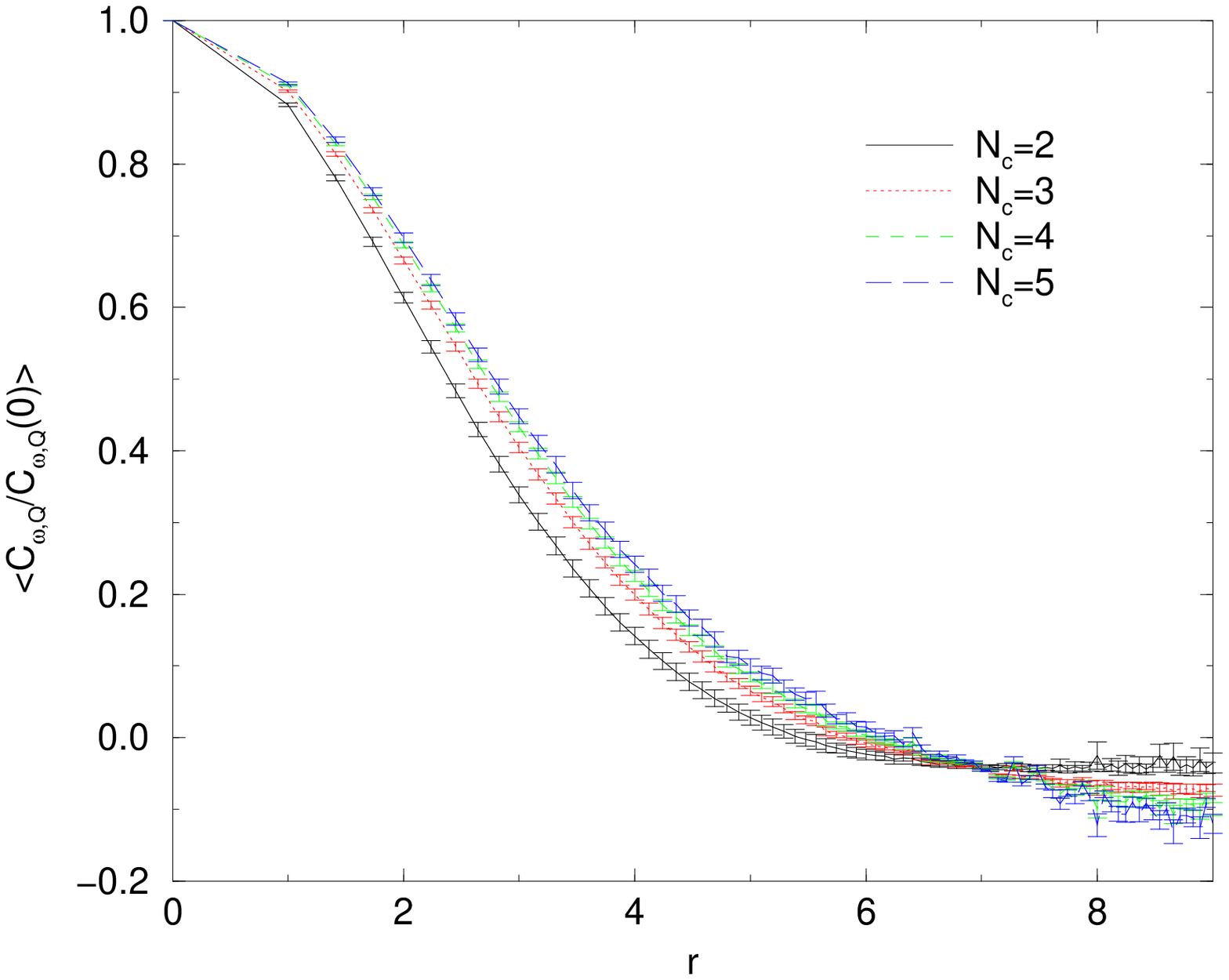} &
\includegraphics[height=6.5cm]{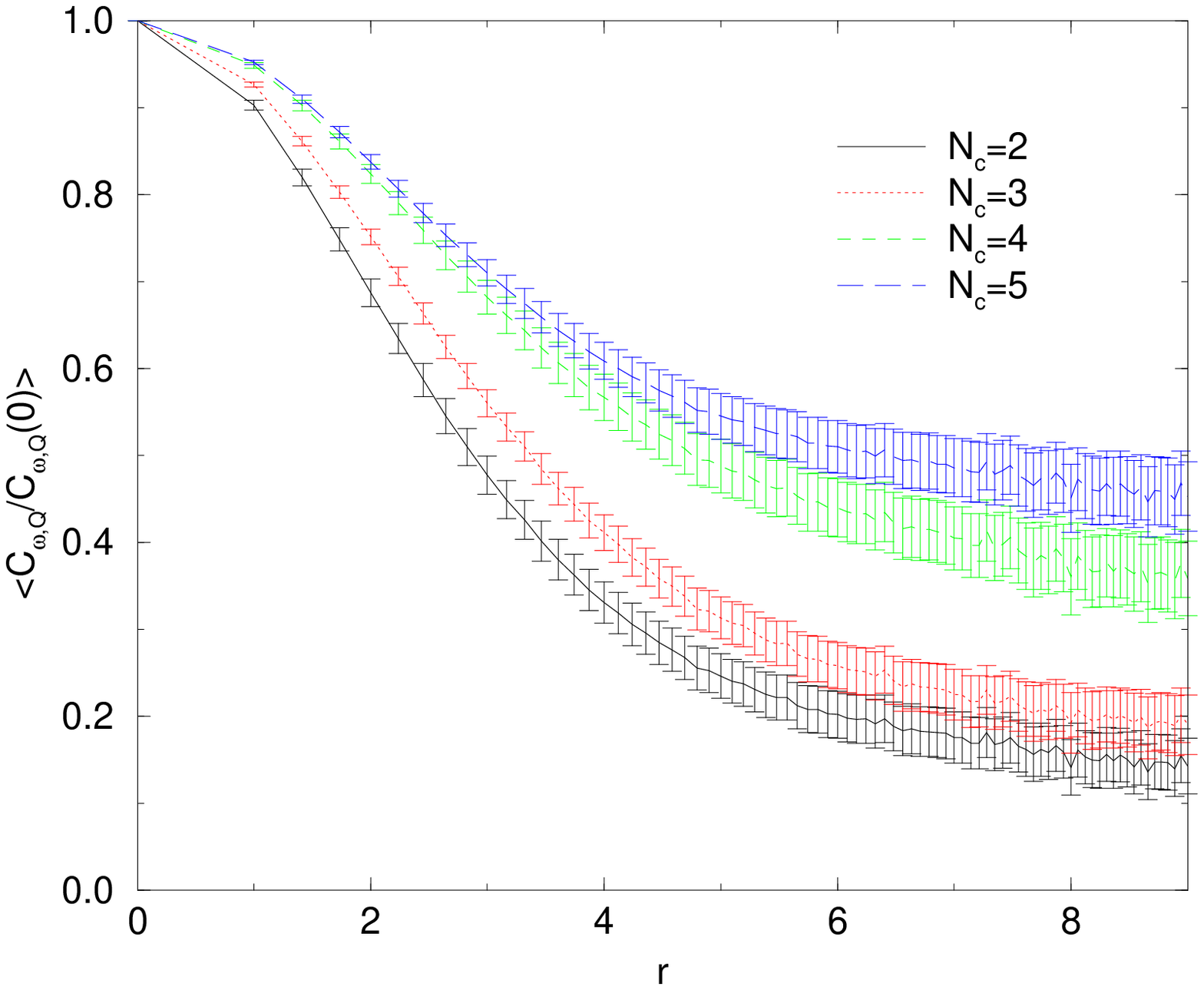}
\end{tabular}
\caption[]{Correlation functions of topological charge density and
pseudoscalar density for the near-zero modes (left column) and the zero
modes (right column). The topological charge density is taken
after 10 cooling sweeps.}
\label{fig:wq_correlations}
\end{figure}

Similarly to eq.~(\ref{eq:Cww}) we construct correlation functions
of the chiral density with the topological charge density obtained
after 10 cooling sweeps,
\begin{equation}\label{eq:Cwq}
C_{\omega,Q}(r) = \frac{1}{V} \int d^4x \omega(x)
 \frac{1}{\Omega_3(r)} \int_{|x-y|=r} d^4y Q(y). 
\end{equation}
The results are shown in figure \ref{fig:wq_correlations} where we show 
$C_{\omega,Q}(r)$ averaged over all the non-zero modes below the
cutoff $\lambda^2 < 0.1$ in the left column and the average over all
the zero modes in the right column.
The features are very similar to the ones observed in the case of the
pseudoscalar density autocorrelation functions. Again the correlators
are peaked at small $r$ showing that the pseudoscalar density is large
where also the topological charge density is large. Again 
$C_{\omega,Q}(r=0)$
decreases for increasing $N_c$ suggesting that the typical regions
with large chirality and topological charge density are less and less
localised. However, the typical sizes of regions where the two densities
correlate change only weakly with $N_c$ for the non-zero modes while
the change is again more distinctive for the zero modes. This can be
seen again by looking at the corresponding correlators normalised by
their value at the origin, cf.~the two lower plots in figure
\ref{fig:wq_correlations}.

Comparing the mixed correlators $C_{\omega,Q}$ with the
autocorrelators $C_{\omega,\omega}$ we observe that the latter are
slightly broader. This is just a reflection of the fact that the
fermionic eigenmodes are generally less localised compared to the
topological charge density due to the different localisation
properties of the probing operators. It is also in accordance with the
size distributions of objects found in the pseudoscalar and
topological charge density, respectively (see section
\ref{sec:instanton_size_distributions}).

All the features observed above are exactly what one would expect from a
model of interacting instantons and anti-instantons. The zero modes
typically couple to the few narrowest (anti-)instantons which overlap 
and interact weakly with each other and with other charges, 
while the non-zero modes couple to
the bulk of broader instantons which interact with each other more
strongly. The fact that the very small instantons are suppressed for
larger $N_c$ while the bulk of instantons is only weakly affected by
going to larger $N_c$ explains the qualitatively different behaviour
of the zero modes and the non-zero modes.

\begin{figure}[htb]
\begin{tabular}{l}
\includegraphics[height=6.5cm]{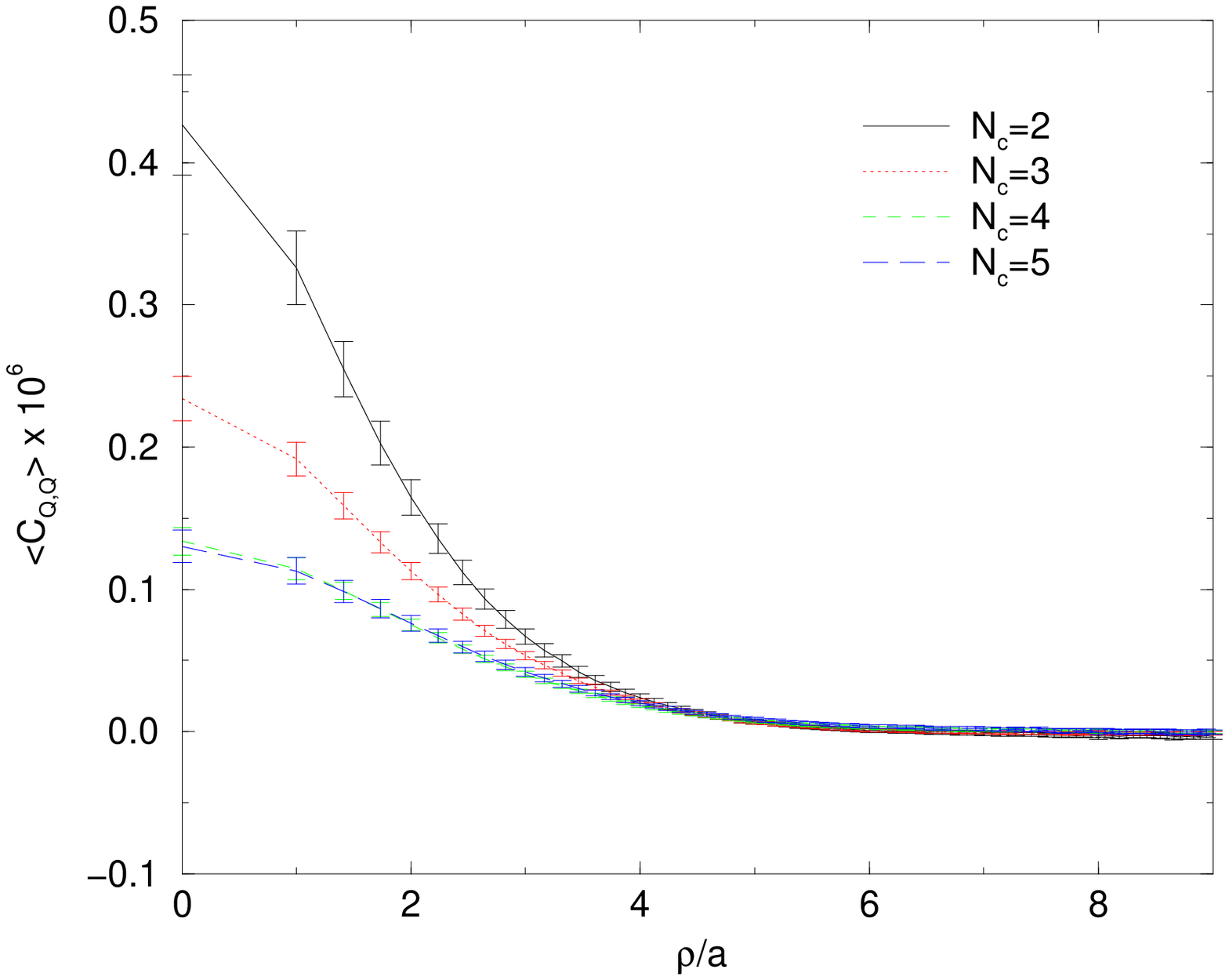} \\ 
\includegraphics[height=6.5cm]{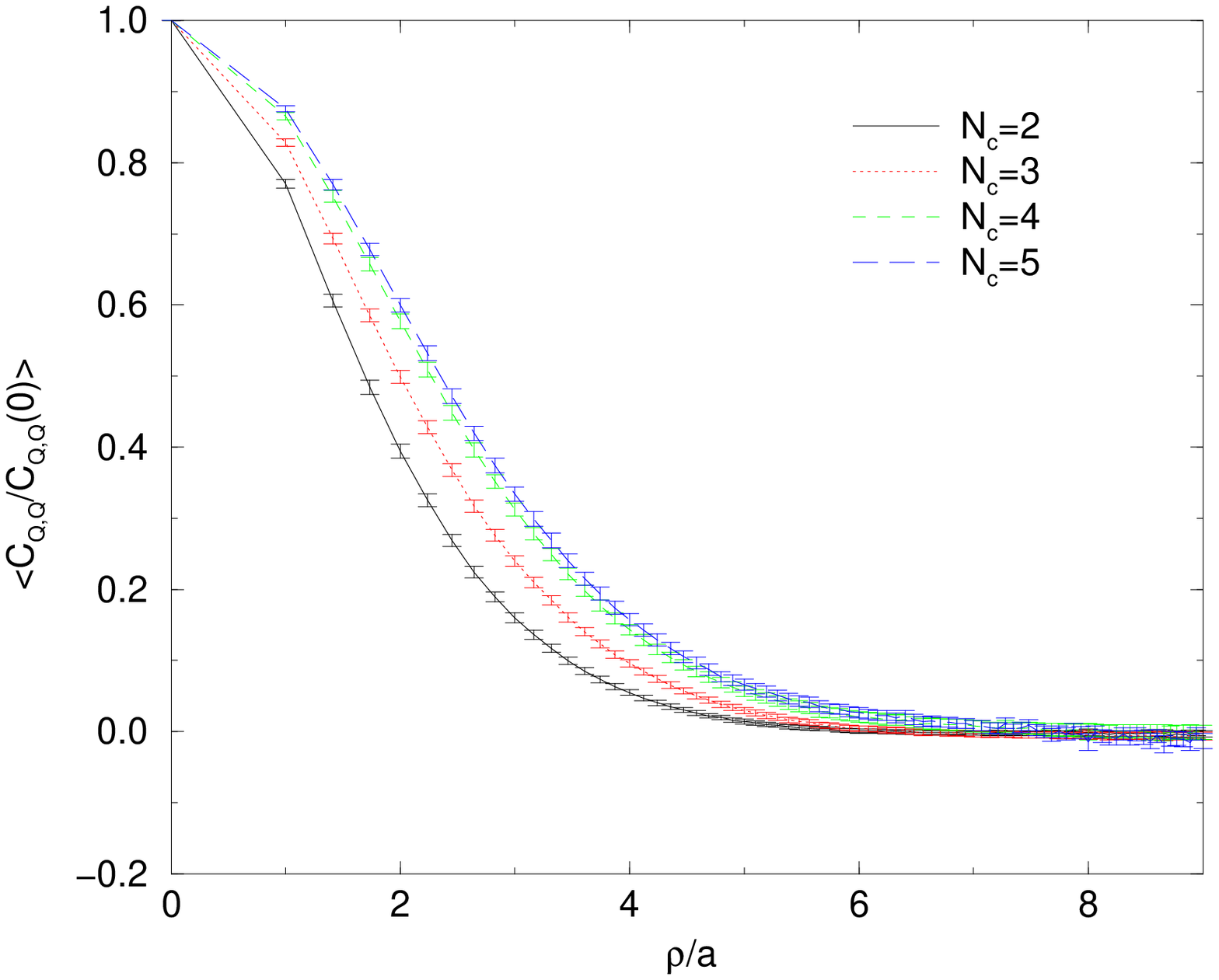} 
\end{tabular}
\caption[]{Autocorrelation functions of the topological charge density
after 10 cooling sweeps.}
\label{fig:qq_correlations}
\end{figure}

For completeness let us now look at the autocorrelation of the
topological charge density. Again we define the correlators as in
eq.~(\ref{eq:Cww}) with the pseudoscalar density $\omega(x)$ replaced
by the topological charge density $Q(x)$. The resulting correlators
 are shown in figure \ref{fig:qq_correlations}.
The characteristic features are very similar to the ones seen in the
autocorrelation functions of the pseudoscalar density of the
near-zero modes (cf.~left column of figure
\ref{fig:ww_correlations}). Again, the peaking of the correlators at
small $r$ indicates that the density is concentrated in localised
regions. However, it is evident that the apparent localisation of the
topological charge is less and less pronounced as we go to larger
$N_c$. This is reflected both in the values of the correlators at the
origin (cf.~table \ref{tab:correlation_normalisation} in the appendix)
as well as in the fall-off of the correlators which becomes slightly
weaker for larger $N_c$. Again, this is most easily seen by
normalising the correlators by their value at the origin, cf.~lower
plot in figure \ref{fig:qq_correlations}.

%\clearpage
%%%%%%%%%%%%%%%%%%%%%%%%%%%%%%%%%%%%%%%%%%%%%%%%%%%%%%%%%%%%%%%%%%
\subsection{Instanton size distributions}
\label{sec:instanton_size_distributions}

To further corroborate our findings in the previous sections we
attempt to generate the size distributions of topological,
instanton-like objects from both the topological charge density and
the (pseudo-)scalar density of the eigenmodes. In order to do so we
identify peaks in the densities and relate the shape of the density
distribution around those peaks directly to the instanton radius
$\rho$.

To be more precise, we assign a size $\rho$ to each peak $\omega(x_0)$
in the (pseudo-)scalar density of the eigenmodes by assuming an
instanton profile
\begin{equation}\label{eq:instanton_eigenmode_profile}
\omega_I(x;x_0) = \pm \frac{2 \rho^2}{\pi^2 (\rho^2+(x-x_0)^2)^3} 
\end{equation}
corresponding to the (pseudo-)scalar density of an eigenmode from a
single instanton configuration. We then make the assumption that the
low-lying eigenmodes are just linear combinations of such instanton
and anti-instanton modes, cf.~eq.~(\ref{eq:mode_mixing}). However, a
priori we do not know the weight with which an (anti-)instanton mode
contributes to a lifted mode. By forming ratios of the (pseudo-)scalar
density at neighbouring points, i.e.~by looking at the shape rather
than just the peak value of the (pseudo-)scalar density we can cancel
out these unknown factors $c_I$. Thus using
eq.~(\ref{eq:instanton_eigenmode_profile}) for the peak and its
neighbouring points we obtain
\begin{equation}\label{eq:instanton_eigenmode_profile_radius}
\rho^2 = \frac{\sqrt[3]{d}}{1-\sqrt[3]{d}}, \quad  d = \frac{1}{8} \sum_{\mu=1}^4 \frac{\omega(x_0 + \hat \mu) + \omega(x_0 - \hat
\mu)}{ \omega(x_0)}.  
\end{equation}

Similarly, we assume for each peak $Q(x_0)$ in the topological charge
density an instanton profile
\begin{equation} \label{eq:instanton_charge_profile}
Q_{I}(x;x_0) =
\frac{6}{\pi^2} \frac{\rho^4}{(\rho^2+(x-x_0)^2)^4}.
\end{equation}
The radius $\rho$ is then calculated from the fall-off of the
topological charge density, i.e.~from the shape of the instanton
profile according to
\begin{equation}\label{eq:instanton_charge_profile_radius}
\rho^2 = \frac{1}{1-\sqrt[4]{d}}, \quad   d =  \frac{1}{8}
\sum_{\mu=1}^4 \frac{Q(x_0 + \hat \mu) + Q(x_0 - \hat \mu)}{ Q(x_0)}.
\end{equation} 

To summarise we essentially apply the same procedure for the
identification of instanton-like objects and for the calculation of
their sizes to both the topological charge density and the (pseudo-)scalar
density of the eigenmodes.

Before discussing the results we should point out the limitations and
the reliability of such an approach. For both the distributions from
the topological charge density and from the (pseudo-)scalar density, the
analysis outlined above is limited by the fact that a peak has to be
prominent enough in order to stand out of the background
fluctuations. This provides a natural cut-off for very broad objects
having a flat profile. This constraint, however, is not rigid and
therefore puts a limitation on the accuracy of the size distribution
for large radii $\rho$, that is, one should exercise caution when
considering the tail of the distribution for large $\rho$.  On the
other hand, due to the finite resolution of the probing operators as
discussed in section \ref{sec:topology}, there is a lower cut-off for
the size of objects which can be seen at all.

Furthermore, a problem which we face by looking at the peaks in the
(pseudo-)scalar density of the fermionic modes is the fact that a
topological object can cause a prominent peak in several of the
eigenmodes, although they might be slightly shifted and distorted. In
order to avoid a multiple counting of those peaks we collect all the
peaks present in the eigenmodes of a given configuration and impose a
minimal distance $r_{\text{min}}$ between any two peaks in order to
count as two separate ones. If two or more peaks lie within this
minimal radius, all are discarded but the one yielding the smallest
radius. While the size distributions from the topological charge
density is obviously not much affected by a change in
$r_{\text{min}}$, the distributions from the (pseudo-)scalar densities
are more sensitive. However, it is interesting to see that a change in
$r_{\text{min}}$ almost exclusively affects only the larger objects
and not the tail of the distribution at small $\rho$.  Another
approach is to use the cumulated density of the
modes in a given configuration, i.e.~to identify the peaks in 
\begin{equation}\label{eq:cumulated_density}
\omega(x) = \sum_{\lambda} \psi_\lambda^\dagger(x) \Gamma \psi_\lambda(x)\, ,
\end{equation}
where the sum is over all the eigenmodes below a given cut-off and
where $\Gamma=\gamma_5$ and $1$ for the pseudoscalar and scalar
density, respectively. In this way, we collect the peaks from just one
density per configuration and so it is less likely to double count
some of the peaks. Moreover, small distortions in the shape of the
peaks will average out in the sum over the eigenmodes.

Taking all these considerations into account we are quite confident
that the size distributions are reasonably accurate in the range of
sizes we are most interested in, that is, in the range $2.0 < \rho/a <
6.0$.

\begin{figure}[htb]
\includegraphics[width=7cm]{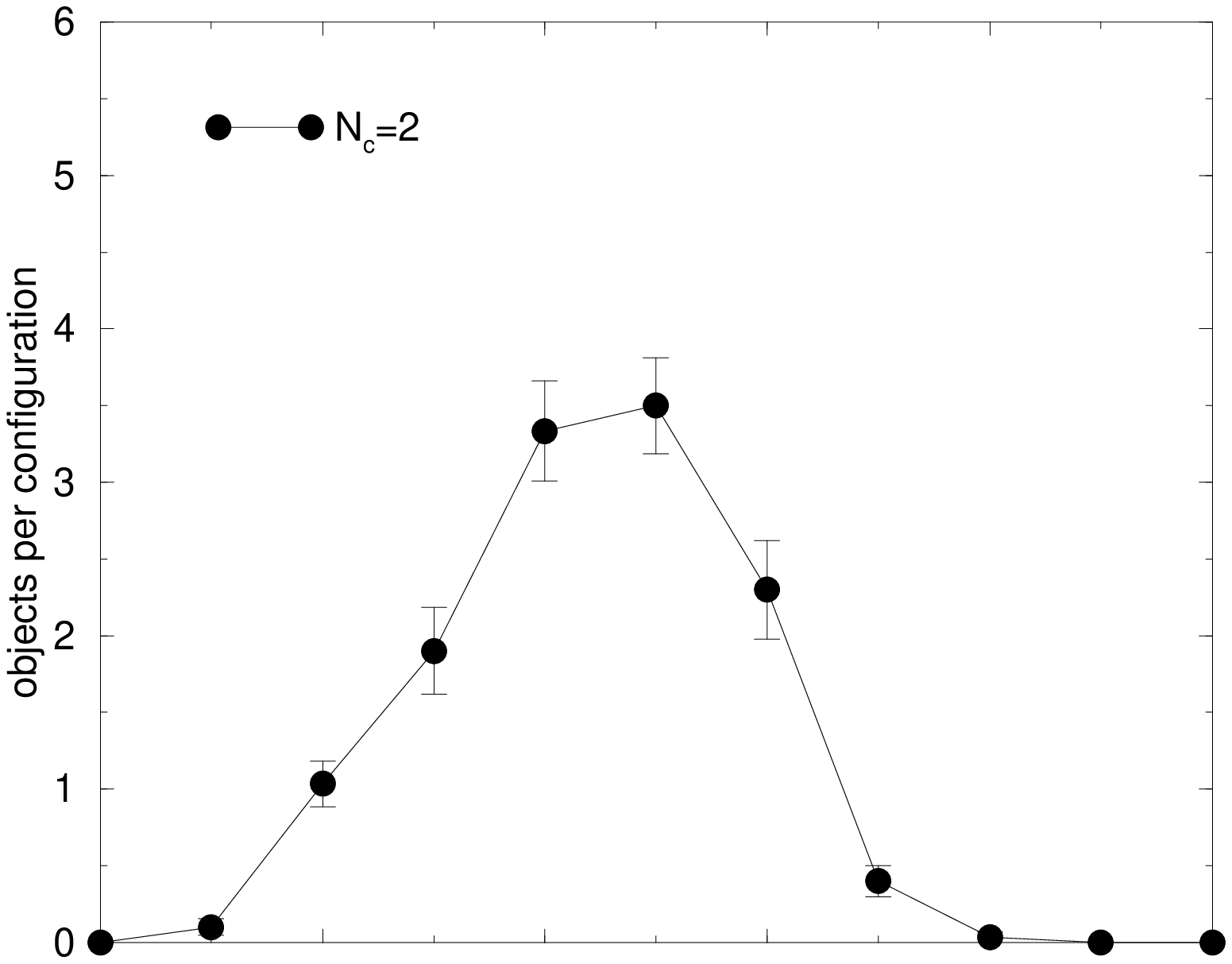}\\
\includegraphics[width=7cm]{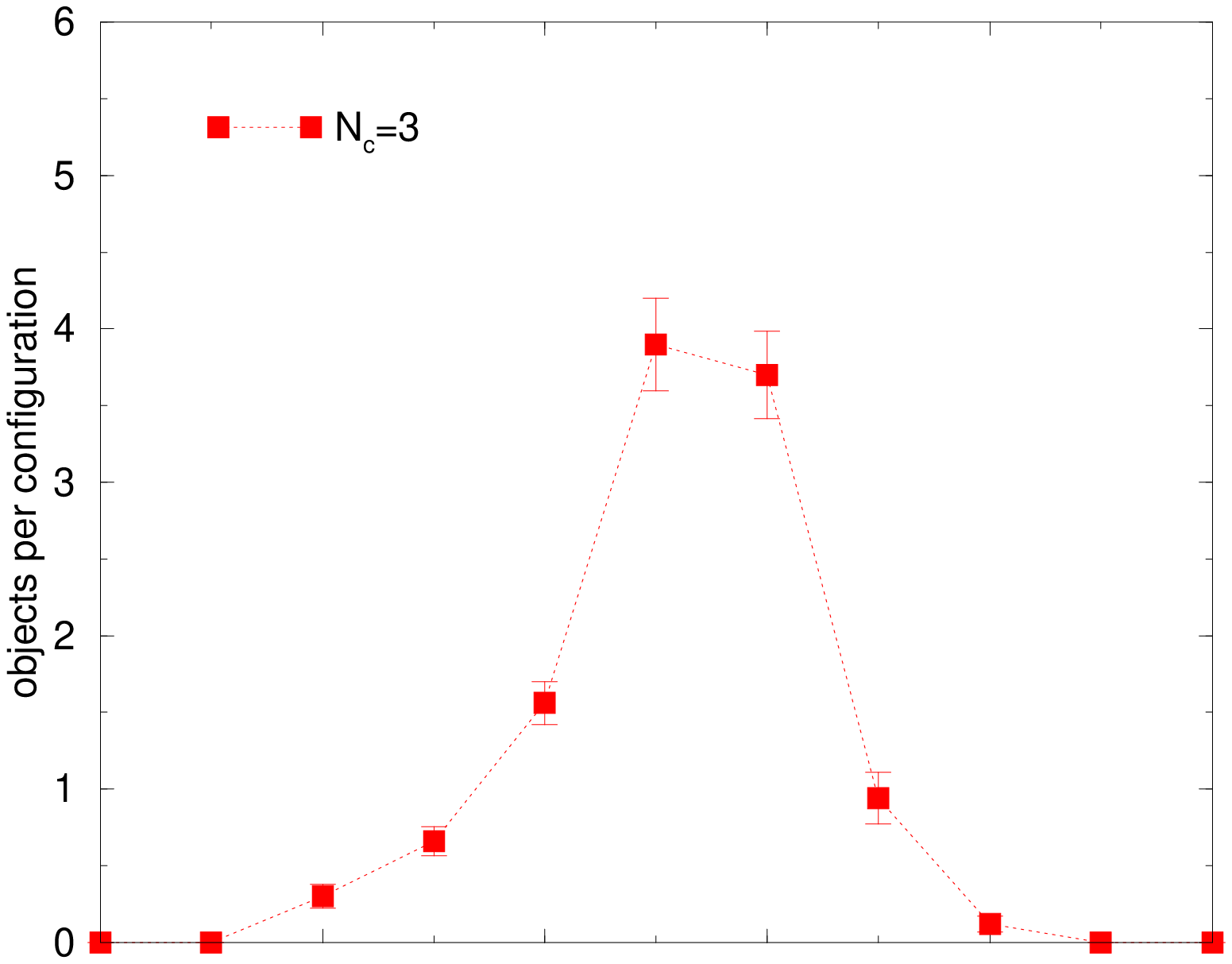}\\
\includegraphics[width=7cm]{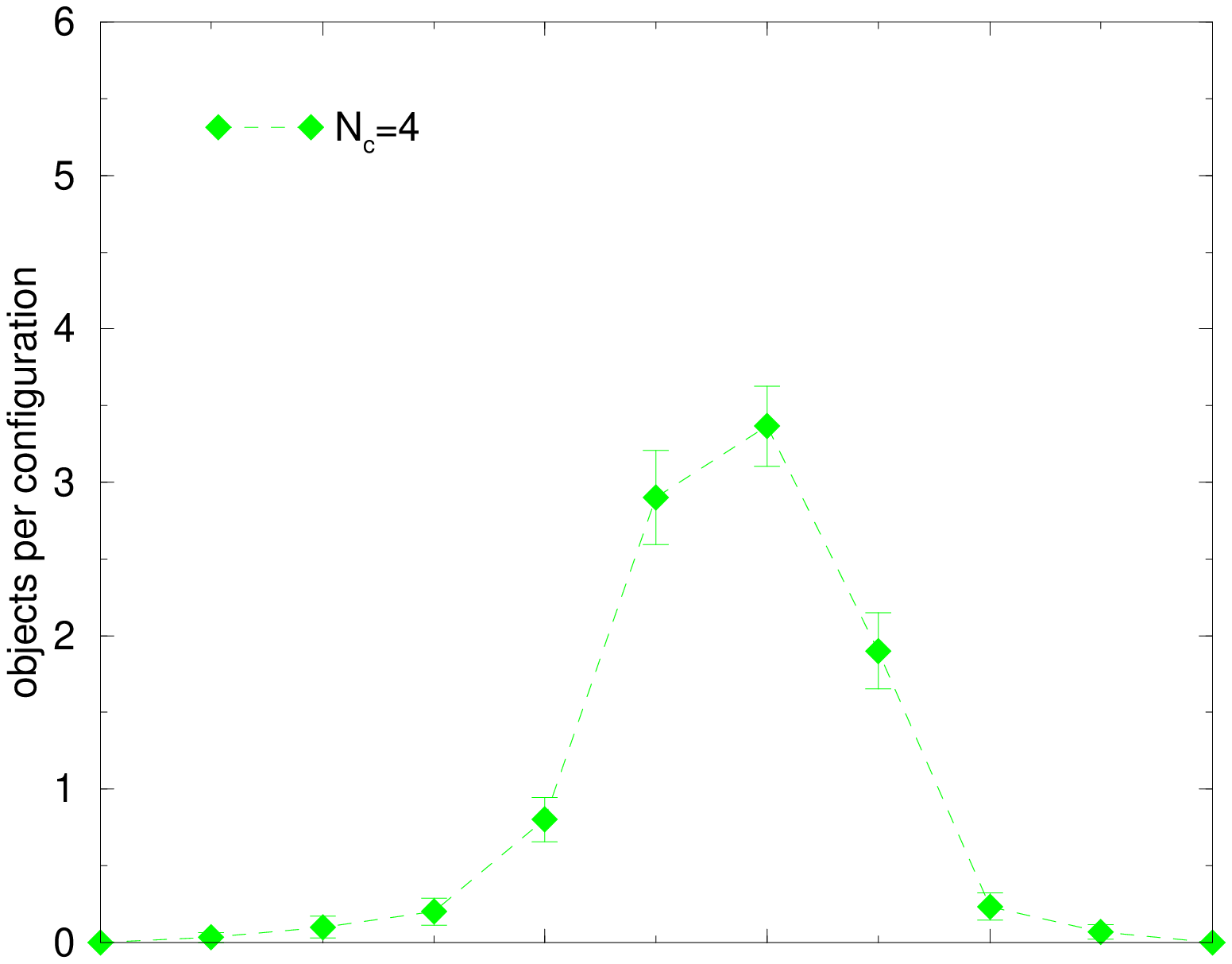}\\
\includegraphics[width=7cm]{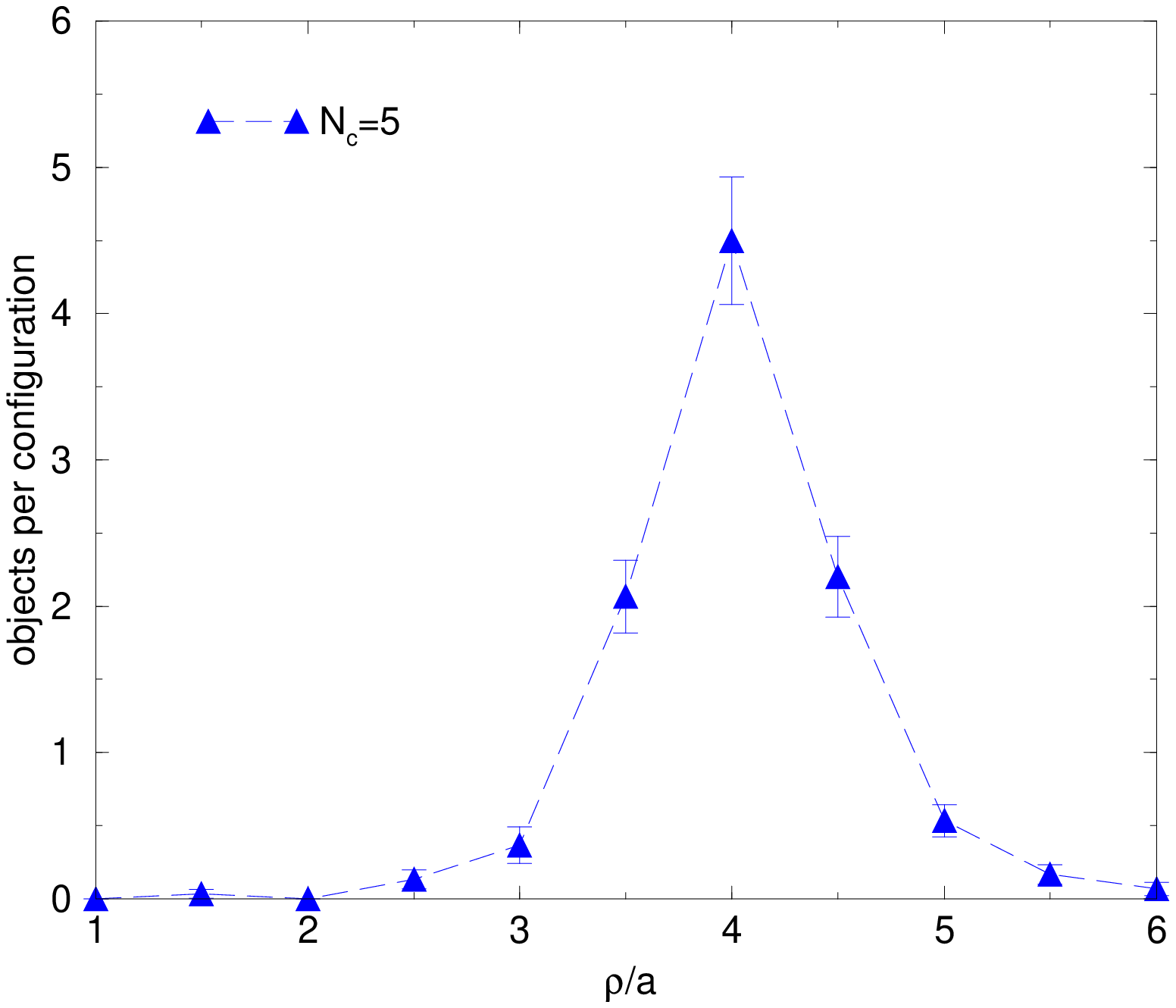}
\caption[]{The size distribution of instanton-like objects as seen in
the topological charge density after 10 cooling
sweeps. The radius is calculated from the shape around the peak value, cf.~eq.~(\ref{eq:instanton_charge_profile_radius}).}
\label{fig:top_charge_size_distribution}
\end{figure}

Figure \ref{fig:top_charge_size_distribution} collects the size
distributions obtained from the topological charge densities for the
different $N_c$. Here we use the topological charge density after 10
cooling sweeps and impose $r_{\text{min}}/a = 2.0$.
There are two important features of the distributions for which a
change is evident as we go to larger $N_c$. Firstly, the peak of the
distribution moves slowly from about $\rho/a \simeq 3.0$ for $N_c=2$ to
about $\rho/a \simeq 4.0$ for $N_c=5$. In physical units this amounts to a
change of roughly 0.12 fm in the size of the bulk instantons,
i.e.~from $\rho \simeq 0.36$ fm to $\rho \simeq 0.48$ fm. It supports
our conclusion from section \ref{sec:correlation_functions} that the
typical size of regions with large topological charge density is
slowly growing and thereby causing only a weak broadening of the
correlation functions of the near-zero modes, which mainly couple to
the bulk of the topological objects.

Secondly, we observe a suppression of small instantons at large $N_c$
which is quite dramatic for instantons smaller than $\rho/a \lesssim
3.0$ and still clearly visible for objects with $\rho/a \simeq
3.5$. Again this supports our conclusion from section
\ref{sec:correlation_functions} where we found that the correlation
functions of the topological charge density with the pseudoscalar
density of the zero modes broaden considerably for larger $N_c$. This
is exactly what one would expect if the zero modes couple mainly to
only a few rather small objects. On the other hand the near-zero modes
are expected to mainly couple to the bulk of instantons not just the
smallest ones. One would therefore expect the correlation functions to
depend only weakly on $N_c$ since the size distribution of the bulk of
instantons is only slightly shifted. The correlation functions in
section \ref{sec:correlation_functions} comply indeed with these
expectations.

\begin{figure}[htb]
\begin{tabular}{cc}
\includegraphics[width=7cm]{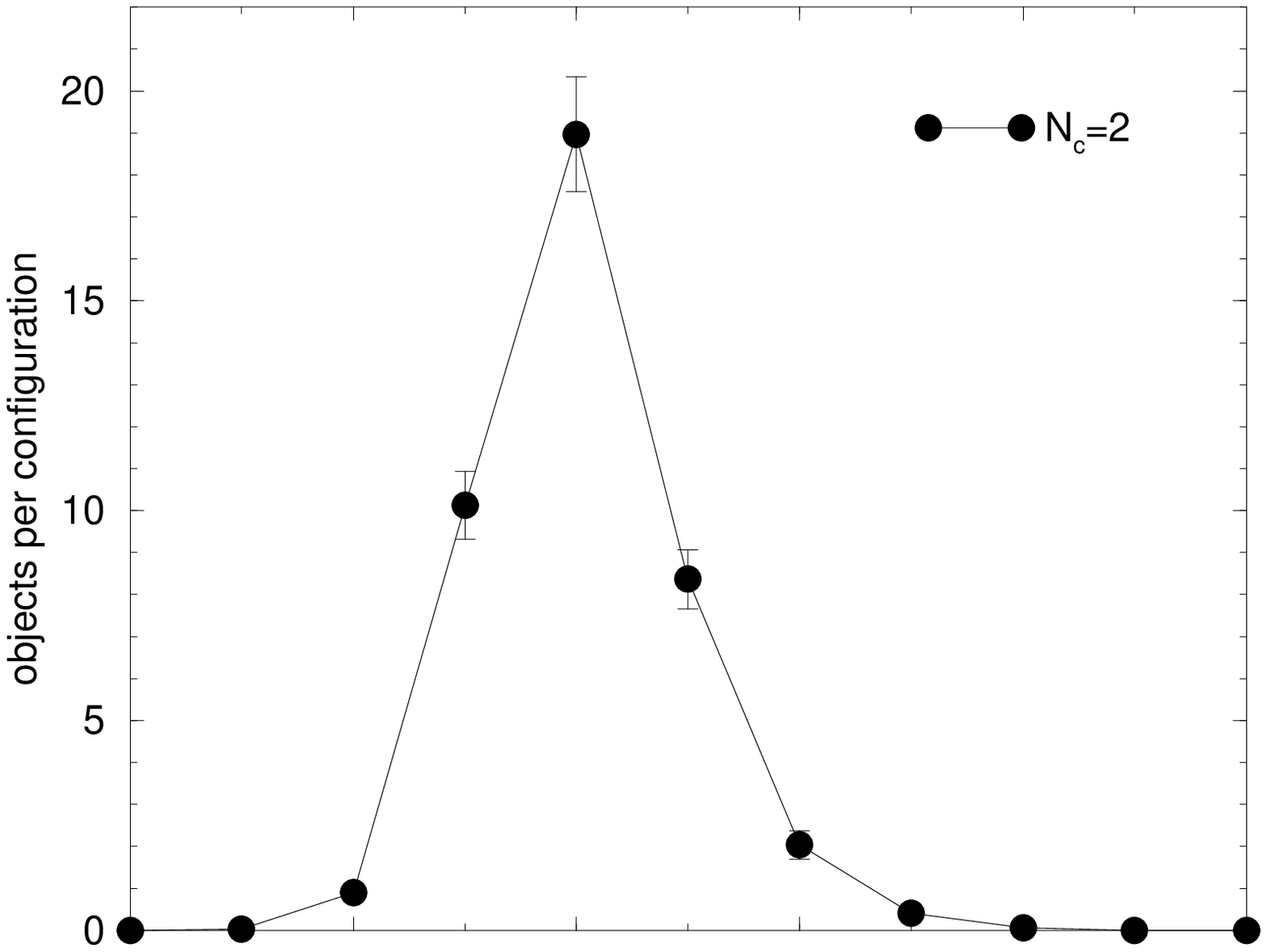} & 
\includegraphics[width=7cm]{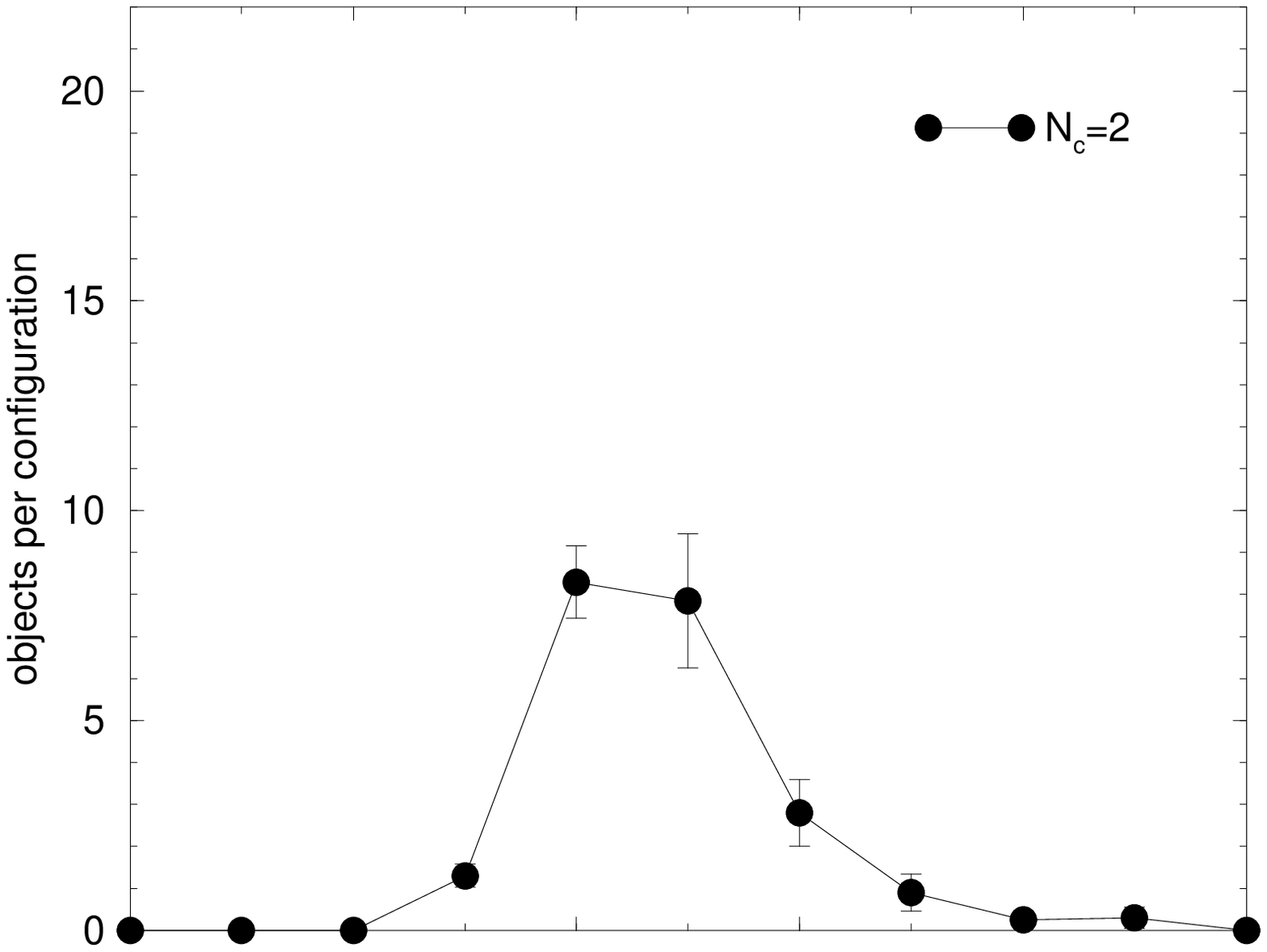}\\
\includegraphics[width=7cm]{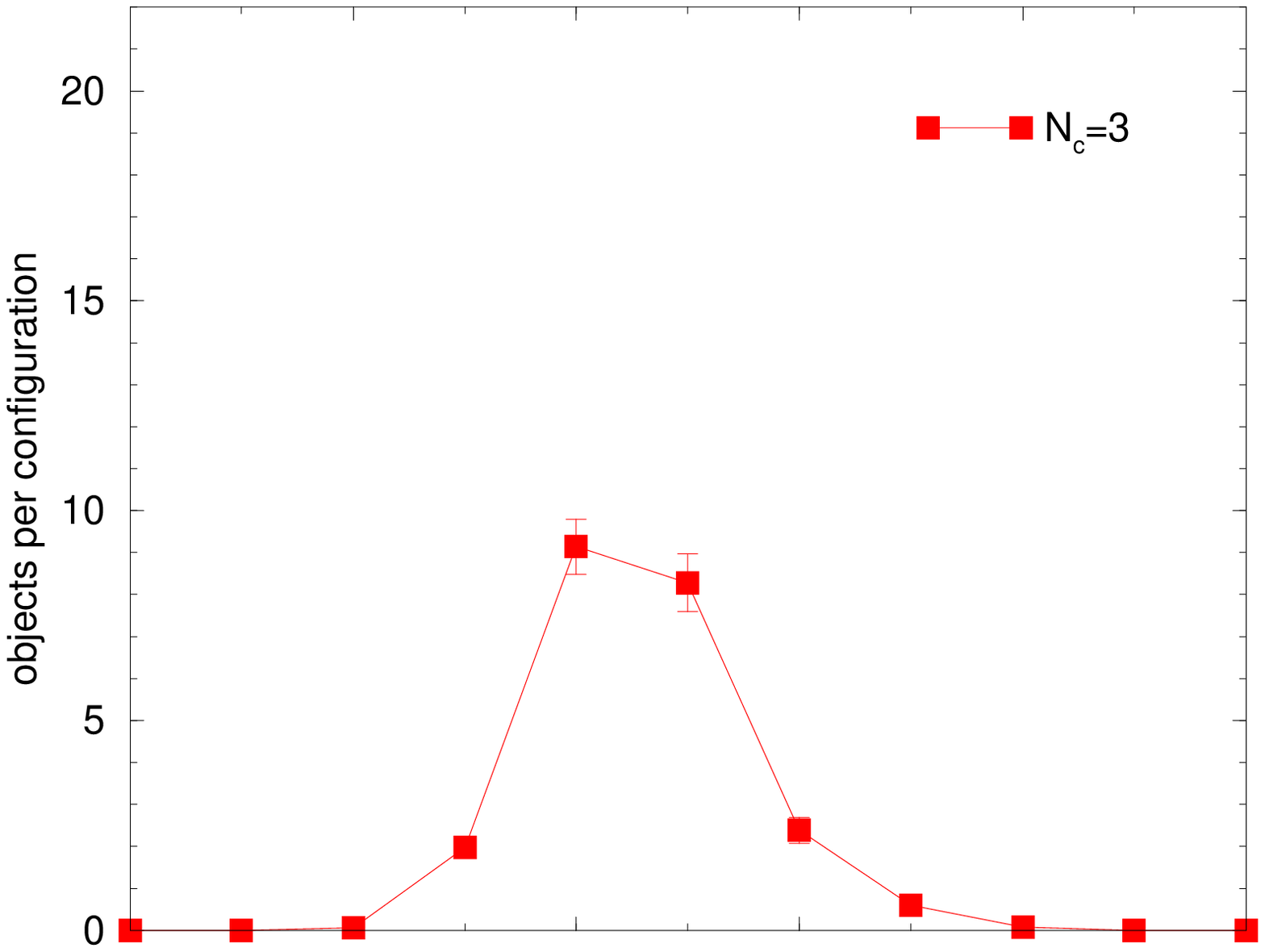} &
\includegraphics[width=7cm]{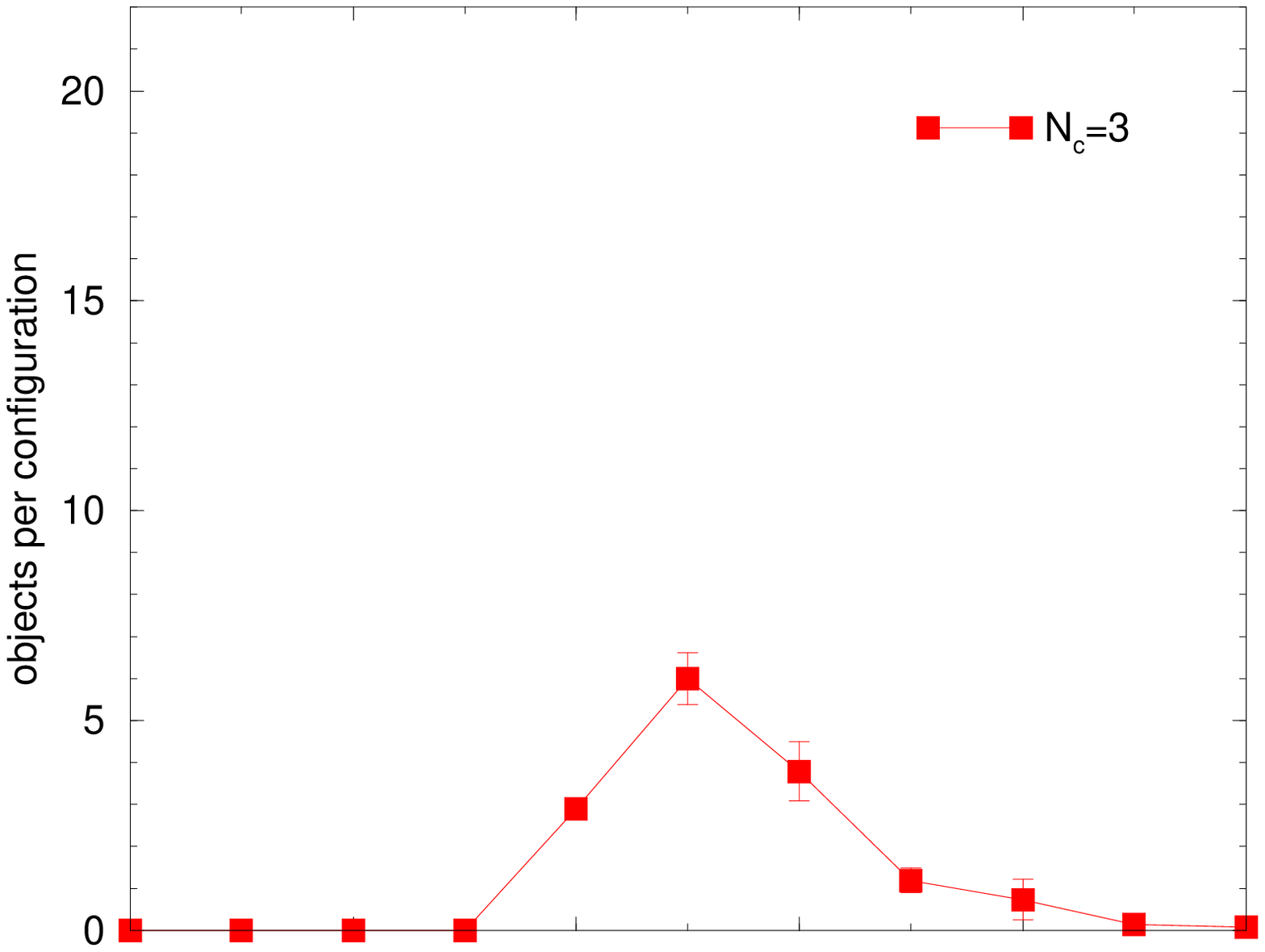}\\
\includegraphics[width=7cm]{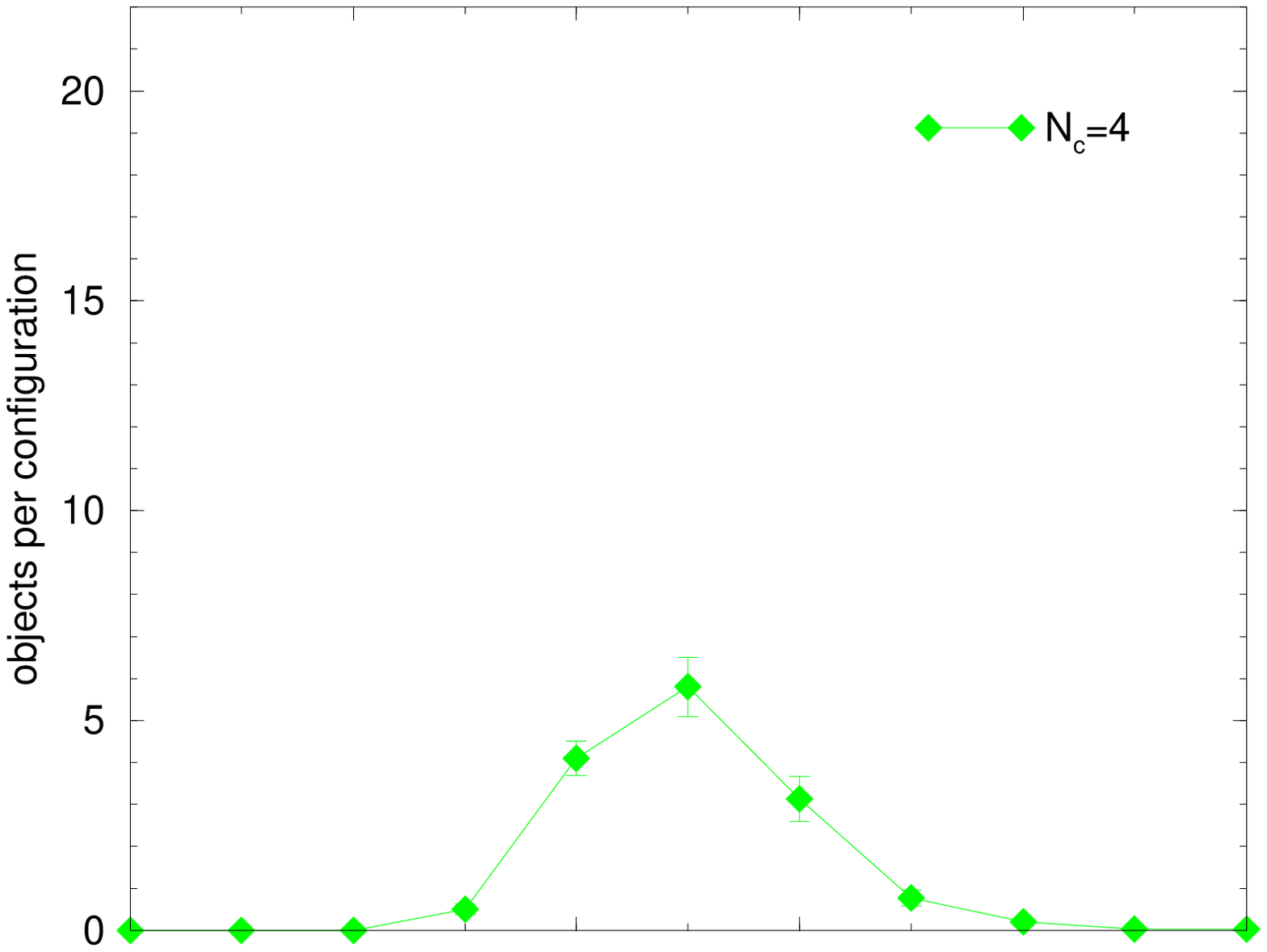} &
\includegraphics[width=7cm]{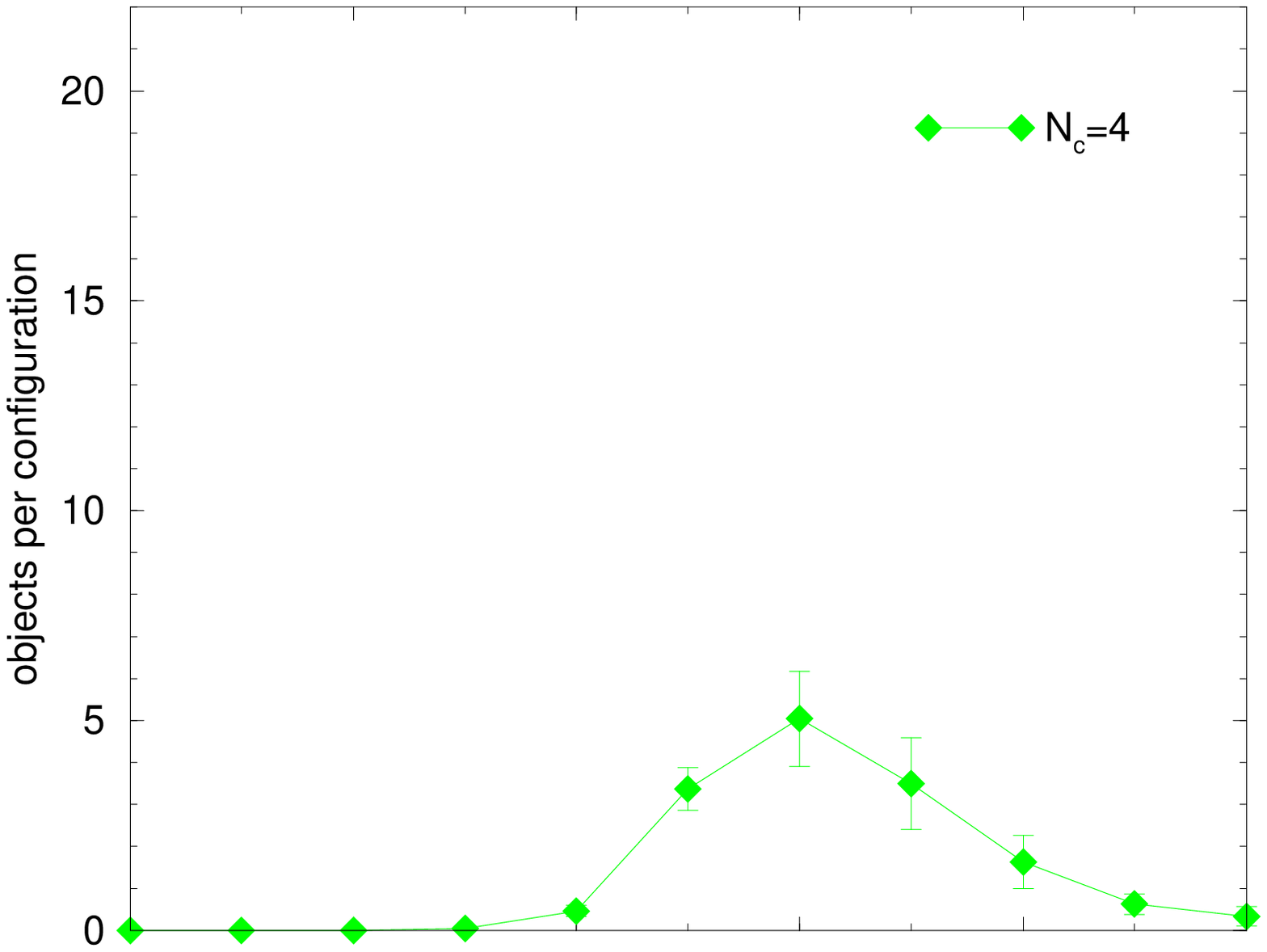}\\
\includegraphics[width=7cm]{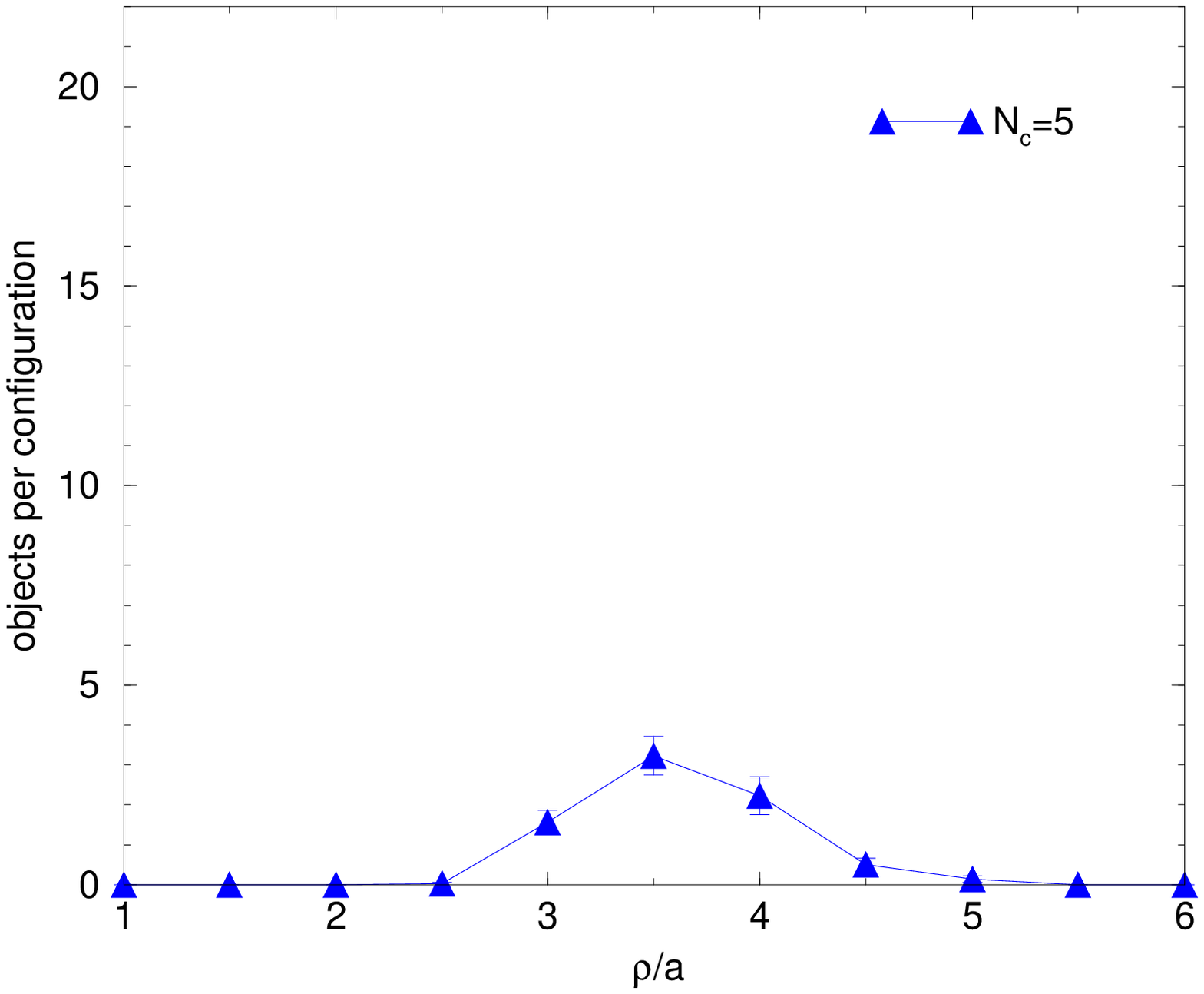} &
\includegraphics[width=7cm]{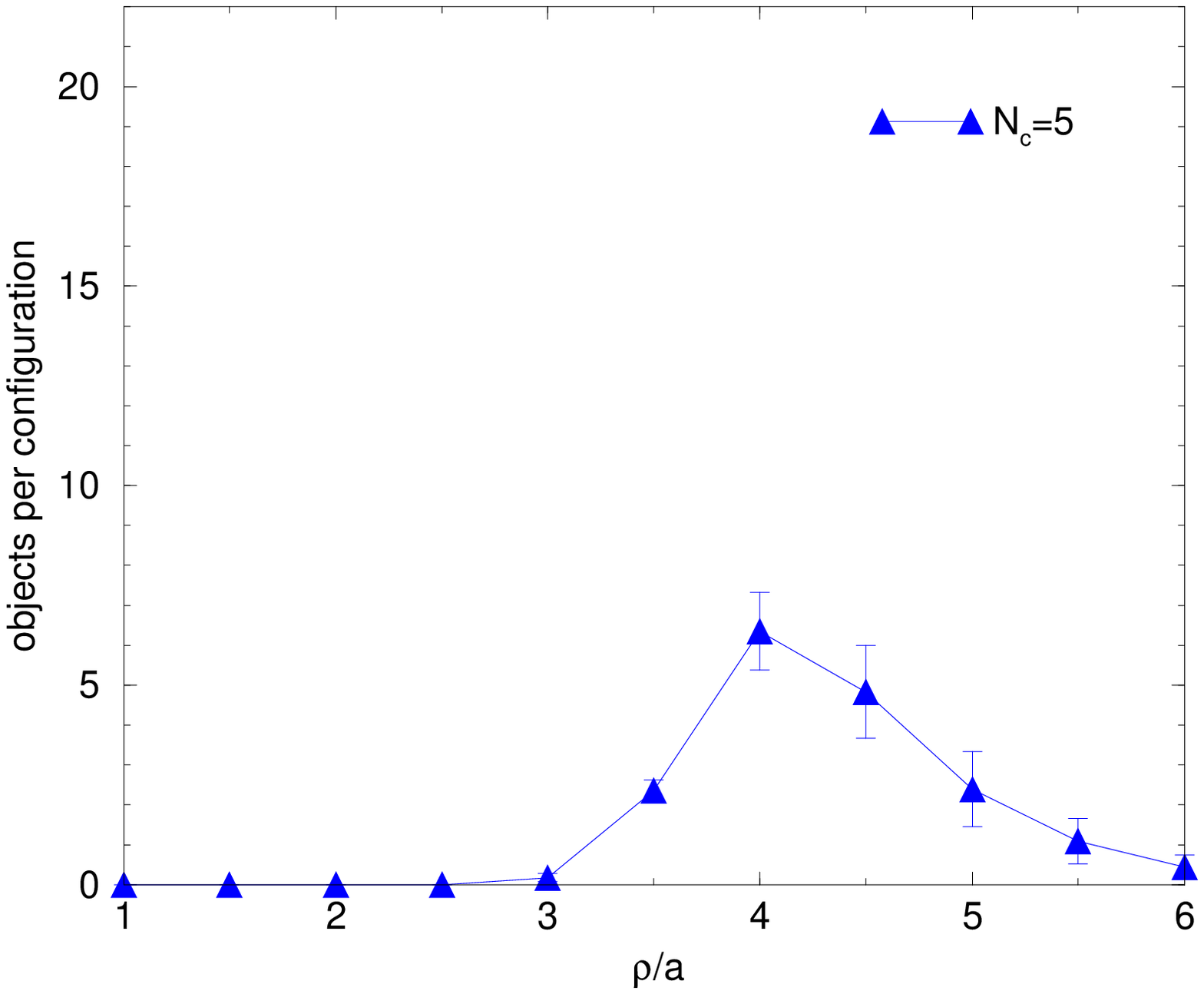} 
\end{tabular}
\caption[]{The size distribution of instanton-like objects as seen in
the cumulated pseudoscalar density of the near-zero modes with
$\lambda^2<0.1$ (left column) and the cumulated scalar density of the
zero modes (right column). The radius is calculated from the shape
around the peak value,
cf.~eq.~(\ref{eq:instanton_eigenmode_profile}).}
\label{fig:add_p_size}
\end{figure}

It is now interesting to look at the size distributions obtained from
identifying peaks in the \mbox{(pseudo-)}scalar densities of the
near-zero modes and the zero modes. In figure \ref{fig:add_p_size} we
show the size distributions for the eigenmodes of the different gauge
groups. The plots in the left column contain the distributions from
the non-zero modes below $\lambda^2<0.1$ while the plots in the right
column show the distributions from the zero modes. Here we use the
cumulated pseudoscalar density, cf.~eq.~(\ref{eq:cumulated_density}).
. The sum is taken over all the non-zero modes with $\lambda^2<0.1$ or
all the zero modes of a given configuration and we impose again
$r_{\text{min}}/a=2.0$. It is important to note that these size
distributions do not involve any cooling or smearing of the underlying
gauge field configurations at all.

Let us first concentrate on the size distributions calculated from the
cumulated pseudoscalar densities of the near-zero modes (plots in the
left column). The two main features
already observed in the distributions obtained from the topological
charge densites are essentially reproduced. Firstly, as we increase
$N_c$ from 2 to 5 the peak of the distribution moves slowly from about
$\rho/a=3.0$ to about $\rho/a=3.5$. So, as we increase $N_c$ the
regions with pronounced chiral density slowly become more
delocalised. Again, this complies with our conclusion from
the autocorrelation functions of the chiral densities in section
\ref{sec:correlation_functions}, where we observed that the typical size of
local chirality regions in the near-zero modes changes only slightly
with increasing $N_c$. Secondly, the suppression of small instantons
at large $N_c$ is again clearly visible. It is most evident for
instantons smaller than $\rho/a \lesssim 3.0$ and still apparent for
objects with $\rho/a \simeq 3.5$. By contrast, the number of objects
with $\rho/a \gtrsim 4.0$ is completely unaffected by an increase in
$N_c$.  In addition to these two features, we also observe that the
total number of objects seen in the chiral densities decreases
significantly as we go from SU(2) to SU(5) indicating that the
chiral densities of the eigenmodes become not only more delocalised but
simultaneously smoother for larger $N_c$.

The two characteristic features are also present in the size
distributions obtained from the cumulated (pseudo-)scalar densities of
the zero modes. The peak of the distribution moves from about $\rho/a
\simeq 3.0$ for SU(2) to about $\rho/a \simeq 4.0$ for SU(5) and we
observe a suppression of objects smaller than $\rho/a \simeq 3.5$ as
we increase $N_c$. Essentially all instanton-like objects with $\rho/a
\lesssim 3.0$ have disappeared by the time we arrive at $N_c=5$ or
even $N_c=4$. 

Of course there is a large freedom in how the distributions are
extracted from the eigenmodes. So, in addition to the distributions in
figure \ref{fig:add_p_size} one can also obtain size distributions
from the scalar densities of the near-zero modes as well as from the
uncumulated scalar and chiral densities of both the near-zero and
zeromodes. Moreover, one can look at the distributions for a large
range of varying parameters. For example, we may consider the peaks
from only the lowest few modes with $\lambda^2<0.03$ in each
configuration instead of all the modes below our cut-off
$\lambda^2<0.1$. The results further support our confidence in the
distributions so obtained: the shape of the distributions remains
exactly the same -- it is only the total number of objects that
changes. In fact it turns out that the total number of
objects seen in the densities differs considerably for the various
distributions. This is already clear by comparing for example the
distributions in figure \ref{fig:top_charge_size_distribution} with
the ones in figure \ref{fig:add_p_size}. By contrast, the qualitative
features, in particular the behaviour at large $N_c$, essentially
persist in all the distributions we looked at.

To summarise we can confidently say that we observe a shift in the
typical sizes of the bulk of instanton-like objects from about 0.36 fm
to 0.48 fm as we increase $N_c$ from 2 to 5 and a suppression of small
instantons with $\rho \lesssim 0.45$ fm at large $N_c$. 
  
%\clearpage
%%%%%%%%%%%%%%%%%%%%%%%%%%%%%%%%%%%%%%%%%%%%%%%%%%%%%%%%%%%%%%%%%%
\section{Conclusions}
\label{sec:conclusions}

There have been a number of lattice calculations, during this past
year, that have addressed long-standing conjectures concerning the
role of topology in driving the spontaneous breaking of chiral
symmetry. The resurgence of interest in these questions has been
largely driven by the recent development of lattice fermion actions,
such as overlap fermions, that possess exact topological zero modes
and good chiral symmetries at finite lattice spacing.

The main aim of this paper has been to perform an analysis for various
SU($N_c$) gauge groups in order to address directly an old puzzle that
has motivated many of these recent studies: instantons are expected to
disappear from the vacuum as $N_c\to\infty$ while chiral symmetry
breaking is expected to persist.

As a preliminary to addressing this question we needed to determine
how well the overlap Dirac operator resolves lattice topology for the
moderately coarse lattice spacings, $a$, with which we work. To
address this question we compared the fermionic topological charge,
$Q_f$, obtained from the number of exact zero modes, with the gluonic
topological charge, $Q_g$, obtained by first cooling the lattice field
and then integrating its topological charge density. In SU(2) and
SU(3) we found that $Q_f$ and $Q_g$ frequently disagree with each
other; but that any such mismatch becomes rapidly less frequent as we
go to larger $N_c$. At the same time we found, as observed earlier,
that the gauge vacuum rapidly loses all small instantons as $N_c$
increases, so that the distinction between physical topology and
lattice artifacts becomes unambiguous even for our modest value of
$a$.  The `clean' character of the large-$N_c$ vacuum enabled us to
show that in fact the disagreements between $Q_f$ and $Q_g$ arise
because the former does not resolve topological charges which are
smaller than about $\rho \simeq 2.5a$ (as calculated after ten cooling
sweeps). This demonstrates that $Q_f$ is indeed a good measure of
lattice topology: it excludes (near-)dislocations but includes all the
physical continuum-like topology, at least once $a$ is small enough
that the corresponding length scales are distinct.

The agreement between $Q_f$, which is calculated directly from the
Monte-Carlo generated lattice fields, and $Q_g$, which is calculated
from the fields only after they have been smoothened, also provides
reassurance that the process of cooling does not distort the
topological content of the lattice fields in some unexpected
way. Indeed we found that the agreement between the fermionic and
gluonic measures goes well beyond the global topological properties
and extends to the qualitative behaviour of the way the size
distribution of the topological charges changes as $N_c$
increases. This study confirms that at large $N_c$ there are
essentially no instantons with $\rho \lesssim 0.45$ fm.

Having established how well the overlap Dirac operator sees topology
on our lattice fields we proceeded to perform local chirality analyses
of the kind that had been previously performed for SU(3). We found
that as $N_c$ increases the low-lying eigenmodes rapidly lose their
definite chirality properties and our results are consistent with this
loss becoming total at $N_c = \infty$.

Since the gauge fields lose their smaller instantons as $N_c$
increases from $N_c=2$ to $N_c=5$, it is natural to ask whether this
might not be responsible for the change in the local chirality
properties of the eigenmodes.  However this does not appear to be the
case: we found that when we excluded from our SU(3) ensemble any
lattice fields with small instantons (the large majority), this did
not change the local chirality properties appreciably.

We also calculated the pseudoscalar densities of the low-lying
eigenmodes and looked at how these are correlated with themselves and
with the topological charge density. All the modes, but the zero-modes
most dramatically, become more delocalised as $N_c$ increases,
consistent with the increasing smoothness of the topological charge
density and, indeed, of the eigenmode densities.

A straightforward reading of our results would be that instantons, at
least those that are readily identifiable as such, do not survive at
large $N_c$ and do not drive the chiral symmetry breaking there. Such
a conclusion must however be regarded as very preliminary; local
chirality provides a limited probe of the properties of the vacuum. We
are now in the process of performing what we hope will prove to be a
more realistic analysis and one which we hope will provide a more
definitive answer to these interesting questions.

%%%%%%%%%%%%%%%%%%%%%%%%%%%%%%%%%%%%%%%%%%%%%%%%%%%%%%%%%%%%%%%%%%
\begin{acknowledgements}
We have enjoyed useful discussions with many
colleagues, and in particular with R.~Edwards, U.~Heller, D.~Leinweber
and H.~Neuberger. N.C. is supported by PPARC grant PPA/S/S/1999/02872
and U.W. acknowledges financial support from PPARC SPG.
\end{acknowledgements}
%%%%%%%%%%%%%%%%%%%%%%%%%%%%%%%%%%%%%%%%%%%%%%%%%%%%%%%%%%%%%%%%%%

%%%%%%%%%%%%%%%%%%%%%%%%%%%%%%%%%%%%%%%%%%%%%%%%%%%%%%%%%%%%%%%%%%
%\clearpage

%\clearpage
%%%%%%%%%%%%%%%%%%%%%%%%%%%%%%%%%%%%%%%%%%%%%%%%%%%%%%%%%%%%%%%%%%

\appendix
\section*{Appendix}

\begin{table}[htb] 
\begin{ruledtabular}
\begin{tabular}{r|rrrr}
  $f_V$ [\%]  & $N_c=2$ & 3 & 4 & 5 \\
\hline
    1.00  &  5.18(12) &  3.32(05) &  2.76(04) &  2.54(04)  \\ 
    2.00  &  8.86(17) &  5.90(08) &  5.01(07) &  4.66(06)  \\
    6.25  & 19.90(28) & 14.72(14) & 13.00(13) & 12.30(12)  \\
   12.50  & 33.09(34) & 25.27(17) & 22.86(18) & 21.76(18)  \\
\end{tabular}
\caption[]{Percentages of the wave function fraction $f_W$ as a
function of volume fraction $f_V$. The first column gives the volume
fraction of included lattice sites with largest scalar density while
the remaining four give the contribution of these points to the total
wave function for different $N_c$. All eigenmodes with $\lambda^2 <
0.1$ are included.}
\label{tab:locality_measure_for_all_Nc}
\end{ruledtabular}
\end{table}

\begin{table}[htb]
\begin{ruledtabular}
\begin{tabular}{r|cccc}
 $f_V$ [\%]    & $N_c=2$   &     3     &     4     &     5     \\
\hline
1.00    & 0.577(33) & 0.482(52) & 0.394(53) & 0.360(56) \\
2.00    & 0.539(28) & 0.447(40) & 0.365(37) & 0.333(48) \\
6.25    & 0.477(19) & 0.383(19) & 0.315(23) & 0.286(31) \\
12.50   & 0.438(15) & 0.343(10) & 0.286(15) & 0.259(24) \\
\end{tabular}
\caption[]{Weighted average value of the chirality angle $\langle |X|
\rangle$, cf.~eq.~(\ref{eq:average_Xn}), for different $N_c$ from all
eigenmodes with $\lambda^2<0.1$. The first column denotes the
volume fraction $f_V$ of included points.}
\label{tab:average_X}
\end{ruledtabular}
\end{table}

\begin{table}[htb] 
\begin{ruledtabular}
\begin{tabular}{r|cccc}
 $f_V$ [\%]    & $N_c=2$   &     3     &     4     &     5     \\
\hline
1.00    & 0.387(25) & 0.276(28) & 0.192(29) & 0.164(29) \\
2.00    & 0.348(21) & 0.243(21) & 0.169(20) & 0.144(24) \\
6.25    & 0.287(14) & 0.189(09) & 0.133(12) & 0.111(15) \\
12.50   & 0.250(10) & 0.158(04) & 0.113(07) & 0.094(11) \\
\end{tabular}
\caption[]{Weighted average value of the chirality angle $\langle |X|^2
\rangle$, cf.~eq.~(\ref{eq:average_Xn}), for different $N_c$ from all
eigenmodes with $\lambda^2<0.1$. The first column denotes the
volume fraction $f_V$ of included points.}
\label{tab:average_X2}
\end{ruledtabular}
\end{table}

\begin{table}[htb] 
\begin{ruledtabular}
\begin{tabular}{r|cccc}
 $f_V$ [\%]    & $N_c=2$   &     3     &     4     &     5     \\
\hline
1.00    & 0.209(17) & 0.110(10) & 0.059(11) & 0.045(10) \\
2.00    & 0.179(14) & 0.091(07) & 0.048(08) & 0.037(07) \\
6.25    & 0.134(08) & 0.062(02) & 0.033(04) & 0.024(04) \\
12.50   & 0.110(06) & 0.047(01) & 0.026(02) & 0.019(03) \\
\end{tabular}
\caption[]{Weighted average value of the chirality angle $\langle |X|^4
\rangle$, cf.~eq.~(\ref{eq:average_Xn}), for different $N_c$ from all
eigenmodes with $\lambda^2<0.1$. The first column denotes the
volume fraction $f_V$ of included points.}
\label{tab:average_X4}
\end{ruledtabular}
\end{table}

\begin{table}[htb] 
\begin{ruledtabular}
\begin{tabular}{r|cccc}
 $f_V$ [\%]    & $N_c=2$   &     3     &     4     &     5     \\
\hline
1.00    & 1.61(25) & 1.55(34) & 1.42(51) & 1.32(55) \\
2.00    & 1.52(21) & 1.46(27) & 1.32(38) & 1.22(48) \\
6.25    & 1.36(15) & 1.26(12) & 1.12(25) & 1.01(33) \\
12.50   & 1.24(12) & 1.11(07) & 0.98(15) & 0.87(26) \\
\end{tabular}
\caption[]{Value of the chirality cumulant $3 - \langle |X|^4
\rangle/\langle |X|^2\rangle^2$ for different $N_c$ from all
eigenmodes with $\lambda^2<0.1$. The first column denotes the
volume fraction $f_V$ of included points.}
\label{tab:average_cumulant}
\end{ruledtabular}
\end{table}

\begin{table}[htb] 
\begin{ruledtabular}
\begin{tabular}{ccccc}
$N_c$ & $\langle C_{Q,Q}(0) \rangle \cdot 10^{7}$ & $\langle
C_{\omega,\omega}(0)\rangle \cdot 10^{9}$ &
$\langle C_{\omega,Q}(0)\rangle \cdot 10^{8}$ & $\langle \tilde C(0)\rangle$  \\
\hline
2 & 4.27(35)  & 1.711(79) & 0.771(61) & 0.29(3)\\
3 & 2.34(16)  & 0.766(23) & 0.534(25) & 0.40(3)\\       
4 & 1.34(10)  & 0.509(14) & 0.351(16) & 0.43(3)\\ 
5 & 1.30(11)  & 0.413(10) & 0.305(14) & 0.42(3)\\
\end{tabular}
\caption[]{Values of the correlation functions $\langle
C_{Q,Q}(0)\rangle, \langle C_{\omega,\omega}(0)\rangle, \langle
C_{\omega,Q}(0)\rangle$ at the origin for the near-zero modes. The
last column gives the ratio defined by $\tilde C(0) = \langle
C_{\omega,Q}(0)\rangle / \sqrt{\langle C_{Q,Q}(0) \rangle \langle
C_{\omega,\omega}(0)\rangle}$. }
\label{tab:correlation_normalisation}
\end{ruledtabular}
\end{table}

\begin{table} [htb]
\begin{ruledtabular}
\begin{tabular}{cccc}
$N_c$ &  $\langle
C_{\omega,\omega}(0)\rangle \cdot 10^{9}$ &
 $\langle C_{\omega,Q}(0)\rangle \cdot 10^{8}$ & $\langle \tilde C(0)\rangle$  \\
\hline
2 & \phantom{1}7.19(35)  & 1.727(163)  & 0.31(4)\\
3 &           4.210(93)  & 1.117(58)   & 0.36(3)\\      
4 &           3.512(66)  & 0.737(50)   & 0.34(2)\\ 
5 &           3.174(47)  & 0.707(53)   & 0.35(3)\\
\end{tabular}
\caption[]{Same as table \ref{tab:correlation_normalisation} but for
the zero modes.}
\label{tab:correlation_normalisation_zeromodes}
\end{ruledtabular}
\end{table}

\end{document}